\documentclass[1opt]{article}
\usepackage{graphicx,amsmath,amsfonts,amssymb,dcolumn,amsthm,euscript}
\def\bea{\begin{eqnarray}}
\def\eea{\end{eqnarray}}
\def\be{\begin{equation}}
\def\ee{\end{equation}}
\usepackage{xcolor}

\newcommand\nn{\nonumber} 
\newcommand{\bq}{\begin{equation}}
\newcommand\eq{\end{equation}}

\def\bar{\overline}
\newcommand\pa{\partial}

\def\d{{\rm d}}

\usepackage{slashed}
\usepackage{amsmath}

\def\demi{\frac{1}{2}}

\def\red{  \color{black}}
\def\blue{  \color{black}}

\def\green{black}
\def\orange{black}
\def\orange{black}

\begin{document}
\title{\textbf{ 
  Leaf of  Leaf Foliation \\and
  \\
   Beltrami Parametrization  in   $d>2$ dimensional Gravity 
}}
\author{\textbf{
Laurent Baulieu
 }
\thanks{{\tt baulieu@lpthe.jussieu.fr}
}
\\\\
\textit{
%
LPTHE, Sorbonne Universit\'e, CNRS
 } \\
\textit{  4 Place Jussieu, 75005 Paris, France  }\\
}

\date{
$ $
%
}
\maketitle
%

\hskip 8cm \textit {  Dedicated to the memory of Isadore Singer,}

\hskip 8cm \textit{ my dear friend and collaborator.}

\vskip 1cm

\begin{abstract}

This work shows the existence of a $d>2$ dimensional covariant ``Beltrami vielbein" that generalizes
the $d = 2$ situation.
 Its definition   relies on   a      sub-foliation   $\Sigma^{ADM}_{d-1}= \Sigma_{d-3}  \times \Sigma_2$  of     the    Arnowitt--Deser--Misner~leaves of~$d$-dimensional    Lorentzian manifolds  ${\cal M}_d$. $ \Sigma_2$    stands  for the sub-foliating  randomly varying   Riemann surfaces in ${\cal M}_d$.  
   The       ``Beltrami $d$-bein" associated  to      any given  generic vielbein   of~${\cal M}_d$ is  systematically  
 determined by  a covariant gauge fixing of the   Lorentz~symmetry of the latter. 
   It       is parametrized by~$\frac {d(d+1)} 2$~independent fields  belonging    to  different categories. Each  one  has  a specific   interpretation. 
   The~Weyl invariant field  sector of the  Beltrami $d$-bein    
 selects     the~$\frac {d(d-3)}{2}$~physical local degrees of freedom of~$d>2$ dimensional gravity.
        The components of the 
        Beltrami $d$-bein are in   a  one to one correspondence with     those of the associated
      Beltrami  $d$-dimensional  metric.    The~Beltrami parametrization of  the Spin connection and of the Einstein action    delivers interesting expressions.   Its use~might~easier    the search of  new    Ricci flat solutions classified by the genus of the sub-manifold $\Sigma_2$. 
       A gravitational ``physical gauge" choice is introduced that takes advantage of the  geometrical specificities of the Beltrami parametrization.    Further   restrictions           simplify     the  expression of    the  Beltrami  vielbein  when  ${\cal M}_d$ has     a given   spatial holonomy. This   point is     exemplified in the case of ~$ d=8$~Lorentzian spaces with   
        $  G_2\subset  SO(1,7)$~holonomy.  The  Lorentzian results presented in this paper can be extended   to   the Euclidean~case.
%
%
%
\end{abstract}

\def\s{{s_{stoc}}}
\def\hl{{\hat l}}

\def\l{\lambda}
\def\ll{{\hat \lambda}}
\def\th    {\theta}
\def\thh{{\hat \theta}}
\def\ee{{\epsilon}}
\def\eeh{{\hat \epsilon}}
\def\dd{\delta}
\def\ddh {\hat \delta}

\def \pa{\partial}
 \def\l{{\lambda}}
  \def\g{{\gamma}}
   \def\T{{t}}
   \def\t{\tau}
  \def \d{\delta}
 \def\d{{\rm d}}
 \def\x{{q}}
 
 \def\ol{\o^{leaf}}
  \def\Rl{R^{leaf  }_2}
\def\Diff{\mathrm{Diff}}

 \def   \hg { {\hat g}}
  \def\g {{\sqrt{g}}}
  
  \def\bx  {
   {  {{\bar \xi}}^{(01)   } } }
   
  \def\bxm {
   {  {{\bar \xi}}^{(01)  \mu} } }
   
    \def\bxn {
   {  {{\bar \xi}}^{(01)  \nu} } }
   
     \def\L  {
   {  {{L }}^{(00)   } } }
   
   \def\Lx{     Lie_{\xi   }  }  
  
      \def\s{  \hat  s   }
          
    \def\hg{    {\hat g}} 
    
       \def\g{    {\sqrt  {g}  }}
 
 \def\v{ {\varphi}}
     \def\s{     s   }
     \def \vx {\vec{x}}
     \def\V{   {{\Phi}    }}
     \def \Vx { \V (\vx)}
     \def\vv{   {\vec {v}    }} 
     \def\vV{   {\vec {V}    }}
  \def \vX {\vec{X}}
 \newpage

 \def \m {\mu^z_{\bar z}}
 \def \mb {\mu_z^{\bar z}}

 \def \mo {\mu^z_{0}}
 \def \mbo {\mu_0^{\bar z}}
\def\N{\hat N}
 
  \def\={&=&}
 \def\w{{_\wedge   }}
  \def\w{{\wedge   }}

 \def\Z{Z}
 
 \def\Zb{{\bar Z}}
 
  \def\p{\pa_z}
  \def\bp{\pa_{\bar z}}
   \def\0p{\pa_{0}}
   
   \def\pZ{\pa_Z}
  \def\bZp{\pa_{\bar Z}}
   
\def \E{E^z}
   \def \Eb{E^\bz}
  \def\bz{{\bar z}  }
  
   \def\zb{{\bar z}  }
   
    \def\c{c^z}
        \def\cb{c^\bz}
    \def\e {e^z}
     \def\eb {e^\bz}
     \def\E {E^z}
     \def\Eb {E^\bz}
     \def\Eoo{E^0}
      \def\o {\omega}
     \def\vp {\varphi}
     \def\vpb {\bar{ \vp}}
     \def\bZ{{\bar Z}}

      \def\Eo{ {\cal{ E}}^z}
      \def\Eob{  {\cal{ E}}^\bz}
       \def\Eoo{E^0}
%
      
        \def\Do{ {\cal {D}}_o  }
  \def\Dz{ {\cal {D}}_z  }
    \def\Dbz{ {\cal {D}}_\bz  }
   \def\MD{\frac{1}{1-\m\mb} }
      \def\Dt{ {\cal {D}}_\tau  }
    
    \def \pbM  {   \begin{pmatrix}     }
     \def \peM  {   \end{pmatrix}     }
     
     \def \bM  {   \begin{matrix}     }
     \def \eM  {   \end{matrix}     }
     
%
%
  \def \id
{ \begin{pmatrix}
0
&
1
 \cr 
1
  &
0
\end{pmatrix}  
}

  \def \Mut
{ \begin{pmatrix}
1
&
\mb
 \cr 
\m
  &
1 
\end{pmatrix}  
}

  \def \Mu
{ \begin{pmatrix}
1
&
\m
 \cr 
\mb
  &
1 
\end{pmatrix}  
}
%
  \def \invMu
{ \begin{pmatrix}
1
&
-\m
 \cr 
-\mb
  &
1 
\end{pmatrix}  
}
%

  \def \invMut
{ \begin{matrix}
1
&
-\m
 \cr 
-\mb
  &
1 
\end{matrix}  
}
  
  \def \tMu
{ \begin{pmatrix}
1
&
\mb
 \cr 
\m
  &
1 
\end{pmatrix}  
}

   \def \invtMu
{ \begin{pmatrix}
1
&
-\mb
 \cr 
-\m
  &
1 
\end{pmatrix}  
}


   \def\DZ{\p-\mb\bp}
  
   \def\DZb{\bp -\m\p}
%
 \def  \mmb {1-\m\mb}
%
%
%
      \def\eps{   \epsilon}
  \def\eps{ { \color{red} { \epsilon}}   }
       \def\eps{   \epsilon}

        {{ \tableofcontents}} 
      \section{Introduction }
     
     A
       pillar  of    $d=2$ gravity  is      
      the          holomorphic   and antiholomorphic factorization  of 
 the Polyakov path integral  that     sums over  all     configurations of Riemann surfaces~\cite{polyakov}. 
Euclidean $d=2$ gravity  is a vast~subject. Its~so-called  
        Beltrami representation    expresses
    the    three  component   of the       bidimensional metric     
             $g_{\alpha\beta}$   in function of   its Beltrami differential  $\m$ and     its   conformal factor  $\Phi$. 
       The  squared infinitesimal  line length on Riemann surfaces   is        represented 
    as the  covariantly   factorized   expression $ds^2 \equiv g_{\alpha\beta} dx^\alpha dx^\beta   = \exp \Phi   ||   dz+\mb  d\bz   ||^2$.  The covariant    change of field  variables 
    $g_{\alpha\beta} \to\m,\mb,\Phi$
      enforces  ab initio all  relevant   path integral    factorization properties when  building the local quantum field theories of
      theories coupled  $d=2$ gravity.
      It clarifies  the      proofs of    many of  their  properties. 
   A   transparent    BRST invariant construction defines the  string theory gauge fixing     by    equating   the Beltrami differential~$\mb$~to the moduli of the string worldsheets  modulo the modular symmetry.    The     explicit holomorphic   and antiholomorphic  factorization  of the Beltrami representation  is useful for examining   
 the  classification of  the bidimensional     conformal anomalies, 
   the determination  of     the heterotic theories and of the  Liouville theory, etc...  See \cite{ Baulieu:1986hw} and  \cite{lazarini}      for   the definition  and  some   applications of  the Beltrami parametrization in  bidimensional  gravity and supergravity.

     The present work investigates    the   possibility   of  generalizing the  Beltrami  parametrization of the  bidimensional gravity in the case of higher  dimensional gravitational theories. It shows    that 
       the   
        $d=2$   case is  the   initial condition of   a    recurrence      that defines    such    a   generalization.
        
       The Arnowitt Deser Misner (ADM) paradigm is of great relevance when studying gravity. It~eventually~determines  the  famous  ADM     metric~\cite{Arnowitt:1962hi}. 
     Its fundamental     principle    is to   consider   any given  Lorentzian manifold~${\cal M}    _d$  as  a   set of  spatial  $(d-1)$-dimensional   leaves~$  \Sigma^{ADM} _{d-1} $~covariantly  ordered  by  the Lorentz time rather  than  as  a disordered set of points.  
     {
     The~present~work  extends this    framework. It   introduces a     refined leaf of leaf sub-foliation that      decomposes  $  \Sigma^{ADM} _{d-1} $ into  $ \Sigma_{2}\times  \Sigma_{d-3}$.  It is quite  obvious that   the  topological structure of  certain     manifolds~${\cal M}    _d$  can     globally complicate the definition of   such a sub-folation of  their ADM leaves but, in this work,     one only considers those that admit   at least locally a leaf of leaf decomposition.
       This~decomposition is the key to      determine    the   generalized    "Beltrami  parametrization"   for   the      $d$-bein and   the   metric of   ${\cal M}    _d$  for   $d>2  $ modulo the reparametrization invariance. Of~course the  appellation    Beltrami field is  taken  in an enlarged sense all across this article. }
     
     The  leaf of leaf  sub-foliation    of~${\cal M}    _d$   covariantly separates   the  coordinates  of~${\cal M}    _d$ as $
      \{ \t,
       z,\bar z,  x^1,.., x^{d-3 }   \}$.   $\t$~is the  real Lorentz time coordinate of ${\cal M}_d$. ($z, \bz$) is       the   complex coordinate  of    $ \Sigma_{2 }$          
        and  the  ${x^i }$'s    ($i= 3,..,d-1$)~are the $d-3$~real~spatial  coordinates   of 
        $ \Sigma_{d-3}$. This geometrical decomposition   goes  one step further than the      simpler ADM leaf decomposition   that     covariantly separates the     
       coordinates  of~${\cal M}    _d$ as    $ \{\t, \vec x\}$ but provides no inner structure for the leaves $\Sigma^{ADM}_{d-1}$.
       The Beltrami metric provided by  the  leaf of leaf decomposition  is  thus   a refinement  of the ADM metric for any given value of $d  >2$.  The fields that determine its  components    fall in different categories  classified   according to their Weyl weights, each one  with  a different gravitational interpretation.  The~Beltrami      $d$-bein    is covariantly   parametrized by the same $\frac {d(+1)}{2}$  fields as the  Beltrami $d$-dimensional metric.
 The whole formalism   can be adapted    to the situation     where  the primary  ADM~type foliation is done along  a   space-like direction. Thus, it might  also  be    of interest   in the context of    Kaluza--Klein   compactifications and also  for studying gravitational solutions  with various possible choices of~the~metric~signature.

       The leaf of leaf framework   $  \Sigma^{ADM}_{d-1} =     \Sigma_{2}\times  \Sigma_{d-3}$    suggests that  the  sub-leaves  $ \Sigma_{d-3}$    can  be perhaps  used to mathematically concentrate the gravitational physical degrees of freedom of gravity
      with possible interactions with  other states on the boundary $\pa \Sigma_{d-3} 
 $ (if  the latter exist).  In fact,   a well identified  set of  the Weyl~invariant   components  of the  fields that compose the    Beltrami $d$-bein  covariantly  expresses    the~$\frac {d(d-3)}2$~physical   gravitational  degrees of freedom.
 Moreover, one  can use the conformal gauge to gauge fix the~inner~metric~of~$\Sigma_2$. 
 The~remaining   fields  of the   Beltrami  parametrization can be identified  
 as    defining 
     the conformal factor, $d-3$~rescaling functions for  the spatial coordinates of~$\Sigma_{d-3}$,           both    ADM    lapse      and   shift functions and   the Beltrami   differential of~$ \Sigma_{2}$.    One can easily  check  $\frac {d(d+1 )  }2=  \frac {d(d-3)}2+ 1+(d-3)+1+(d-1)+2 $.

   The quite  appealing   physical interpretation of all the fields that  covariantly compose   the $d$-dimensional~Beltrami parametrization   could   help  for deciding   which field variables should be    quantized in  non perturbative quantum gravity.   Getting  ab initio the $\frac {d(d-3)}2$~gravitational physical degrees of freedom as a subset of the~$\frac {d(d+1 )  }2$~local  fields that parametrize the    components of the Beltrami  $d$-metric       is a progress   as compared to   their definition   in the ADM~formulation. York~shows  in  \cite{York} that for the latter  the~$\frac {d(d-3)}2$ physical  gravitational degrees of freedom are
 the  equivalence classes of  the   components of  the ADM leaf inner $(d-1)$-metric  defined  modulo   reparametrization and Weyl transformations.  The~difference is striking. In~the~leaf of leaf framework, a~local~definition~of~the~gravitational  field components that compose the  observables is enforced from the beginning  as a well  identified subset  of the   fields that   parametrize   the  generalized Beltrami  $d$-metric and are considered as fundamental fields.    In~the    ADM~formulation,  the~construction   of the observables implies a  non trivial    BRST invariant  quantum field theory gauge fixing process for extracting   the relevant Weyl invariant parts of  the ~ADM~components of the metric  as they are defined by York (for instance  by    using  an unimodular~gauge~choice~\cite{luca}).

   The  paper is written in         a bottom  to  top approach. It first shows 
      how the $d=2$~Beltrami parametrization  generalizes in  $d=3$.  Going   from the case $d=3$       to the  case  $d=4$          is    more involved   than going from~$d=2$  to  $d=3$. The technical reason is that  the off-diagonal part of the  Beltrami  vierbein  is  more complicated   
      than   that of the  Beltrami  dreibein.     Once this point is resolved and one succeeds in computing
  the  four dimensional Beltrami parametrization,       the logics   of the  inductive process   becomes   transparent. It allows  to systematize    the    construction of the    Beltrami  $d$-bein and   the corresponding  Beltrami metric   for all values  $d>2$.

   The  $d>3 $  situation   is        of greater     relevance for physicists since  both   for~$d=2$ and   $d=3$ the little group of~$SO(1,d-1)$~is too small to admit the propagation   of  Spin~$2$~particles. 
     The  four dimensional case~opens~new~perspectives. The~sub-foliation of its  three dimensional ADM leaves $\Sigma_3$ by   a    Riemann surface~$\Sigma_2$~ is simplest though non trivial since it provides  a  one-dimensional sub-manifold   $\Sigma_1$ and  one  may  heuristically   consider  that  both  physical helicity states of the graviton    propagate along  $\Sigma_1$.  In this particular case, the complex coordinates~$z$~and~$\bz$~of the~sub-foliating surface   $\Sigma_2$  share some resemblance with    the      light-cone coordinates  $x^\pm$ giving a simpler  and more symmetrical expression for the Beltrami~vierbein as shown in section 5. 
     
In fact,     the basic result of this  work     is   the generic    Beltrami parametrization of the $d$-dimensional  metric  that   is   established in     \eqref{ds2d}.  It reads   as follows :
\bea\label{para}
ds^2\=-N^2 \Big ( 
d\t   +\sum_{i=3}^{d-1}  \mu^\t _i dx^i \Big )^2
+
\exp \Phi  \Big   { |} \Big { |}
dz+\m d\bz +\mu^z_3 d x^3 +\ldots \mu^z_{d-1}  dt^ {d-1}+ \mu^ z_\t d\t
\Big { |}\Big{|}
^2
\nn\\
&&
+\sum_{i=3}^{d-1} \sum_{  j=3}^{d-1}    {N^i}^2 
 \Big
(
\mu^i_3dx^3
+... 
+\mu^i_{j-1}dx^{j-1}
+dx^i
+\mu^i_{j+1}dx^{j+1}+
... 
+
\mu^i_{\t }d\t
\Big
)^2. 
\eea
In this expression, one has  the antisymmetry properties $\mu^i_j= - \mu_i^j$ and  $\mu^i_\t =-\mu_i^\t $ and an easy counting shows that the metric \eqref{para} is  indeed  parametrized by   $\frac{d(d+1)}{2}$ independent  fields.  

   The found metric \eqref{para} coincides with  the standard Euclidean  bidimensional  Beltrami metric $ds^2= \exp \Phi || dz+\m   d \bz||^2$  when   $d\t=dx^i=0$.    Some
factorization properties occur    because   of   the      $z\leftrightarrow \bz$~symmetry    symmetry of    the    Riemann surfaces  
$\Sigma_{2}$  that    sub-foliate   the ADM leaves  $  \Sigma^{ADM}_{d-1}$ of    ${\cal M}_d$. 
The   paper   establishes the   transformation laws  under the   Weyl and   reparametrization symmetries  of   all  the  Beltrami fields   that compose  the metric~\eqref{para}. They  are expressed       under the form  of their BRST symmetry transformations. The latter  are  obtained by       generalizing    the  simpler    algebraic   technics    currently used in the bidimensional case \cite{Baulieu:1986hw}.

 The  $\frac {d(d-3)}{2}$  fields $ \mu_i^m \equiv   ( \mu_i^z, \mu_i^\bz,   \mu_i^j)$   build  a   geometrically  meaningful
   Weyl invariant  subset  of the fields     that figure in    the  definition of the Beltrami metric~\eqref{para}. 
 The claim  is that  their excitations~can  be   locally identified  (at least perturbatively)     
        as the  above mentioned  $\frac {d(d-3)}{2} $~gravitational  physical degrees of freedom that possibly  propagate  in   $d>2$~dimensional Lorentzian manifolds \footnote{For    $d=2$  and $d=3$, 
    the ghost loops  of   semi-perturbative quantum gravity   give opposite contribution to the closed loops   of  all   propagating       metric  field components  and no room is left for the propagation of  gravitational physical degrees of freedom.  Thanks to the BRST symmetry, this property  remains  true   whichever   (consistent)  gauge   choice one  uses to fix the path integral zero modes due to the reparametrization invariance. 
   For  $d>3$, there are  extra   loop contributions    for the $\frac {d(d-3)}{2}$ physical  degrees of freedoms. The   cutting rules of those   loops  induce   the    particle interpretation of   $d>3$ gravity theories.   
   The  covariant sub-foliation of  the  ADM  leaves  as  $\Sigma_{d-1}^{ADM}=\Sigma_{d-3}\times \Sigma_ 2 $  directly parametrize  these  gravitational  $\frac{d(d-3)}2$  physical degrees of freedom. 
    The latter point can be verified  by   expressing the Spin connection and  the Einstein  action in function of the Beltrami fields and by checking the resulting  propagators. This improves the  York  classical analysis 
        \cite{York} that identifies the gravity degrees of freedom in a non local way, as the   equivalence classes of the ADM leaf  metrics  defined modulo  Weyl invariance. In fact    \cite{luca}  underlines  that  stochastic quantization  of gravity   indicates quite naturally  the property  that the gravity physical observables are   defined by  the functional of the metrics defined modulo Weyl transformation and   that  a   BRST invariant gauge fixing of gravity  in an unimodular gauge  allows one to represent  observables   as functionals of    unimodular metrics. The use of the  covariant generalized Beltrami parametrization allows one to bypass this construction.}. 
       It   suggests that the dynamical gravity physical observables can be postulated as  being the functionals of these    local
 fields.  This proposition is consistent with York's predictions \cite{York}.
%

  The  generalized  Beltrami metric   \eqref{para} and   the associated  Einstein action  is  best understood  by using the vielbein  and Spin connection first order formalism.
  The paper shows that  
  the number ~$d^2$~of the components of  a generic $d$-bein $e^a_\mu$ can be covariantly     (i.e. by preserving the $\Diff_d$ symmetry)  reduced  down  to~$\frac{d(d+1)}2$~independent components  
   by       gauge fixing    the ~$ \frac {d(d-1)}{2}$~local freedoms offered   by  the  Lorentz gauge symmetry       $  SO(1,d-1)\subset  SO(1,d-1)\times \Diff _d$ of the  complete gravitational  local   symmetry.   
    By doing this appropriately, one  determines    the covariant  Beltrami  $d$-bein  that is to be displayed   in  \eqref{nbein}-\eqref{beld}. Its generic expression  remarkably   generalizes     the  Beltrami zweibein  formulae as  originally written  in~\cite{Baulieu:1986hw}.  
    The     $d$-Beltrami metric~\eqref{para} is then     computed     from  the  standard   quadratic relation between a vielbein and a~metric.   The   one to one   relation  between the  $ \frac {d(d+1)}{2}$~components of the   Beltrami  $d$-bein   and   of          the    Beltrami $d$-metric as well as  their  relationship   with  the ADM fields deepen the geometrical understanding of all     fields that parametrize the    Beltrami metric~\eqref{para}. 
   
  The   Spin connection in the Beltrami parametrization 
  is obtained      by  solving     covariant algebraic  constraints on the torsion 2-form
$T=de+\o\w e$ when $e$ is  the  Beltrami  vielbein.  The Einstein  action can then  be expressed as an algebraic   quadratic expression of the Spin connection modulo boundary terms.  In fact, the Beltrami parametrization      expresses the Einstein action under a form      analogous to   that of its  ADM expression but  with some further refinements. This paper concretely computes  the Beltrami Spin connection as well as the  Einstein  action  in  the three dimensional  case  and establishes the relevant linear  equations   satisfied by the  Beltrami Spin connection  in    four dimensions. These results  illustrate  the general methodology  for using the Beltrami parametrization.


            The   Beltrami   metric   formula   \eqref{para}   further simplifies if the space~like part of ${\cal M}_d$   has  a given  holonomy.    Extra~freedoms  occur   in this case and  add up to   those of the local Lorentz invariance. They   offer       more possibilities  to  reduce the number of independent fields involved in the Beltrami parametrization.  
   This paper  gives the  example of the
      $d=8$ space-times  with holonomy
        $G_2 \subset  SO(1,7)  $ where $G_2$ is  the   smallest exceptional rank 2 group.  $G_2 $ has   $14$ generators, so  that the  $28$ freedoms offered by      the  $SO(1,7)$ gauge symmetry get   enhanced  into $28+14 =42$ freedoms, allowing one to express  the $ d=8$ Beltrami metric  in a   simpler form. 
       The~formal resemblance  between  this   restricted $ d=8$ Beltrami metric  with that of   a generic   four dimensional   Beltrami metric   reminds other similarities         known to  exist     (though in a  very  different  context)  between     $d=4$      and $ d=8$ topological quantum field theories~\cite{bks}. Other examples are under investigation.


      The paper is organized as follows.
      Sections   2 and  3  are useful for a better   self-consistency of the whole   presentation. Section   2  summarizes  the geometrical  BRST methods for a better mastering of  the  reparametrization  symmetry at the quantum level.    
   Section   3  is a    
            reminder      of   the~$d=2$~Euclidean gravity  Beltrami parametrization methodology and fixes notations that   are to  be generalized in higher dimensions.  
The new results are presented in the further sections. 

 Section~4~details how the  $d=2$ Beltrami parametrization can be generalized   in   three dimensions. 
 It   displays  in this case the   computation of the    Spin connection  and of the  Einstein  action in the Beltrami parametrization.
 
   Section~5~explains the    four dimensional  case and its new features as compared to the case $d=3$ of  section~4.     5.1  and 5.2  also suggest  that using of the   Beltrami~parametrization might easier    the search of  new  four dimensional Ricci flat solutions. 
    When $\Sigma_2$ has  genus zero  the  Ricci flat solution of 
   \eqref{para}   is quite obviously  the Schwarzschild solution. When  the  sub-foliating   manifold $\Sigma_2$ has   the topology of a torus, the    Beltrami~parametrization  
       suggests that  the  so-called  axisymmetric Weyl metric  \cite{Weyl}    is a particular case of   
   the  
    Beltrami metric \eqref{para}    when  $\gamma$ is a complex constant  and      one imposes the existence of~two Killing vector~fields.
   
   Section 6~proves~\eqref{para} by computing  the  generic  $d$-dimensional  covariant Beltrami  $d$-bein and afterward  the Beltrami $d$-metric.  It~expresses   
   various considerations about the  physical  relevance   of the    sub-foliation of~ADM~leaves according to  $  \Sigma^{ADM}_{d-1} =     \Sigma_{2}\times  \Sigma_{d-3}$ and   identifies     the     $\frac    {d(d-3)}2$ propagating gravitational  physical    degrees of freedom  as the      subset  of    fields $\mu^m_i\equiv (\mu^z_i$,   $\mu^\bz_i$,  $\mu^j_i)$  for  $3\leq  i,  j\leq d-1$.    
It~explains~that      the rest of the  Beltrami   fields   $\Phi$,~$N$ and~$(\mu^z_\t,  \mu^\bz_\t $, $\mu^i_\t) $, $N^i$  and ~$\m,\mb$   in~\eqref{para}    are   respectively  related to  the conformal factor field, the  ADM   lapse  and         shift vector,  $d-3$ independent dilatation   factors  for each  spatial coordinates~$x^i$~of~$\Sigma_{d-3}$ and  the   Beltrami differential of $\Sigma_2$
at fixed $x^i$ and $\t$. 
    A~gravitational ``physical gauge choice" for the reparametrization invariance   is presented   that takes advantage of the  geometrical specificities of the Beltrami metric. In~this~gauge,   the Einstein action only depends  on the $\frac    {d(d-3)}2$ gravitational  physical    fields
and   the time lapse and shift functions.

 Section 7 indicates       further simplifications   of the Beltrami parametrization    that  may  occur   when the Lorentzian  manifold ${\cal M}_d$ has a spatial holonomy
 ${\cal G} \subset SO(1,d-1) $. This is exemplified  
    in the     case $d=8$
  with $G_2\subset SO(1,7) $~holonomy.  
   
   The conclusion     speculates about   about  a possible   stringy origin  of  the   Riemann surfaces $\Sigma_2 $ that  sub-foliate the ADM leaves  of pure gravity for     tentatively giving  better  perspectives  about the gravitational   path integral~definition.
%
 
  Appendix A    computes   the        linear equations that    determine    the  four  dimensional    Spin connection $\o(e)$    in~function~of  the Beltrami vierbein.  
  
   Appendix  B  solves them  for     three  dimensional   case.  

         \def\ix{i_\xi}
       
       \section{Reminder  of the  classical ghost  field unification  for gravity  }
        This   section    is  a reminder of  the     geometrical method for     determining   the  gravitational  BRST symmetry that is used in the further sections for  computing    the   BRST  transformation rules  of $d$-dimensional        Beltrami fields.

            The   gravity fields  are     the  vielbein     $e ^a$  and   the Spin connection~$\o^{ab}$  in      first order       formulation.  
            (Here and elsewhere   the  flat  indices $a,b, \dots$ are  Lorentz indices that run from $1$ to $d$ for ${\cal M}_d$.) Their field strengths are  the  torsion $T=de +\o\w e $    and the  Lorentz  curvature   $R=d\o+\o\wedge \o$.
          $\xi^\mu$ and   $\Omega  ^{ab}$  are   the anticommuting  reparametrization    vector ghost and the Lorentz symmetry 0-form ghost, respectively.   The     exterior derivative is 
        $d=
       dx^\mu   \pa_\mu $. 
         \cite{ml}~proves that  the    nilpotent~BRST~symmetry operation  $s$ of general relativity is defined  by      three    covariant  and geometrical ``horizontality   conditions".  Their     dependence on      $e$ and $\o$ and on  their    ghosts~$\xi $~and~$\Omega$~is  through     the ``ghostified" vielbein    $\tilde e^a\equiv \exp \ix  e^a    = e^a   +\xi^\mu e^a_\mu $ and        the ``ghostified"    Spin connection  
       $\tilde\o^{ab}   \equiv\o^{ab}+\Omega^{ab}$.  The gravitational BRST symmetry is in fact defined  as  follows~:  
       \bea \label{cbrst}
       \tilde e & =&  \exp \ix  \ e 
       \nn\\
       \tilde  T\ & \equiv& (d+s)  \tilde e +(\o+\Omega  )\tilde    e
       =\exp \ix  \  T 
       \nn\\
       \tilde  R& \equiv&  (d+s)     (\o+\Omega  )   +  (\o+\Omega  ) \w    (\o+\Omega  )   
       =\exp \ix   R=\exp\ix   (d\o+\o\w\o)   . \eea
   The  BRST equations \eqref{cbrst}   hold true independently  of the definition of the Einstein action and its possible gauge fixing   \cite{ml}.   If one imposes the covariant constraint
      $T=d\o+\o\w e=0$ the second line of \eqref{cbrst}  implies 
\bea \tilde  T=0. \eea
By combining    both   constraints~$ \tilde e= \exp\ix   e $ and~$  \tilde  T=T$,  one finds that the ghost number two component   of  the equation  $  \tilde  T=T$ implies
       \bea \label{sxi}
       s\xi= \xi^\nu\pa_\nu \xi  =Lie_\xi \xi \equiv \demi \{\xi,\xi\}.
       \eea
      One has the following identity   \cite{ml}
       \bea\label{expix}
       \exp -\ix (d+s)    \exp \ix   =d+s-Lie_\xi  + i_{s\xi-\xi^\nu\pa_\nu \xi}
     \quad 
       {\rm where  }    \quad   Lie_\xi \equiv [i_\xi, d]
      .\eea
    and 
            \bea   \label{hats} \exp -\ix  (d+s)    \exp \ix  =   d+\hat  s   \quad {\rm  where}      \quad \hat s \equiv s-Lie_\xi.\eea 
        Both properties    $s^2=0$  and   $\hat s^2=0$ are equivalent  since  $ s\xi= \xi^\nu\pa_\nu \xi$.
The relation   
 $ \exp- \ix (\o+\Omega) = \o+\Omega -\xi^\mu\o_\mu$ suggests   the   field   redefinitions  $\hat \o\equiv\o +\hat \Omega$ with 
         $\hat \Omega \equiv\Omega -\xi^\mu\o_\mu$.

         Left multiplication of       \eqref{cbrst}  by $\exp-\ix$      determines  the following   equivalent  reformulation  of the  gravitational   BRST     equations    
        \eqref{cbrst} that  
          directly define the action of  operation $\hat s$     on all fields       \bea \label{shat}
           \hat e\=e\nn\\
       \hat T  & \equiv&  (d+\hat s    )  e +(\o+\hat \Omega  ) \w  e
       = T=0
       \nn\\
       \hat R\    & \equiv&   (d+\hat s) (\o+\hat \Omega  )    +   (\o+\hat \Omega  )   \w  (\o+\hat\Omega  )   
       =   (d\o+\o\w\o)   =R.
           \eea
                The   $\xi$ dependence  is hidden in \eqref{shat}  owing to   the field redefinition $\Omega   \to \hat \Omega$ and $s\to\hat s$.   A superficial  look at the BRST symmetry operation $\hat s$ in ~\eqref{shat}~indicates that the   gravitational   BRST algebra  is formally that of   a  flat~space  gauge symmetry for  the     Lorentz invariance   with the redefined Lorentz   ghost $\hat \Omega \equiv \Omega-i_\xi \o$. 
    The  so-called ``covariant   BRST equations"    of  the Diff$\times$Lorentz symmetry are   $s\o=-D\Omega+ \ix R$  and   $s \Omega =- \Omega \Omega +\demi \ix \ix R$ and they   derive from    \eqref{cbrst}. 
    They   are    equivalent to  the simpler ones 
     $\hat s\o=-D\hat \Omega $    and 
     $ \hat s \hat \Omega =- \hat\Omega  \hat\Omega$,
     $\hat s e=-\hat \Omega e$ that   derive from  \eqref{sxi}.

      Proving  these properties boils  down to an appropriate use   of the operation $\exp i_\xi$  and  of the closure of the Poincar\'e Lie algebra.             The Appendix B of  the second reference in~\cite{ml} 
      (that also considers    the determination of the BRST symmetry of gravity  possibly coupled to gauge symmetries including  the cases  of      supergravity)
 analyses the structure of the graded algebra built by  the ensemble of the  generalized derivation operators  $d, s, \ix, Lie_\xi =[\ix,d], i_{s\xi},\hat s,$  etc... . This graded algebra      enlightens the     role  of the operation $\exp i_\xi$ when analysing the BRST structure  of theories coupled to gravity. 
        When one goes beyond the case of genuine gravity,  one     gets    a  gravitational nilpotent operation    $\hat s=s-Lie_\xi -i_\phi$, with
     $   \phi= {s\xi-\xi^\nu\pa_\nu \xi}\neq0$ and      $  s\phi=\{ \phi,\xi\}    $. 
     The definition  of the operation  $\hat s$   is only determined  by the internal  gauge symmetries of the system including local supersymmetry if it is involved.  The  reparametrization symmetry dependence is  systematically   hidden  in all formula thank's to the 
     appropriate
     use of  the   graded operation~$\exp i\xi$.
           \eqref{shat}  is in fact for the simpler example of pure gravity for which  $\phi=0$.   
          When  supergravity  is involved,  the  non vanishing ghost number~2~vector field  
     $\phi^\mu=i\bar \kappa \gamma^\mu \kappa$ occurs  where     $\kappa$ is 
  the  commuting spin 1/2 ghost field  of local supersymmetry. The non vanishing of  $\phi $  in  supergravity is due to the  modification 
  of the torsion $T $  by addition of the gravitino  field $\Psi$ dependent 
  term $\demi i \bar \Psi\gamma\w \Psi$,  giving  $     T\equiv     de+\o\w e + \demi i \bar \Psi\gamma\w \Psi$ and one has  
  a ghostified gravitino
  $\hat \Psi = \Psi +\kappa$~\cite{ml}.
   Having $\phi\neq 0$ implies the existence of   an extra generator   $i_\phi$  that  enlarges  the size of    purely      gravitational    BRST  super-algebra. It~modifies the~BRST~transformation of  the reparametrization ghost   $\xi^\mu$  in   agreement with  the fact  that the anticommutator    of two supersymmetries is a reparametrization. 
    In fact, many of the results presented in this paper  can be generalized to supergravity   ending up with  a $d>2$ "superBeltrami"  parametrization of  the gravitino that will be presented~elsewhere. 
  
   So, in general,  
   the set of  the graded differential operators, $d,s, \ix,Lie_\xi $, etc..  gets completed by the contraction operator  $i_\phi$,   $Lie_\phi
   =[i_\phi, d]$, etc... 
    The graded  differential operator  $i_\phi$   decreases   the form-degree by one  and increases  the ghost number by two,  so
    the value of  its  bi-grading is  equal to one.  Thus $i_\phi$  combines   consistently with $d$,~$s$~and~$Lie_\xi$.  One gets the following  generalization of ~\eqref{sxi}
     \bea \label{sxis}
       s\xi\= \xi^\nu\pa_\nu \xi   +\phi =  \demi \{\xi,\xi\} +\phi \equiv    \demi   Lie_\xi \xi +\phi  \nn\\
        s\phi \=  \{\xi,\phi\}\equiv  Lie_\xi  \phi          \quad \iff\quad  \hat s\phi=0.
       \eea
     The relation  \eqref{expix} becomes 
     \bea\label{expixsg}
       \exp -\ix (d+s)    \exp \ix   =d+s-Lie_\xi  + i_{\phi}.
       \eea 
     One has then  in full generality   that 
     \bea
    s^2=0\quad \iff\quad  \hat s^2  =(s-Lie_\xi  )^2   =Lie_\phi 
     \eea
 as    analysed in  \cite{ml}.

 The   fields that  define   in this work the generalized  Beltrami parametrization   of  the $d$-dimensional   metric and vielbein as well as  the      associated         ghosts (that are field redefinitions of the standard   reparametrization symmetry ghosts  $\xi^\mu$)    are to  be     conventionally called   ``Beltrami fields".      They   undergo the reparametrization BRST symmetry in  their own    non trivial way. 
 Getting their  BRST transformations   by    blindly      combining       the 
   standard  tensorial  gravity field BRST transformations    and the  mapping    between the Beltrami field and the  components 
    $g_{\mu\nu}$ of the metric would be an overcomplicated  
   task.  In contrast,  one can    elegantly
   determine    the   
    BRST transformations      of   all   Beltrami fields        by genuinely adapting    the  geometrical BRST horizontality       equations~\eqref{cbrst}  and  \eqref{shat}    to  the context of the  Beltrami parametrization.   The nilpotency of the    BRST transformations  acting on    the
    $\frac {d(d+1)}{2}$
         fields  that  parametrize both the 
    $d$-dimensional
    Beltrami 
     vielbein and  Beltrami   metric  is  ab-initio   warranted  in the geometrical  construction. It follows that     the     BRST transformations of the Beltrami fields         so directly      determined  by the geometrical method   are automatically  the correct~ones.       \def\ol{\omega}
    \def\Ol{\Omega}  
 \section{Reminder of the Beltrami parametrization   for the  Euclidean  $d=2$ gravity } 
  This section is a reminder  of  the known   Euclidean  $d=2$  Beltrami parametrization and its methodology for bidimensional gravity  \cite{ Baulieu:1986hw}.    It  defines notations   that  are useful  in the further  sections that are devoted to   the construction of the     $d>2$  Beltrami parametrization. 
    \subsection{Beltrami zweibein and $d=2$ metric       }   
   The fundamental  classical  field of  Euclidean $d=2$ gravity is   the bidimensional Euclidean metric $g_{\alpha\beta}$. It~represents   all possible Riemann surfaces that  must be   also  classified according to their   genus.    In real Euclidean coordinates $(x,y)$,   the   metric $g_{\alpha\beta}$ defines the   invariant infinitesimal squared length element   
 \bq \label{pna}
 ds^2=  g_{\alpha\beta} dx^{\alpha}   dx^{\beta}  =     Adx^2 +2Cdxdy + Bdy^2
 \nn
 \eq
  where  $AB- C^2>0$, $A>0$,  $B>0$.
 In complex coordinates one has $  z= x+iy$ and $\bz =x-iy$ and   the ``Beltrami  zweibein"~$e= (e^z, e^\bz) $  that is used  in
 \cite{Baulieu:1986hw}
  is   defined as 
 \bea\label{bel2}
\begin{pmatrix}  
 e^z   \cr      e^\bz
\end{pmatrix} =
\begin{pmatrix}   \exp \vp& 0  \cr  
 0&\exp  \bar  \vp
\end{pmatrix}  
\begin{pmatrix}  
 dz+\m   d  \bar   z  \cr       d\bar z+ \mb   d   z
\end{pmatrix} 
\equiv 
\begin{pmatrix}  
  \exp \vp  \    E^z   \cr      \exp \vpb\  E^\bz
\end{pmatrix} 
  \eea  
  so that $\E \equiv   dz +\m d\bz$ and $\Eb  \equiv   d\bz +\mb d\bz$. 
  The bidimensional  field   $\m$ is the Beltrami differential and~$\mb$~is~its complex conjugate.
  In a path integral formulation,  $\m$ and $\mb$ are treated as independent fields and one can consistently defines the
  reparametrization invariant  path integral
  measure $[d\vp][d\vpb][d\m][d\mb]$ provided  no   conformal anomaly occurs.
  \eqref{bel2} determines the following expression of  $g_{\alpha\beta}$   
  \bea\label{g^ab}
g_{\alpha\beta}=\demi \exp (\vp+\vpb)  \tMu \id \Mu=     
   \exp (\vp+\vpb) \begin{pmatrix}
\mb
&
 \frac     {1+\m \mb} 2
 \cr 
   \frac    {1+\m \mb} 2
  &
  \m   
\end{pmatrix}.
\eea
 One has therefore  the so-called Beltrami expression  for the line element
 \bea \label{MuMu} ds^2 =\exp (\vp+\vpb) | dz +\m d\bz|^2. \eea
One may call \bea \Mu\eea the $d=2$  $2\times2$ Beltrami matrix.   
The paper will generalize this matrix and its role in higher dimensions.
     The~unimodular part  $\hat g_{\alpha\beta}$   of   $g_{\alpha\beta}$    
and  the  inverse $g^{\alpha\beta}$ of  $g_{\alpha\beta}$ are    
   \bea
\hat g_{\alpha\beta}=\frac {i}{1-\m\mb}
\begin{pmatrix}
2\mb
&
    {1+\m \mb}{ }
 \cr 
      {1+\m \mb}{}
  &
 2 \m   
\end{pmatrix}, \quad \quad \det \hat g_{\alpha\beta} =1\nn
\eea
   \bea\label{g_ab}
   g^{\alpha\beta}=    
2 \frac {    \exp -(\vp+\vpb) }{(1-\m\mb)^2} \invMu \id \invtMu=     
 2 \frac {    \exp -(\vp+\vpb)  }{(1-\m\mb)^2} \begin{pmatrix}
-2\m
&
     {1+\m \mb}
 \cr 
      {1+\m \mb}
  &
  -2\mb   
\end{pmatrix}.
\eea\def\o{\omega}
The   zweibein \eqref{bel2} transforms tensorially  under the  $U_{Weyl}(1)\times U_{rotation}(1)\times \Diff_2$ gauge  symmetry. 

The~rotation  gauge symmetry $U_{rotation}(1) $      can  be gauge fixed  by imposing   $\vp=\vpb=\Phi/2$ in a way that preserves  the       $U_{Weyl}(1)\times  \Diff_2$ ~gauge symmetry. Modulo the rotation symmetry,  the   Beltrami  zweibein  \eqref{bel2} is thus   covariantly     determined    by the same    three  fields   $\m,\mb,\Phi $  as the  Beltrami $d=2$ metric.  The factorization between the holomorphic and antiholomorphic sector is obvious  in  \eqref{bel2} and  \eqref{g^ab}. 
The transformation laws under both   reparametrization  and the Weyl   symmetries   of 
$\m,\mb,\vp,\vpb$ will be shortly displayed. 
\subsection {  $d=2$  Beltrami Spin connection and $d=2$ curvature}
\def\gD{{\mathfrak{D}}}
   The Weyl invariant part of the   ``Beltrami zweibein"   \eqref{bel2}  is 
\bea\label{belrot}
\begin{pmatrix} \E\cr \Eb 
\end{pmatrix} 
 =\Mu
 \begin{pmatrix} dz\cr d\bz 
\end{pmatrix}.
\eea
The relevance of the Beltrami matrix  $\Mu$ suggests defining   the     differential operations 
  \bea\label{D2}\red 
   \pbM 
    D_z\cr D_{\bz}
    \peM
    \equiv 
    \pbM 
    \DZ\cr \DZb
    \peM.
   \eea
   $( D_Z,   D_{\bz})$ is basically  the dual basis of $(\E,\Eb)$ in a Cartan moving frame.
   The  exterior differential operator is  
   \bea d\equiv dz\p+d\bz \bp= \frac 1\mmb\E D_{z}   + \frac 1\mmb\Eb D_{\bz} . \eea
   The  Abelian 1-form  Spin connection   $\o= \ol_z  dz  +  \ol_\bz d\bz$  can   be   equivalently expressed as     $\o= \ol_Z  \E +  \ol_\bZ \Eb$~according to 
   \bea
\begin{pmatrix} \ol_Z    \cr \ol_\bZ  
\end{pmatrix} 
 =\frac 1 \mmb \invtMu
 \begin{pmatrix} \ol_z    \cr \ol_\bz  
\end{pmatrix} \quad  \leftrightarrow \quad 
\begin{pmatrix} \ol_z    \cr \ol_\bz  
\end{pmatrix}
= \red \Mut
 \begin{pmatrix} \ol_Z    \cr \ol_\bZ.   
\end{pmatrix}.
\eea 
The  $d=2$ gravity    Spin connection       satisfies the  algebraic  vanishing torsion condition
  \bea
  \pbM 
  T^z   \cr T^\bz 
   \peM
  \equiv
   \pbM  
    d\e -\o\w \e   
     \cr  
     d\e +\o \w\eb
   \peM
   =
   \pbM  
    \exp \vp 
     (    d\vp  \w  \E   +d \E   -\ol\w   \E)   
     \cr  \exp \vpb
    ( d\vpb  \w  \Eb   +d \Eb  +\ol\w   \Eb
   \peM
   =0
  \eea
whose  solution is     
\bea  \label{nablaphi}
\pbM   \ol_Z \cr  \ol_\bZ \peM
     =    \frac 1 \mmb   \pbM     { -  D_z \vpb+			        { {\bp \mb}} }{    {   }  }   \cr
       \cr   {  D_\bz \vp-       {\p \m} }{   }          
     \peM
   \eea
   \bea  \label{Spin2}
\pbM   \ol_z \cr  \ol_\bz \peM
 \=      
        \tMu 
        \pbM  \ol_Z\cr
       \ol_\bZ     
       \peM
       =-\frac 1 \mmb
       \pbM       { D_z \vpb -\mb D_{\bz}  \vp 
       -     {   \bp \mb   + \mb \p \m  } } { }
       \cr
       { - D_\bz  \vp +\m D_z  \vpb
      +     {   \p \m  -  \m \bp \mb } } { }     
       \peM.
   \eea
%
%
  The     Abelian  exact 2-form    Riemann  curvature  is 
   \bea\label {R2}
   R   =d\o=  dz  \w d\bz    \Big [ -
    \bp  (\frac   {D_{z}  \vpb -\mb D_{\bz}  \vp }\mmb )    
     -
      \p  (\frac   {D_{\bz}  \vp -\m D_{z}  \vpb }\mmb )   
   +\bp( \frac{\bp \mb - \mb \p\m }{\mmb})
   +\p( \frac{\p \m - \m \bp\mb }{\mmb})
 \Big]
.
\eea   \def\={&=&}
 One has   
    $
R _{z\bz}   \sim   \p\bp (\vp+\vpb)   -\p^2\mb-\bp^2\m)$
at the  first  non trivial order in $\m$ and $\mb$. 

One can gauge fix  the    Lorentz $U(1)$ symmetry and impose    $\vp=\vpb$. This gives
 \bea\label {R2}
  R _{z\bz} =     (\p,\bp ) \frac 1\mmb \pbM-  \m    &\frac  {1+\m\mb}2\cr \frac  {1+\m\mb}2   &-\mb \peM
       \pbM     \p  \cr \bp   \peM\Phi 
   {\red{   +  }}\bp (\frac{\bp \mb - \mb \p\m }{\mmb})
    {\red{+}}\p( \frac{\p \m - \m \bp\mb }{\mmb})
\eea
where $\vp=\vpb\equiv \Phi/2$. The  Beltrami Laplacian has surged  in the right hand side   of the last  equation. This   expression  of the curvature   is   relevant for a    Beltrami formulation of    conformal anomalies and Wess and Zumino terms in $d=2$ gravity~\cite{Baulieu:1986hw}.

\def \redt{{\red{\tilde}}}
\def \redt{{\red{black}}}

\def\Ol{\Omega{}}
\def\hol{{ \hat \omega} ^{leaf}}
\def\hOl{  {\hat {\Omega}}      }
\def\hol{{ \hat \omega} }
\def\hOl{  {\hat {\Omega}}    }
\def\O{  {  {\Omega}}    }
\subsection {$d=2$ gravitational Beltrami BRST symmetry}

 The standard definition of  the action  of  the nilpotent graded differential $s$  for the   BRST symmetry    of bidimensional gravity is    
 $sg_{\alpha\beta} =Lie_\xi  g_{\alpha\beta} +2\Omega_W   g_{\alpha\beta} $,  $s\Omega _W=  \xi^\beta \pa_\beta  \Omega _W  $   and $s\xi^\alpha=  \xi^\beta \pa_\beta \xi^\alpha$  where $\Omega _W$    is  the     Abelian   ghost of the Weyl symmetry.
 The nilpotency of  $s$ is equivalent to the closure and Jacobi identity of the Lie algebra of the rotation$\times$Weyl$\times \Diff_2$ local symmetry.
 The  reformulation   of the BRST symmetry transformations  within the framework of  the Beltrami parametrization  it to   provide      a   covariant 
 separation between the holomorphic and antiholomorphic    sectors  that   is enforced ab initio.
 
 The determination of  the BRST symmetry acting on      the  $d=2$  Beltrami  fields $\vp$, $\vpb$, $\m$ and $\mb$ is  directly  obtained  by adapting   the   generic  method displayed  in  section 2 to the Euclidean bidimensional case.  One  defines   
       the following  ghost and  classical field   unification of   the Beltrami~fields  \cite{ Baulieu:1986hw}~: 
  \bea
      E^z = dz+\m   d  \bar   z     &\to&   \tilde E^z \equiv  dz+\m   d  \bar   z+c^z  \nn\\ 
 \Eb = d\bz+\mb   d      z  &\to&  \tilde\Eb \equiv  d\bz+\mb   d    z+c^\bz \nn\\
d &\to& \tilde d\equiv  d+s.
\eea   
Consistency implies that   $\c$ and $\cb$ are related to  the  standard reparametrization ghosts  $\xi^z$ and $\xi^\bz$      by the   following  $\m$ and $\mb$ dependent  field redefinition     :
     \bea\label{belg}
  \pbM c^z \cr c^\bz    \peM
    \equiv \Mu
 \pbM \xi^z \cr \xi^\bz    \peM.
 \eea

 The   zweibein  $(\e,\eb) = (\exp\vp    \E,\exp\vpb\   \Eb) $   transforms   under the reparametrization   symmetry, the rotation  and the Weyl  gauge symmetries. Its   ghost unification     involves the conformal factors as  follows :
 \bea\label{unE}
  \pbM \tilde e^z \cr \tilde e^\bz    \peM
  \equiv
  \pbM
  \exp \vp &0\cr0&\exp\vpb
  \peM
 \pbM   \tilde \E  \cr \tilde \Eb   \peM
=\pbM
  \exp \vp &0\cr0&\exp\vpb
  \peM
 \exp i_\xi 
 \pbM     \E  \cr   \Eb   \peM
 \eea
 The exterior differential $d$ and the graded  BRST symmetry operation are unified as :
\bea\label{last}
{ \hat  d }  \equiv     \exp - i_\xi     \tilde d  \exp i_\xi  = d+s-Lie_\xi.
 \eea
   This     defines  the modified  nilpotent BRST operator $ \hat s\equiv s-Lie_\xi$   as generally explained  in section~2 \footnote {One may recall that the 
     equivalence  between the nilpotency  properties of $s$ and $\hat s $  is obvious from \eqref{last} as well as  the property   $s\xi^\alpha=  \xi^\beta \pa_\beta \xi^\alpha$. See \cite{ Baulieu:1986hw}  for the supersymmetric generalization of this in the specific  two dimensional case.}.   
   
  The  Abelian       Spin connection 1-form   $\ol$   that is the gauge field for the    bidimensional   Euclidean rotations  is    ghostified  by addition  of its   anticommuting   0-form   ghost    $\Ol$. This  defines the graded~1-form~$  { \tilde \o } \equiv {   \o } + \Ol$. Section~2~explains  that the existence of 
    the  operation  $\exp i_\xi$  naturally leads
  to the    {\red{      introduction of   the~graded 1-form $    \hol\equiv
 \exp-  i_\xi \  \tilde{ \o }    =  \ol+\hOl $}}, where     
  \bea \hOl= \Omega  -i_\xi \ol    = \Omega - \xi ^\alpha    \ol _\alpha . \eea
 Using      the redefined ghost   $\hOl$ instead of  the  Lorentz ghost $\O$  often simplify    formula and  
  defines the   covariant  graded differential operator $\hat D \equiv \hat d +{\hat  \o}   $. (An  analogous  treatment  applies  to the Weyl ghost  $\O_W$ by generalizing   $\hat D  =
  \hat d +{\hat  \o} 
  \to \hat d +{\hat  \o}  +{\hat  \o}_{W } $, but there   is no need to mention the Weyl invariance  in what follows.)
 
 As already mentioned, the  nilpotency  of the BRST symmetry   associated to   the  rotation$\times\Diff_2$ symmetry is nothing but the consequence of   to the~closure and  Jacobi identity of its    Lie algebra.   It    can be checked  by brute force, but the generic  geometrical construction  of section  2 warrantees  that the action of $s^2$ vanishes on all the  fields since it imposes conditions that are compatible with Bianchi identities.

 The unification between  the ghost and  classical field   within  the Beltrami parametrization framework   therefore conveniently  and consistently  defines  the   BRST symmetry  of all the  Beltrami fields for  the 
  ${\rm rotation}\times\Diff_2$ symmetry.  They read as follows   
 \bea
 \label{TT}
 \red
 \tilde T^z  & \equiv &\tilde D\tilde e    ^z =\tilde d \tilde e    ^z -\tilde \o \w   \tilde e^z  =  \exp i_\xi  T^z=0\nn\\
 \red
 \tilde T^\bz      & \equiv & \tilde D\tilde e    ^\bz =  \tilde d \tilde e^\bz +\tilde\o \w  \tilde e^\bz  =  \exp i_\xi  T^\bz=0\nn\\
 \red
{\hat { R}}  
 & \equiv &  \hat d   \hol      =    R, 
 \eea 
  Let us check the consistency of  \eqref{TT}.   
 Both   vanishing torsion conditions  $T^z=T^\bz=0$  are covariantly compatible with the  Bianchi identities $0=\hat D\w  \hat T=\hat R\w\hat e$ and  $\hat D\w \hat R=0$.     The  BRST transformations determined by the  ghost number decomposition  of  \eqref{TT} remain consistently  true when 
  $\ol$ is  an independent field or when it is  expressed   as  the solution  $\ol(e)$
  of the  condition $T=0$. Both equivalent   nilpotency   relations $(s+d)^2 =s^2=0 $ and 
   $(\hat s+d)^2 =\hat s^2=0 $  
 are  direct consequences  of  the   Bianchi identities  
    \bea \red\label{trucg}
    \tilde  D  \tilde T  \= {\tilde  d}^2   \tilde   e  +\tilde R    \w   \tilde e       \nn\\     
  \red    \tilde  D   \tilde  R  \= {\tilde  d}^2     \o   . \eea
     Since  one imposes    $ \tilde  T=T=0$ and $\tilde R=R$,   the     components  with ghost number larger than one  in the  right hand side of  \eqref{trucg}     must vanish. This implies     $\tilde  d^2     \tilde e = \tilde  d^2   \tilde  \o  =  0$.    Thus $s^2     \hat e = s^2   \hat  \o  =  0$ while  
   $s\xi^\alpha= \xi^\beta \pa_\beta  \xi^\alpha$  is a consequence of $\tilde  T =0$.

           Once the action of $s$ has been determined, the    reparametrization and rotation  symmetry  infinitesimal transformations of all the Beltrami classical fields  are    recovered  by 
 replacing  all  the ghosts by infinitesimal parameters in the  BRST transformations of the classical fields. 
      
 The terms with ghost number one  and two   of the horizontality condition 
 ${\hat { R}} ={  { R}} $   directly   determine   the  transformations  of    $\ol$    and its ghost $\hOl$ under the operation   $\hat s =s- Lie_\xi $.  They are
 \bea
 \hat s\ol\= -d\hOl \nn\\\  \hat s \hOl \=0.\eea
The latter equations determine the  action of $s$ on $\o$ and $\O$ by using the relationship between $\o$ and $\tilde \o$. 

Both                  $z \leftrightarrow  \bz$  symmetric  horizontality equations for $\tilde T$  in 
\eqref{TT}
decompose~as~follows,  where one uses the definition    \eqref{unE} of $\tilde e$ :
  \bea\label{brso}
  { \tilde T}^z   
   \=  \exp \vp \Big (  (d +s) (\E+\c)   +  ((d +s) \vp - \ol -\Omega)\w(\E+\c)  \Big) = 0
   \nn\\
{ \tilde T}^z   
   \=  \exp \vpb \Big (  (d +s) (\E+\c)   +  ((d +s)\vpb + \ol +\Omega)\w(\E+\c)  \Big) = 0.
    \eea
    The first  and the second equation     provide separately     the      BRST transformations of $\m,c^z, \vp$  and    of 
      $\mb,c^\bz,\vpb$ with a minimum effort. Indeed,
   both  ghost number~$1$~components    of \eqref{brso} that are  proportional   to $dz$ and $d\bz$     give respectively
   \bea \label{weilbrst}
  s \vp \=    \Omega    {\red{- } }  \c \ol_z        {\red{+ } }      \c\p \vp   {\red{+ } }  \p\c  
  \nn\\
   s \vpb \= -\Omega    {\red{+ } }   \cb \ol_\bz  {\red{+ } }     \cb\bp \vpb    {\red{+ } }  \bp\cb.
   \eea
   One has trivially $\tilde \E \w   \tilde \E   =   {\tilde E}^\bz\w  {\tilde E}^\bz=0  $. Thus,
the 
      multiplication  of   \eqref{brso}  by  $\tilde \E$ and  $\tilde \Eb$  gives a pair of equation with no      dependence on  $\o$, $\Omega$ and  $\vp$ and $\vpb$. They~are
    \bea\label{ede}
    {\tilde E}^z  \w (d+s) \w {\tilde E}^z\= 0  \quad  \iff  \quad  (d+s)  (\E +\c) =(\E +\c)\w \p(\E +\c)
    \nn\\   
    {\tilde E}^\bz \w  (d+s)  {\tilde E}^\bz\= 0 \quad  \iff  \quad  (d+s)  (\Eb +\cb) =(\Eb +\cb)\w\bp(\Eb +\cb).
    \eea   
  The   expansion in all possible  form-degrees and ghost numbers   of
   \eqref{ede}   determine    the   BRST transformations of all fields belonging  to the         Weyl invariant  sector of the zweibein.
   The    terms with ghost number zero  express    both   trivial   identities
         $d\E- \E\p\E=0$ and $ d\Eb = \Eb \bp\Eb=0$. The terms  with  ghost number one  and two express         the    $z\leftrightarrow \bz$ factorized BRST transformation laws of the Beltrami differential and the associated  ghost :
   \bea \label{smu}     s\m \=\red \bp \c  +\c\p \m-\m\p\c\nn\nn\\  sc^z   \=    c^z \p c^z \nn \\
   s\mb\=\red \p \cb +\cb\bp \mb-\mb\bp\cb  \nn\\ sc^\bz   \=    c^\bz \bp c^\bz. 
   \eea
 Both    equations     \eqref {ede}       
   $
   (s+d ) \tilde E^z -\tilde\E\p \tilde\E = 0 $ 
   and
    $(s+d )\tilde \Eb -\tilde\Eb\bp \tilde\Eb=0$      
  imply     that     $(s+d)^2\tilde\E =\tilde \E \w \tilde\E=0$ and  $(s+d)^2\tilde\Eb =\tilde \Eb \w \tilde\Eb=0$ since  the exterior product of 
 a  1-form by itself vanishes.  This     is the simplest  explicit check      that    $s^2=0$  on $\m,\mb,\c,\cb$.
 Its brute force verification  is   by   iterating  twice the $s$~operation  on all components of~\eqref {smu}.   
   Needless to say that the computation  of the BRST transformations~\eqref{weilbrst}~and~\eqref {smu} by combining   the  Beltrami changes of  field variables~\eqref{g^ab} and \eqref{belg}  with   
  $sg_{\alpha\beta} =Lie_\xi  g_{\alpha\beta} +2\Omega_W   g_{\alpha\beta} $  and $s\xi^\alpha=\xi\cdot \pa \xi^\alpha$
  is    tedious and time consuming as compared to the simplicity of  their derivation    by the geometrical method.

The above   derivation of the   nilpotent  $d=2$ gravity BRST transformations       uses    geometrical arguments  without giving  reference   to an invariant Lagrangian.  It   gives an interesting    perspective on the symmetries  of   $d=2$ quantum  gravity.    
Once the $s$ operation has been built, the determination of the classical 
   BRST invariant  Polyakov classical action  function of a string field $X$ of conformal weight zero  comes as a         second step. It is  defined as the $X$ dependent  ghost number zero part of the cohomology of $s$, namely
   $
 I_{Polyakov}(\mu,X) =  \int dX \w  ^*dX= \int d^2z \ 
 G_{\mu\nu} (X)
   \sqrt {g} g^{\alpha\beta} \pa_\alpha  X^\mu \pa_\beta X^\nu 
 =\int d^2z \   G_{\mu\nu} (X) \frac {1}{1-\m\mb}
 (\pa _z -\mb \pa_{\bar z})X^\mu   (\pa_{\bar z} -\m \pa_{  z})X^\nu$.~The    left (respectively right) conformal anomaly is the  ghost number one  part of the cohomology of $s$  that is 
 $\int  d^2 z  \pa_z  \c \pa_z^2 \m $ (respectively  $\int  d^2 z  \pa_\bz c^\bz  \pa_\bz ^2 \mb $).
 The    simplicity of these expressions    supports the idea  that  $\m,\mb, \Phi$   are the fundamental field variables of bidimensional gravity  both  at the classical level and  at the  quantum level  
 and one can verify that  $[d\Phi] [d\m][d\mb]$ is a consistent 
 classical measure for defining the Polyakov  path integral formulation. One can afterward gauge fix in a  BRST invariant way  the Beltrami differential  equal to some background ${\m}^{bg}$ (eventually a moduli) by addition of a $s$-exact term. ~The~gauge fixing action is actually part    of the trivial cohomology with ghost number zero  of  the BRST operation  modulo $d$-exact terms.  ${\m}^{bg}$  (respectively ${\mb}^{bg}$) identifies itself as the source of  the energy momentum component $T_{zz}$  (respectively 
 $T_{\bz\bz}$)  and    the modular invariance can be enforced  in string theory  in a way that respects the factorization  
 $z\leftrightarrow \bz$~\cite{bb}. All this  clarifies  both mathematically and physically   the occurrence and the     classification of bidimensional   gravitational  anomalies in a left-right factorized way, the construction  of Liouville Polyakov actions,~etc... . 
 More 
 details about this can be found in~\cite{Baulieu:1986hw}~that also also refers to  the supersymmetric generalization of all these results
 including for the case of left right asymmetric~theories.

     What should be remembered from this section is that  the Beltrami     field decomposition  is a covariant    revelatory of  the $d=2$ gravity factorization  properties.    The    elegancy    of         BRST symmetry equations~\eqref{ede}   that neatly separate
     both the  Weyl  invariant and non-invariant sectors of the theory
      justifies    the generalization    of   the  "Beltrami"  denomination    all around this  paper for $d>2$.  The  Beltrami differential~$\m$  is  to be completed in higher dimensions  by generalized entities such as $\mu^z _j, \mu^z_\t$,   $\mu^\bz _j, \mu^\bz_\t$,    $\mu^i_j, \mu^i_\t$  that  parametrize the Weyl invariant part of $d$-beins.    The $d=2$ situation looks   eventually  as    the  extremal   case   with no Lorentz time of a more general Lorentzian   $d>2$-dimensional framework.  
     The redefined     
ghosts $\c$ and $\cb$  \eqref{belg} deserve  being  referred to  as    $d=2$  ``Beltrami ghosts"  because of  ghostified unifications
 $dz+\m  d\bz  \to dz+\m  d\bz  +\c$ and $d\to d+s$. These~notions~and in particular   the Beltrami   ghostified unification will be      generalized      for   $d>2$ dimensional  gravity  theories that also involve Beltrami ghosts for the reparametrization symmetry.
 
   The     next    sections      introduce and detail  the  ``Beltrami parametrization"   for the cases of  3 and 4 dimensions  and afterward for  all $d>4$ dimensions.  
   The generalization  for      $d> 4$  is quite  abruptly          formulated       since   the combination of  both  cases  $d=3$~and~$4$~basically solve most of  the   technical subtleties  one must overcome to generalize the two dimensional case.
The key point is that  the generic possibility of a generalized  ``Beltrami parametrization"   for the metrics of 
 Lorentzian $d$-manifolds ${\cal M}_d$ is the consequence of  the possible sub-foliations
  $\Sigma^{ADM}_{d-1}   = \Sigma_{2}\times \Sigma_{d-3}   $
   of  the $d-1$  dimensional   ADM leaves   $\Sigma^{ADM}_{d-1}$ of   ${\cal M}_d$  by    Riemann surfaces. This~paper~calls this description     a leaf of leaf structure for  the formulation of the gravitational 
 interactions in   ${\cal M}_d$.  This quite refined  structure     cannot be guessed so easily  in dimension~$d=2$ since in this limiting case the gravity field  has no physical  dynamics and there is  obviously nothing to be sub-foliated. 

    

   \section{    $d=3$  Beltrami   gravity  }
    { The smallest dimension     for having a locally non trivial Einstein action is equal to three. For $d=3$  the sub-manifold $\Sigma_{d-3}$
 of  the leaf of leaf foliation   reduces to a point.  This  simpler case   is however  quite instructive   to    get a suggestive understanding   of  the   basic features of the Beltrami parametrization for all values of $d>2$. A~deeper insight
  is to be reached   in the next section for the case $d=4$      that     is the lowest dimension for which       $\Sigma_{d-3}$ is  a  non trivial  space.

The       possible couplings  of    $d=3$     gravity   to  matter are of great relevance  although     they  cannot generate  physical   propagating gravitons.  
  Moreover,   the      $d=3$   genuine   quantum gravity   QFT  has   non trivial mathematical interest  as  beautifully discussed  in  \cite{WittenCS} and  many other     publications including recent ones.  
    Some new   information about  the stochastic quantization   of~(Euclidean)~$d=2$~and~$d=3$~gravity      can be    reached by using the~$d=3$~Beltrami parametrization  determined   in this section.  This      will be explained  in a separate paper.

      In this section,   the   Beltrami parametrization of   $d=3$     gravity   is     built  with a  unified  notation   
      that treats at once  both     Euclidian and       Lorentzian   cases.
          All formula will be indexed by a  parameter $\eps=\pm 1$.  The~values~$\eps =-1$ and~$\eps =1$~are for  the Lorentz    and   the     Euclidean cases, respectively.

   \subsection {   Notations for   the   $d=3$    Beltrami gravity   and its     bidimensional leaves  
   }

   The     flat Lorentz (or Euclidean) indices   of  $SO(2,1)$ (or $SO(3)$)  Lie algebra  of the Lorentz (or rotation) gauge symmetry of the   $d=3$     
   Lorentzian (or Euclidean)  gravity    can   be  expressed either with  the real indices       $a= 0,1,2$ or    with   the     complex ones 
     $a= 0,z,\bz $ according to  the relationship  $z=x^1+ix^2$. 
 The couple  ($z$,    $\bar z$)     stands   for      the      complex coordinate  of   the        bidimensional   ADM  leaves $\Sigma_2$ of  ${\cal M} _3$, foliated  by the third direction with the  real~coordinate $ x^0$. The latter  (that   is    often denoted  as $t$ in this section)  is the extrinsic  Lorentz time  coordinate        in the Lorentzian case and 
 the  third ``spatial extrinsic"  coordinate    in the Euclidean case.

   The      fully 
    antisymmetric tensor  $\epsilon _{abc}$  allows one to identify any given   antisymmetric   tensor $M^{ab }$   to its     ``dual"~expression  $   M_a $ such that       $M^{ab }  \equiv   i  \epsilon ^{abc}M_{ c } $.  Thus,    the      dreibein $e^a$    and the Spin connection~$  \o ^{ab} $   can be expressed  as a pair of   1-forms    $\o ^a $ and  $e^a$  where 
   $   \o ^{ab}  = i \epsilon ^{abc}  \o _{c}   $, both valued in the fundamental representation of $SO(2,1)$ in the Lorentzian case   (or  $SO(3)$ in the Euclidean case).
    Upper  indices are lowered by the   $\eps=\pm 1$~dependent invariant flat metric~$   \eta  _{ab}  $. 
   
    The   2-form    field strengths  of $e  $  and   $\o$  are      the  torsion $T \equiv de  +g \o\w e$  and    the Riemann curvature   
 $R \equiv d\omega  +g \o\w\o$.   $g^2>0$ is  basically the gravitational constant.   One  chooses  from now  on   $g=1$ (equivalently, one can absorb $g$ in a redefinition of $\o$).
 
 When one expresses the spin connection as    $   \o ^{ab}  = i \epsilon ^{abc}  \o _{c}   $ and  one uses  real  indices $a=0,1,2$
 for the Lie~algebra of   $SO(2,1)$  (or $SO(3)$),
 the  three    components  of both $T$ and $R$ are    
   \bea
   \label{TR}
 \begin{matrix}
 T^0&=&de^0 +i (\o^1\w e^2 -\o^2 \w e^1) \\
 T^1&=&de^1 +i \epsilon (\o^2\w e^0 -\o^0\w  e^2) \\
 T^2&=&de^2 +i   \epsilon (\o^0\w e^1 -\o^1\w  e^0)
 \end{matrix}
 \quad\quad\quad
\begin{matrix}
 R^0&=&d\o^0 +  i \o^1\w \o^2  \\
 R^1&=&d\o^1 +  i \o^2\w \o^0 \\
 R^2&=&d\o^2 +  i \o^0\w  \o^1.
  \end{matrix}
 \eea 
  (The factors $i$ are  a mere consequence of the  definition  of $   \o ^{a}  $).
 
 If one uses  complex   Lie algebra indices  $a=0,z,\bz$,  the  dreibein and  the  Spin connection   read respectively  as 
$(e^0, \e, \eb )$ and $(\o^0, \o^z,\o^\bz)$.    By definition of 
   $z=x^1+i x^2$ and $\bz=x^1-i  x^2$ one has 
\bea\begin{pmatrix}
\o^z
 \cr 
    \o^\bz  
\end{pmatrix}=
 \begin{pmatrix}
1
&
i
 \cr 
1
  &
  -i   
\end{pmatrix} \begin{pmatrix}
\o^1
 \cr 
    \o^2 
\end{pmatrix}
\quad \quad \quad \quad \quad\quad
\begin{pmatrix}
e^z
 \cr 
    e^\bz  
\end{pmatrix}= 
 \begin{pmatrix}
1
&
i
 \cr 
1
  &
  -i   
\end{pmatrix} \begin{pmatrix}
e^1
 \cr 
    e^2 
\end{pmatrix}.
\eea 
The  torsion  and Riemann curvature    \eqref{TR}  are  then  expressed  as     
\bea
 \label{TRz} 
\begin{matrix}
 T^0&=&    de^0 -\demi \o^z \w e^\bz    + \demi   \o^\bz \w  e^z   \\ 
 T^z&=&     de^z -  \eps \o ^0\w e^z    +  \eps \o^z\w  e^0    \\
 T^\bz &=&       de^\bz +\eps  \o^0 \w e^\bz -\eps  \o^\bz \w e^0
 \end{matrix}
 \quad \quad  \quad \quad 
 \begin{matrix}
 R^0&=&   d\o ^0-  \demi  \o^z\w \o^\bz   \\
 R^z&=&d\o^z - \eps \o^0\w  \o^z    \\
 R^\bz&=&d\o^\bz + \eps\o^0 \w \o^\bz.    
  \end{matrix}
 \eea
The      Bianchi identities  are      
\bea
\bM
dT^0 \=\demi \o^z \w T^\bz    - \demi   \o^\bz \w  T^z 
  -\demi R^z \w e^\bz       + \demi   R^\bz \w  e^z   
  \\
dT^z \=\eps  (    \o^0  T^z      - \o^z  T^0  -R^0 \e  +R^z  e^0)
\\
dT^\bz \=\eps   (   -\o^0  T^\bz      + \o^\bz  T^0  +R^0 \eb  -R^\bz  e^0
\eM
\quad
\quad
\quad 
\bM
dR^0&=& \demi  \o^z\w R^\bz    - \demi   \o^\bz \w    R^z 
   \\
 dR^z&=& \eps(  \o^z\w  R^0    -  \o^0\w  R^z )  \\
 dR^\bz&=&  \eps (  \o^\bz \w R^0   +  \o^0 \w R ^\bz).
\eM
\eea
   The Einstein action is the integral  of the scalar curvature  density   $ \sqrt {-g} R(g_{\mu\nu})$.  In first order formalism, it is  a function of the dreibein $e$ and the Spin connection $\o$.  One has  the following     relations  \bea\label{einstein}
I_{Einstein }=\int d^3x \sqrt {-g}  R  (g_{\mu\nu}) \sim \int \epsilon _{abc} R^{ab}(\o)\w  e^c =\int 2 \eps e^0 \w  R^0 + e^z \w  R^\bz  +e^\bz \w  R^z.
\eea
The global  Lorentz (or   rotation) invariance  of the last term follows from  the fact that  the   flat   metric~is
\bea \eta_{ab}  = \pbM 0&1& 0\cr  1&0&0 \cr 0&0& 2  \eps  \peM.\eea
when one 
 uses complex indices. The        invariant scalar product    is
$A\cdot B\equiv A^a B_a  =\eta_{ab} A^aB^b= 2  \eps  A^0B^0 +    A^z B^\bz    +   B^z A^\bz  $       for        both      $A$ and   $B$   valued in the   fundamental representation  of either  $SO(2,1)$ ($\eps =-1$) or    $SO(3)$ ($\eps =1$).

The equivalence    between the second order and  the first order  Einstein  action in  \eqref{einstein} holds  true modulo the  vanishing torsion  condition $T^a=0$.  
The  $\o$-linear   equation $T^a=de^a+\o^{ab}\w e_b=0$
 determines  all components ~$\o^{ab}_\mu$~of  the  Spin connection 1-form~$\o$~as a  function of the dreibein  components $e^a_\mu$. Then, the relation between    $g_{\mu\nu} =e^a_\mu e^b_\mu    \eta_{ab}$ and $ e^a_\mu$ 
 holds true modulo 
   any given   Lorentz (or rotation)  gauge     transformation of~$e^a_\mu$. 
 Eventually,     any given well-defined  and complete  gauge choice  for  the   Lorentz (or rotation)  gauge symmetry  consistently determines~$e^a_\mu $ and   $\o^{ab}_\mu$    as  functions of  the metric   $g_{\mu\nu}$ and of  its derivatives  
  \footnote{
In general, the condition $T^a$  can be   imposed   as   a covariant constraint   independently  on  the gravity action that one uses, but  it must be often modified   by addition of a term $G^{abc }e_b \w e_c$, giving the condition
$T^a=G^{abc }e_b \w e_c$.   Finding the value of the Spin connection $\o(e,G)$ that solves  the  latter equation  amounts     to add  $G^{abc }e_c $ to the found solution  $\o^{ab}(e)$ for     $de +\o\w e=0$.  This manipulation  takes into  account possible couplings of gravity  to matter and auxiliary fields and $\cite{ml}$ explains the determination of the BRST symmetry with such gneralizations.  For  the  genuine Einstein  gravity the constraint $T^a=0$ is   an equation of  motion of the    Einstein  action  $\int  \epsilon _{abc} R^{ab}\w  e^c $. }.  
For a simpler expression of the $d=3$  Einstein action,  the 
part integrations  of  the    terms  that involve $R=d\o+\o\w\o$  in the right hand side of \eqref{einstein},
followed by the use of the constraint $de=-\o \w e$,  imply  $\int e\w (d\o+\o\w\o)\sim \int d e    \w  \o+ e\w\o\w\o  \sim\int e\w\o\w\o$,
modulo the  boundary term   $\int    d(e^a \w\o _a)$. The Lorentz indices must be adequately  contracted in these expressions  for the various    terms  related by the symbol $\sim$  such  as $e\w\o\w\o$. Once it is done,  the   $d=3$     Einstein action 
$\int \epsilon _{abc} R^{ab} \w e^c$
 can   be   identified  with     the following Lorentz (or rotation) invariant quadratic form of the Spin connection  
     \bea \label{e3}
     \label{Ea} I_{Einstein }=\int \epsilon _{abc} R^{ab} \w e^c
     \sim    \int   e^0\w \o^z(e)\w\o^\bz (e)  +e^z  \w \o^ \bz(e) \w   \o  ^0(e)+    e^\bz  \w \o^ 0 (e)\w  \o^ z(e).  
     \eea
 \eqref{e3} has no explicit  dependence  on the parameter $\eps=\pm 1$   that distinguishes  the     Lorentzian and    Euclidean cases.
 The  dependence  on $\eps$ is in fact hidden in the definition   \eqref{TRz}   of  the curvature $R$  and of  the torsion equation~$T=0$  that one must solve to compute the Spin connection $\o(e)$. The   quadratic  expression  \eqref{Ea}  of $I_{Einstein }$ 
 made  more   explicit in  the further equation\eqref{Einstein3d}) is   of course   in agreement with the  standard  expression   of  the  genuinely metric dependent Einstein~action     as  the     integral  of a quadratic form in the Christoffel coefficients modulo a boundary term.
 
}

  \subsection{   Beltrami      dreibein  }
 
 One now defines    ``Beltrami       dreibein"  as   the following  restricted triplet of 1-forms  $\e,\eb,e^0$ that are   covariantly  parametrized  by  the  left  and right conformal factors 
  $\exp \vp$ and $\exp \vpb$  and    the  Weyl invariant 
  fields $\m,\mb,\mo,\mbo$~:
    \bea\label {bel}
 \e  &=& \exp \vp \ (dz+\m   d  \bar   z    +     \mo dt)    \equiv      \exp \vp    ( E^z  +\mo dt )
 \equiv \exp \vp\ \Eo \nn\\
 \eb  &=& \exp \bar\vp \ (d\bar z+ \mb   d   z   +\mbo dt) \equiv  \exp \vpb (  E^\bz+\mbo dt)   \equiv \exp \vpb\ \Eob 
  \nn\\
  e^0  &=&  Ndt \equiv   \exp(\vp+\vpb)  \N dt.
\eea 
 The  Weyl symmetry  distinguishes  the  four     fields    $\m,\mb, \mo,\mbo$ and  the  three  fields  
 $ \vp, \vpb, N$   as belonging to  two different categories.    The former are Weyl invariant and the latter are not. 
It must be noted that the   definition~$e^0 =N dt$   for the 
  Beltrami dreibein     is an early signal  
  that the    first order formalism dreibein  parametrization~\eqref {bel}     anticipates  the   ADM paradigm and $N$ is closely related to the ADM  lapse.
 
The Beltrami dreibein that is defined by     \eqref{bel}  is parametrized  by seven independent   fields while a  generic and unconstrained  dreibein   is parametrized by  nine independent  fields. The definition   $e^0  =  Ndt$ amounts to  
   both  gauge fixing  conditions $e^0_z= e^0_\bz=0$  for such a generic dreibein.  It  can be performed  as a         partial gauge fixing  of the     $  SO(2,1)  \times  \Diff _3$ (or   $  SO(3)  \times  \Diff _3$ of the     local  symmetry that       preserves      a remaining  $U(1)\times \Diff_3$ local symmetry.  
     The  latter $U(1)$  symmetry can be further fixed by imposing  
 $\vp=\vpb=\Phi   / 2$. Eventually, the  Beltrami dreibein is parametrized by the six fields
  $\m,\mb, \mo,\mbo,  \Phi, N$ with a genuine~$   \Diff _3$~covariance.  The  fate of these  six three dimensional  `` Beltrami  fields" is to   parametrize the six  independent components of the  Beltrami  $d=3$ metric and the  Einstein action.

In fact, both conditions    $e^0_z= e^0_\bz=0$ in \eqref{bel}     exhaust in  a reparametrization invariant way   two  of the three  local freedoms    that are allowed     by the  local   Lorentz  (or rotation) gauge symmetry of the dreibein. But then, 
    consistency requires that    both  components of   the  Lorentz ghost ~$\Omega^z$~and~$\Omega^\bz$ must be constrained.  Indeed   $e^0_z= e^0_\bz=0$ must be imposed in a  way that respects the  BRST symmetry  of  the $\Diff_3\times SO(1,2) $ (or $\Diff_3\times SO(3) $)~local invariance that determines the classical physics, that is, one must  have  $se^0_z= se^0_\bz=0$    where~$s$~is the nilpotent   BRST symmetry operator that will be shortly computed in section 4.6. The definition of the $s$-transformations of the vielbein is such that 
the     Lorentz   ghosts  $  \Omega ^z$ and $  \Omega ^\bz$  must equate    certain    functions  of the diffeomorphism  ghosts when $se^0_z= se^0_\bz=0$.
    Analogously, when   the    third       freedom corresponding to the gauge  transformations  around the $x^0$ axis in the tangent space  is    used   to fix  covariantly $\vp=\vpb \equiv \frac  \Phi 2$,    the consistency  with the BRST symmetry  of this third  constraint  (namely the equation $s( \vp-\vpb )=0$)  fixes    the value~of~$  \Omega^0$
     in  function  of the diffeomorphism ghosts.  
      The      formulae that detail the values of     $  \Omega ^z, \  \Omega ^\bz$   and~$\Omega^0$   in  function   of the  Beltrami reparametrization ghosts are  expressed   right after~\eqref{sN} in section 4.6.  
      It must  be noted that  these formulae can be obtained in quantum field theory  as the equations of motion of the Lorentz antighost, when the gauge fixing
      $e^0_z= e^0_\bz=\vp-\vpb=0$
       is enforced  by adding to the classical gravitational action   an $s$-exact term.
       Altogether, the symmetry of the first   order gravity      reduces to    the genuine  $\Diff_3$ symmetry  of the     second  order gravity when it expressed in function of  the   Beltrami dreibein  \eqref{bel}  with  $\vp=\vpb=\Phi   / 2$ but it keeps the memory of the Lorentz gauge symmetry expressed in terms of the effective Lorentz ghosts that are  well-defined  functions of the reparametrization ghosts. 
 The  one to one   correspondence between the  six fields 
  $\m,\mb,\mo,\mbo, \Phi, N$  and  the  six components of the $d=3$     metric $g_{\mu\nu}$ will be shortly established 
 in    \eqref{ds3333}.  
  
The  first order formalism  that introduces     the Spin connection~$\o^{ab} $~that gauges the Lorentz (or rotation) symmetry and the vielbein $e^a$~as independent fields      greatly enhances  the  comprehension  of the  local symmetries of gravity since it enhances
the   $\Diff _3 $ symmetry of  the Einstein  action $\int d^3x \sqrt g R$ 
 into   $  SO(2,1) \times \Diff_3$ (or $  SO(3) \times \Diff_3$). In particular, it expresses quite simply~\eqref{Ea} whose structure  would be difficult to derive in the second order formalism by using the relation between the Christophel coefficients and the Spin connection. The  use of the     first order formalism    doesn't  change the physics due to the   covariant constraint  
$T=de+\o\wedge e$,  but the  
vielbein has more field  components  than the metric,  the difference being 
equal to   $\frac {d(d+1)}{2}$ that is the number  freedom offered by the Lorentz gauge symmetry.      A    quantum field theory  cost is     to be paid  in first order formalism:~a~consistent  and  BRST invariant gauge~fixing of the Lorentz  (or rotation) gauge symmetry of  the dreibein  must be done    to concretely  eliminate   its   redundancy of   field components  as compared to the  lower number of field components  of  the metric.  This gauge fixing must preserve    the reparametrization invariance.
  A~quite~obvious~advantage of  introducing the Beltrami parametrization of the dreibein~\eqref{bel}  is of providing such a geometrically   consistent   and   
$\Diff_3$ covariant   gauge fixing of the local  Lorentz (or rotation)  invariance.  It   fits the ADM framework and moreover it    provides a quite interesting 
     expression of the  Spin connection as it is  to  be verified~shortly.   
 
 Having explained these points, one can   go on.

 Consider   the parametrization \eqref{bel} with  $\vp\neq \vpb$.
 No   gravity~fluctuation can occur that  leads    to the singular value  $N=0$.  Therefore the   following     triplet of     one-forms 
  \bea\label{3bein} 
\pbM
 \Eo \cr
 \Eob\cr
dt
\peM\equiv
\pbM
 \exp -\vp\   \e \cr
  \exp -\vpb\  \eb\cr
dt   
\peM
=\pbM
\bM
1 &\m
\cr
\mb &1
\eM
& 
\bM
\mo
\cr
\mbo
\eM\cr
\bM
0
&
0
\eM
& 
1
\peM\pbM
dz\cr
d\bz\cr
dt
\peM,
\eea
determines an alternative and   meaningful basis for all $d=3$ differential form that is equivalent to the  standard  basis made of  $dz,d\bz$ and $dt$.

The definition   \eqref {3bein}  that generalizes  the bidimensional case  \eqref{MuMu}  introduces  the     following     generalized     three by three~``Beltrami matrix"  ${\cal M}$   and    its  inverse~${\cal M}^{-1}$ :
  \bea \label{trois}
  {\cal M} \equiv  \pbM 
\bM
1 &\m
\cr
\mb &1
\eM
& 
\bM
\mo
\cr
\mbo
\eM\cr
\bM
0
&
0
\eM
& 
1
\peM 
  \quad  \quad  \quad \quad 
{\cal M} ^{-1}=
\pbM
\frac 1\mmb\invMu & 
- \frac 1\mmb\invMu
 \pbM
 \mo
\cr 
 \mbo
 \peM
\cr
\bM
0
&
0
\eM
& 
1
\peM.
\eea
One has the following expression of  the  exterior differential operator $d=  dt\pa_0  + dz\p  + d \bz \bp  $
\bea\red   d   = {\cal E}^0 \Do  + \Eo \Dz  +  \Eob    \Dbz 
\eea
where $\red \Do$,   $\Dz $,  $  \Dbz $
are
  \bea  \label{Dz}\red 
 \pbM
 \Dz
 \cr
 \Dbz
 \cr
 \Do
 \peM
  =
 \frac 1 \mmb
 \pbM 
   \p- \mb\bp    
\cr 
\bp- \m\p       \cr
(\mmb)\pa_ 0-  (\mo- \m\mbo)\p
-
\pa_ 0-  (\mbo- \mb\mo)\bp
\peM.
  \eea       
  
   Analogously,    the    Spin connection   
       $
\o     =
dt  \o_0   + dz    \o_z +d\bz    \o_\bz    
 $ 
 decomposes  as 
 \bea\label{sco}
 \o= {\cal E}^0  \o_0   + \Eo   \o_Z   +\Eob  \o_\Zb  
 \eea  
 where   
$\red
(
\o_Z,
\o_\Zb,
\o_0)
=
(
 \o_z,
 \o_\bz,
\o_0
) {{\cal M}^{-1}}
$.
Using the  $d=3$     basis   $
\Eo$,    
$\Eob$,  
${\cal E}^0$ instead of  the basis  $dz$, $ d\bar z$, $ dt$ is quite~convenient. It corresponds  to a specific~choice of a Cartan $d=3$      moving frame. It
 easies         the   computation  of the~nine~components    $\o^a_{\mu}(e) $  of~$\o$~stemming   from the  three 2-form vanishing   torsion conditions $T^a=de+\o\w e=0 $   if one chooses to express them  as   the equivalent nine 3-form
  equation    $\epsilon _{abc}   e^a\w T^b=0  $. 
         Appendices    B displays the resolution of these~equations.
         Once $\o(e)$  is    determined as a function    of the Beltrami dreibein,
  the~quadratic~formula \eqref{Ea}   advantageously    determines       the  Einstein action without having to  compute     the derivatives of the Spin connection $\o(e)$ or the Christoffel symbols and its derivatives (see both further sections~4.4~and~4.5).

Each  one of the fields that parametrize the Beltrami vielbein \eqref{bel} has its  own specificity.
Both~Weyl and~Lorentz (or rotation) gauge symmetries  operate    on    
   $\vp+\vpb$  and  $\vp- \vpb$ by       shift operations.  The former~is by  the Abelian  parameter of  the Weyl symmetry  and  the latter~is  by   the  third  parameter of  the Lorentz (or rotation)    symmetry that must be gauge fixed to impose   $\vp=\vpb$. 
    The fields     $\m(z,\bz,t)$ and $\mb  (z,\bz,t)$    are  Weyl and 
  Lorentz  (or rotation) invariant and they stand for the Beltrami differential of    each  bidimensional  ADM~leaf~$\Sigma_2$ 
   for any given value $t$ ($t$  is    often denoted as   $x^0$ in this section). 
  The $2\times2$  matrix   within   the upper left  part of ${\cal M}$ 
  is  a  Beltrami  $d=2$ matrix   as in  \eqref{belrot} for  the $t$ dependent Beltrami differential      $ \m(z,\bz,t)   $. 
    $ N$  and  $(\mo,\mbo) $   are to  be   shortly  identified   as    the ADM    time  lapse  function  and       shift  vector of    the       ADM leaves $\Sigma_2$  of ${\cal M}_3$. 
The~BRST~transformation for the   Lorentz$\times \Diff_3$  (or rotation$\times \Diff_3$) 
local symmetry  of the Beltrami fields that is  computed      in section 4.6 checks that 
    $N$ and  $(\mo, \mbo)$    transform
  at fixed value of $t$ time  $t$ as    a scalar and  a vector field in $\Sigma_2$.

\subsection {          $d=3$   Beltrami  metric         } 
One may redefine the rescaled function   $  \N\equiv   \exp-\frac {\vp+\vpb}{2}N $ that is      Weyl independent. 
The         relation     $ g_{\mu\nu}
=  e^a_\mu \eta_{ab}e^b_\nu$    determines    the   $d=3$     ``Beltrami "  metric   
in function   of  the six independent fields $\m,\mb,\mo,\mbo, \hat N, \Phi=\vp+\vpb$. One gets
\bea\label{truc}
 g_{\mu\nu} \= 2\eps e^0_\mu e^0_\nu  +e^z_\mu e^\bz_\nu  + e^\bz_\mu e^z_\nu 
\nn\\  
\=
 \exp    (\vp+\vpb) \begin {pmatrix}
     \begin {matrix}2\mb & 1+\m\mb\cr  1+\m\mb  &2\m   \end{matrix} 
&
 \begin {matrix}  \mbo  +\mb\mo  \cr \mo +\m\mbo
\end{matrix}\cr 
\begin {matrix}   \mbo  +\mb\mo & \mo +\m\mbo    \end{matrix} 
& 2\eps \hat N^2  +2\mo\mbo 
\end{pmatrix}\nn\\
\=
 \exp (\vp+\vpb ) \begin {pmatrix}
 \tMu \id \Mu
& \tMu
 \begin {pmatrix}  \mbo  \cr \mo
\end{pmatrix}\cr 
\begin {pmatrix}  \mbo & \mo    \end{pmatrix} \tMu
& 2\eps \hat N^2  +2\mo\mbo 
\end{pmatrix}.
\eea

  A    slightly different  (but obviously equivalent)   matricial  computation directly provides    the infinitesimal 
 Lorentzian (or rotation) length of the infinitesimal line element  $ds^2= g_{\mu\nu}dx^\mu dx^\nu $   according to a    formula 
  that neatly separates   the fields   $N$ and  $\mo,\mbo$  that compose   the lapse  function and the shift vector. One has~indeed
\bea \label{ds2}
ds^2\=
\pbM dz \ d\bz\  dt 
\peM
\pbM
 \Mut  
&  
\bM  0  \cr 0
\eM
\cr 
\bM \mo & \mbo\eM
&1 
\peM   
\pbM
\exp\vp
&0&0
\cr
0&\exp\vpb&0
\cr
0&0&N
\peM
\begin {pmatrix}
\bM   0  & 1\cr  1  &0 \eM
& \begin {matrix}  0  \cr 0
\end{matrix}\cr
 \begin {matrix}  0 & 0
\end{matrix}
&
2  \eps
\end{pmatrix}
\pbM
\exp\vp
&0&0
\cr
0&\exp\vpb&0
\cr
0&0&N
\peM
  \pbM
 \Mu  
&  
\bM  \mo  \cr \mbo
\eM
\cr 
\bM 0 & 0\eM
&1 
\peM
\pbM dz\cr
 d\bz
 \cr dt
\peM
\nn\\
\=
\pbM dz \ d\bz\  dt 
\peM
  \exp(\vp +\vpb)
\pbM
 \Mut  
&  
\bM  0  \cr 0
\eM
\cr 
\bM \mo & \mbo\eM
&1 
\peM   
\pbM  
 0  &1  &0 \cr
 1&    0  & 0
\cr
    0 & 0
&
2\eps \hat N^2
\end{pmatrix}
  \pbM
 \Mu  
&  
\bM  \mo  \cr \mbo
\eM
\cr 
\bM 0 & 0\eM
&1 
\peM
\pbM dz\cr
 d\bz
 \cr dt
\peM.
\eea
The latter computation implies
\bea \label{ds33}
\demi ds^2\=  {
\exp (\vp+\vpb) \Big  (    \eps \hat N^2 dt^2 +   ( dz+\m d\bz+\mo  dt)(  d\bz+\mb d\bz+\mbo  dt)\  \Big )  }
.\eea
The Beltrami metric  is   
 independent~on~$\vp-\vpb $, as it must be the case.     The    third Lorentz freedom has been  left free in 
   \eqref{ds33}.   It can be   further   gauge fixed with    $\demi \vp=\demi \vpb\equiv \Phi$, giving 
   \bea\label{ds3333}
\demi ds^2=\red 
   \eps N^2 dt^2 +   ( dz+\m d\bz+\mo  dt)\exp \Phi (  d\bz+\mb d\bz+\mbo  dt ).
   \eea
   
\eqref{ds3333}  expresses  any given  generic three dimensional metric in function of   the six Beltrami fields 
   $N,\Phi,\m,\mb,\mo,\mbo$.  It quite evidently    generalizes     the bidimensional Beltrami metric 
   $\exp \Phi|| d\bz+\mb d\bz   ||^2$  that is reproduced for   $dt=0$.
   
 Some extra care is needed  to           precisely check the relation     between         $N$ and $
  =
  (\mo,\mbo)$ 
 with   the    ADM  lapse  function and the  shift vector field.
The  standard  ADM formula in real coordinates~is
 \bea ds^2= \eps    N^2dt^2+ (dx^\alpha+N^\alpha dt )g_{\alpha   \beta} (dx^\beta+N^\beta dt), \eea
  where $g_{\alpha   \beta} $ is   the $d=2$  leaf inner metric.  The   $d=3$  ADM   metric is  thus  
  \bea\label{admf}
     g_{\mu\nu} =
    \begin{pmatrix}
    g_{\alpha   \beta} &N_\alpha   
    \cr
    N_   \beta&   \eps    N^2+N_  \gamma    N^{\gamma   }
     \end{pmatrix}
     \quad
      g^{\mu\nu} =
     \begin{pmatrix}
    g^{ \alpha   \beta } +  \eps \frac {N^\alpha N^{\beta} }{N^2 }  &  -  \eps \frac {N^\alpha}{N^2}
    \cr
    - \eps \frac {N^\beta }{N^2}
 &  \eps \frac {1 }{N^2}
     \end{pmatrix}.
  \eea
  Since one uses      complex coordinates  in  the   leaves,      $g_{\alpha   \beta} \to 
\pbM  g_{zz}&   g_{z\bz}\cr       g_{z\bz}&   g_{\bz\bz}  \peM$.   The two by two   $g_{\alpha   \beta}$ figuring in
the    upper left  corner of \eqref{truc} can  be written as in section~2 according to   \bea
g_{\alpha\beta}=
 \exp \Phi
   \tMu \id \Mu 
 \  {  \rm with }\ 
g^{\alpha\beta}=\frac {\exp -\Phi}{(\mmb)^2}
 \invMu
 \id  
 \invtMu .
\eea 
Then, by comparing   of    \eqref{admf} and  \eqref{ds2}        one       finds  that 
both     ADM   shift  1-form  and shift vector $N_\alpha $    and   $N^\alpha$ are the following functions of  the dreibein parameters defined by~\eqref{bel}  
\bea
N_\alpha= \exp \frac { \vp+\vpb} {2} 
   \tMu
 \begin{pmatrix}
 \mbo   
 \cr
 \mo  
 \end  {pmatrix}
\quad \quad
  N^\alpha\equiv g^{\alpha\beta} N_\beta
=
 \frac {1}{\mmb}
 \invMu  
 \begin{pmatrix}
 \mo   
 \cr
 \mbo  
 \end {pmatrix}.
\eea
 This explains      how    the Beltrami type pair  $(\mo,\mbo)$     introduced    in \eqref{bel}  can be identified as   the (Weyl invariant)   shift  vector of   the  $d=3$     gravity ADM leaves.
 The generic  $d$-dimensional relation  between the ADM lapse and shift  functions and the Beltrami fields  is in \eqref{lapse} and
  \eqref{shift}.
   \subsection{       $d=3$   Beltrami   Spin connection } 
  \def\o{\omega} 
  \def\mo{\mu_0^z}
  \def\mob{\mu_0^\bz} 
    The  2-form torsion free condition  $T^a=0$     determine     nine  covariant linear equations that fix  the  nine field components
    $\omega^{ab}_\mu\sim  \omega^{a}_\mu $ of   the Spin connection $\o$ as functions of  the dreibein components.  Their content   can be    equivalently expressed as  the nine independent     3-form equations $ 
\epsilon_{abc}    e^a\w T^b=0$.  Both systems  $T^a=0$ or     $ 
\epsilon_{abc}    e^a\w T^b=0$   can be used to  determine the  the Spin connection  $\o^a_\mu$   in  function of the   Beltrami dreibein field  components. It is convenient to  choose the basis    $
\Eo$,    
$\Eob$,  
${\cal E}^0$  defined in  \eqref{3bein} for expanding the three components of   $\o^a$  as in     \eqref{sco}. One thus  parametrizes
the Spin connection  $\o$ as
         \bea  
\pbM
\o^z\cr
\o^\bz\cr
\o^0
\peM
\equiv 
\pbM
\o^z_Z
&
\o^z_\bZ
&
\o^z_0
\cr
\o^\bz_Z
&\o^\bz_\bZ 
&
\o^\bz_0
\cr
\o^0_Z
&
\o^o_\bZ
&\o^0_0
\peM
\pbM
\Eo\cr
\Eob\cr
Ndt
\peM
\eea
   Appendix B   solves the linear $\o$ dependence of the   three     equations $T^a=de^a+\o^a_b \w e^b=0$ in $ d=3$.  The result  is
 \bea
 \label {Tva3} 
\pbM
 \o ^z_Z= \eps\frac {\red    \exp \vp }{2 N }    (          \Do(\vp+\vpb)       - \frac{  \nabla_z\mo  +     \nabla_\bz\mob }\mmb )
&
\o ^z_\bZ=  \eps{\frac 1 N} \frac { \exp \vp  }
       {\mmb }  (\pa_o \m-  \nabla_\zb \mo )
       &
       \o ^z_0 = - 2{   \exp -\vpb} \   \Dbz  N 
\cr
  \o_\Z^\bz  =   \eps {\frac 1 N}  \frac{ \exp \vpb   }
      {\mmb  }   (-\pa_o \mb + \red \nabla_z \mbo )
&
\o^\bz_\bZ= \eps
\frac { \red \exp \vpb }{2 N }    (      -   \pa_0(\vp+\vpb)             +\frac {\nabla_\bz\mob  +       \nabla_z\mo }{\mmb}  )
& \o ^\bz_0=   2{\exp -\vp } \    \Dz  N
\cr
\o ^0_Z       = \eps (-\Dz \vp          +\frac   { \bp\mb}{\mmb})
&
\o ^0_\bZ=\eps (
\Dbz \vpb -\frac   { \p\m}{\mmb})
&
\o^0_0=
 \eps   \red  \demi (   \frac{ \nabla_z\mo  + \nabla_\bz\mob}{\mmb}  +  \Do (\vp-     \vpb))
\cr
\peM.
\nn\\
\eea
The Spin connection     components   \eqref {Tva3}    involve the   operation   $\nabla $       defined  in Eq.~\eqref{nablamuo} of         Appendix B whose action is defined as follows : 
\bea  \nabla_\bz \mo   &= &   \bp \mo + \mo  \p\m 
    -\m  \p \mo \nn\\
  \nabla_z \mob  &= &     \p \mo + \mob  \bp\mb 
    -\mb  \bp \mob
    \nn\\
    \red   \nabla_z \mo   &= &  \red \p \mo - \mbo  \p\m 
    -\mb (  \bp \mo   -\pa_0 \m)
    \nn\\
    \red   \nabla_\bz \mbo   &= &  \red \bp \mbo - \mo  \bp\m 
    -\m (  \p \mbo   -\pa_0 \mb).
    \eea
     One may observed   that   $ \nabla_\bz \mo  $ and   $ \nabla_\bz \mo  $
 are  formally  identical to  the      $\Diff_2$-BRST transformations  $s\m$  and   $s \mb$~\eqref{smu} where  one  replaces the    ghosts  
  $\c $ and  $\cb$ by the fields  
  $\mo$  and   $\mob$. 
 
      \def\DD{\mathbb{D}}
 The Spin connection  components are  independent on the  derivatives with respect     to $t$ of the lapse  $N$   and the shift vector components  $\mo$ and $\mob$. It trivially follows that  these field     don't have      a conjugate momentum at the classical level as can be checked   from the 
 expression  of  the $d=3$ Einstein action that is to be  shortly written in   \eqref{Einstein3d}.  
  \eqref{Tva3}~gets~simpler by gauge fixing   the  third      freedom of the    Lorentz invariance      with $\vp=\vpb\equiv \demi \Phi $.     
  Then,  the  Beltrami Spin  connection~becomes 
\bea
 \label {ointuitif3} 
\o^a_\mu=
\pbM
 \o ^z_Z=  \eps  {\frac 1 {2N}}\DD _{0} \Phi  
&
\o ^z_\bZ=   \eps  {\frac 1 N} \DD_o \m     
   &
       \o ^z_0 =   -  2\DD_\bz  N 
\cr
  \o_\Z^\bz  =    -   \eps {  \frac 1 N}\DD _0\mb  
&
\o^\bz_\bZ=-  \eps  {\frac 1 {2N}}\bar {\DD} _{0} \Phi
& \o ^\bz_0=   2        \DD_z  N 
\cr
\o ^0_Z       = -   \eps  \demi \DD_{z} \Phi       
&
\o ^0_\bZ= \eps \demi \DD_{\bz} \Phi 
&
\o^0_0= \eps 
 \demi     \nabla \cdot \mu_0 
\cr
\peM
\eea
where the definition   of the operator  $\DD$ is to  be read off from \eqref{Tva3}.
 
\subsection{ 
 $d=3$     gravity Einstein action in the Beltrami parametrization}
 \eqref{Ea} expresses   the  $d=3$     Einstein action   as  
\bea 
I_{Einstein} = \int L_{Einstein}dt\w dz\w d\bz= 
  \int     e^0 \w \o ^z \w  \o^\bz 
    +  \e \w \o ^\bz\w\o^ 0    +\eb     \w \o ^\bz\w\o^ 0
\eea
 modulo boundary terms proportional to   $\int d(\o^a\w e_a)$. Since   $e^0=Ndt$, one has 
\bea
I_{Einstein} =
 \int dt dz \bz (\mmb)   N
  \Big(  {\red {- }}
\o^z  _   \bZ  \o^\bz_Z{\red +}\o^z  _   Z         \o^\bz_\bZ
+ \frac {  \exp\vp }{N}  (\o^\bz _\bZ       \o^0_0   -\o^\bz _0       \o^0  _\bZ)
+   \frac { \exp\vpb}{N}   (\o^z _Z       \o^0_0  -\o^z _0      \o^0  _Z)
\Big).
\eea
For $\vp=\vpb$, one thus  gets
 the following   expression of the Beltrami  Einstein action modulo the  boundary term~$\int d(e^a\w\o_a)$~:
\bea \label{Einstein3d}  I_{Einstein}=
 \int   dzd\bz dt  ( \mmb ) &  \red \Big[ \exp \Phi
  \frac {{1} }{4N}  ( \Do\Phi -  \frac {  \nabla_z\mo +   \nabla_\bz\mbo  }{\mmb})^2 
 - \exp \Phi   \frac{ 1}{N(\mmb)^2}
  (\pa_0\m -\nabla _\bz \mo)(\pa_0\mb -\nabla _z \mbo)
  \nn\\& \red \quad 
 +\eps  \Dz N  ( \Dbz \Phi -2\frac {\p \m}{\mmb})
    +\eps \Dbz N  ( \Dz \Phi -2\frac {\bp\mb}{\mmb}  )
    \Big ].
  \eea
 
This expression has the   ADM structure. The terms proportional to~$1/N $ and $N$ (after a part integration) correspond respectively  to  the   ADM kinetic energy and potential energy. The latter is    the product of the   lapse function  by  the intrinsic curvature       of the $d=2$ leaf modulo boundary terms.

\subsection {  $d=3$     gravity  Beltrami BRST symmetry  }
\def \et{{\hat e}}
\def \ot{{\hat \omega}}
\def \Ot{{\hat \Omega}}
\def \O{{  \Omega}}
\def \O{{ \hat  \Omega}}

\def \Et{       { \cal  E  }    }
\def \Eth{      \tilde  { \cal  E  }    }
\def \Th{       { \tilde  T  }    }
\def \ds{d+s}
\def\dt{  {\hat d}}
One considers the following  
  ghost and classical field unification     to derive the     BRST transformations of  all the fields that compose the  Beltrami dreibein :
   \bea\label{troix}
   \tilde   e^0 \=   N dt +c^0\nn\\
 \tilde \Et ^z\=   dz+ \m d\bz   +\mo dt  +\c  \nn\\
   {   \Eth}^\bz \= d\bz+ \mb dz+\mob dt+\cb\nn\\
    \ot^a\=\o^a +\Ot^a  \nn\\   
   \tilde d  \=\ds.
   \eea 
 \eqref{troix} generalizes the bidimensional  Beltrami  classical and  ghost  field unification  of Section 2.  It  is  compatible   with   the   set-up of Section 3 that defines the BRST symmetry under the form of geometrical horizontality equations. 
   The three first   equations  of \eqref{troix}  define  the     Beltrami  ghosts $ c^z,c^\bz,c^0$  in function of   the three standard  $d=3$     reparametrization ghosts 
  $ \xi^z,\xi^\bz,\xi^0$  such   that       $s\xi=\xi^\mu\pa_\mu \xi$. The relation is provided  by using 
    equation \eqref{cbrst}  for  $d=3$. One has indeed
      \bea 
   \label{tilde}
\tilde e^0 \=\exp i_\xi  e^0 =      Ndt +c^0 +N\xi^0  \nn\\
\tilde e^z \=\exp i_\xi  e^z \equiv \exp \vp\  \tilde  \Et ^z \nn\\
  \tilde e^\bz \=\exp i_\xi  e^\bz \equiv  \exp \vpb\  \tilde  \Et ^\bz    
   \eea
    \eqref{tilde}  provides the relation between the ghosts $c$ and $\xi$     by expanding  and    $\exp i_\xi= 1+i_\xi +\demi 
    i_\xi  i_\xi +...$... The generic   gravitational  BRST horizontality conditions  \eqref{cbrst} and~\eqref{shat}   taken at     $d=3$ imply
  \def\sL{s-Lie_\xi}
\bea\label{magicalbrst}
 \Th ^0\=    (\ds) \tilde  e ^0 
 -     \frac    {\exp\vpb   }{2}   \tilde \o^z\w \Eth^\bz
  +   \frac    {\exp\vp   }{2}    \tilde\o^\bz \w\Eth^z=0
  \nn\\
   \Th^z&=&    (\ds) \tilde e ^z -\eps  \tilde\o ^0\w \tilde e  ^z    +  \eps \tilde \o^z \w \tilde e^0 =0
   \nn
   \\
    \Th^\bz&=&    (\ds) \tilde e ^\bz +  \eps \tilde  \o ^0\w \tilde e^\bz    - \eps\tilde   \o^\bz \w \tilde e^0=0
\nn\\\red
 \hat R^0&=&   (d+\sL)  \ot ^0-  \demi  \ot^z\w \ot^\bz   =R^0 \nn\\\red
 \hat R^z&=&(d+\sL)\ot^z - \eps \ot^0\w  \ot^z =  R^z  \nn\\\red
 \hat R^\bz&=&(d+\sL)\ot^\bz +\eps \ot^0 \w \ot^\bz  =R^\bz.
\eea
The following subsections compute the  nilpotent BRST  $s$-transformations  of    the Spin connection,  of  all  the  Beltrami fields that compose           the dreibein and    of   the associated  ghosts by expanding  the bi-graded equations~\eqref{magicalbrst}~in form degree and ghost number. 
 
\subsubsection{  BRST transformations    of the $d=3$     Spin connection  and   its  Lorentz  ghost }

The  BRST equation $\hat R=R$  ensures    that 
 \bea\label{lolo}
 \hat s \o \=-d\O-[\o,\O]\nn\\
 \hat s\O \=-\demi [\O,\O]
 \eea 
 (Remember that $\hat s=  \sL  $.)
Both the  $s$ and $\hat s$  BRST transformations of the Spin connection   are   the same   wether    $\o$ is  an independent field  or   
  it the function of $e$   $\o=\o(e)$  that solves the covariant constraint   $T= de +\o\w e=0$. 

\subsubsection{ BRST symmetry of the  $d=3$     lapse $N$    and its  ghost $c^0$ }
%
\def\O{\Omega}  \def\O{ \Omega}
The BRST transformations of $N$ and $c^0$  are determined   by the     components with     form degree  equal to zero and  ghost numbers equal  to one  and two,  respectively, of the following  BRST horizontality constraint  :
\bea\label{T0h}
 \Th ^0=    (\ds) (Ndt + c^0) 
 -
 \frac    {\exp \vpb   }{2}
  (  \o^z    +  \O ^z)  
     \w     (d\bz+ \mb dz   + \mob dt   + \cb   )  
  + 
   \frac    {\exp\vp   }{2} 
    ( \o^\bz     +   \O^\bz  )
    \w     (dz+ \m dz   +\mo dt   + \c   ) =0.
     \eea
     One gets in this way   
   \bea\label{sN} 
   s N  \=  \red  \pa_0 c^0 +\demi ( \exp \vp    (  \o^\bz_0 \c  -  \Omega^z \mo) + 
   \exp \vpb    (  \o^z_0 \cb  -  \Omega^\bz \mbo)
   )
   \nn\\
   sc^0\= \red  \demi   (  \exp \vp \ \c   \O^\bz   - \exp \vpb \   \cb   \O ^z).
   \eea
  {  
    It is mentioned   right after \eqref{bel}  that       
      both   Lorentz  ghosts   $  \O ^z$ and  $ \O^\bz$  in \eqref{sN}  are functions of the Beltrami reparametrization ghosts in order that  the Beltrami   condition
   $e^0=Ndt$  be compatible with the nilpotent  BRST~symmetry equations    as they are defined in   \eqref{T0h}. 
   One is  now on position to quantitatively detail their expressions.

   Indeed, both 
        ghost number one  components  of $  \tilde T^o=0$ in  \eqref{T0h}  that are      proportional  to $\Eo$ and  $\Eob$ 
        must vanish  giving the constraint 
     $
  \hat \Omega^\bz  = \exp-\vpb \  \Dz   c^0+  \exp(\vp -\vpb)     \o_Z ^z (e) c^\bz 
   $
   and
    $
  \hat \Omega^z  = \exp-\vp \  \Dbz   c^0-   \exp(\vpb -\vp )    \o_Z^\bz  (e)  c^z  
   $.
  Analogously,    one must have $s\vp=s\vpb$ when one uses the third Lorentz freedom to impose 
   $\vp=\vpb=\Phi/2$ that gives the simpler formulae 
    $
  \hat \Omega^\bz  = \exp-\demi \Phi \Dz   c^0+    \o_Z ^z (e) c^\bz 
   $
   and
    $
  \hat \Omega^z  = \exp-\demi \Phi      \Dbz   c^0-     \o_Z^\bz  (e)  c^z  
   $.
    Then,~both   form-degree zero  and ghost number one  components of  the  BRST horizontality constraint  $\tilde T^z=\tilde T^\bz=0$ \eqref{Tzh},  which  are to  be shortly displayed in   \eqref{svp}, imply  that the third Lorentz ghost
  is constrained by  
       $
 2 \O^0 =  \bp \cb -\p \c +  \demi ( \cb\pa_\bz \Phi  - \c\pa_z)\Phi 
  +\eps(\o^0_z(e)\c +\o^0_\bz (e)  \cb )
    -  \eps \exp-\frac \Phi 2( \o^\bz_\bz(e)  +  \o^z_z(e) c^0 
  )
  $.  These  expressions of the   Lorentz ghost  components $   \O^0,  \O^z,  \O^\bz $
  in function of the reparametrization ghosts  $\c,\cb$ and $ c^0$ are to  be consistently  used        within   the BRST transformation laws  of the Spin-connection \eqref{lolo}   in the Beltrami parametrization scheme~when $\vp=\vpb$.
   }
   
 \subsubsection{   BRST symmetry  of        Weyl invariant    fields 
 $(\m, \c,\mb,\cb, \mo,  \mob)$~and~of $(\vp,\vpb)$   }
 
 The BRST variations of  the   fields   $\m, \c,\mb,\cb, \mo,  \mob$~and $\vp,\vpb$   derive  from the various term stemming from  the decompositions 
 in  form degree and ghost number  of   both   BRST horizontality constraints
 for the $z$ and  $\bz$~components of the torsion :
 \bea\label{Tzh} 
 \Th^z&=&    \tilde d  \tilde e  ^z -\eps  \tilde   \o ^0\w \tilde e ^z    +  \eps  \tilde  \o^z \w \tilde e^0  = 
    \exp\vp
    \Big (  
    (  \tilde d  \vp     - \tilde \o ^0 )\w\Eth^z   + \tilde d       \Eth^z +   \eps  \exp -\vp   \  \tilde  \o^z \w \tilde e^0  
     \Big ) 
     \nn\\
     \= 
      \exp\vp
    \Big (  
    (( \ds ) \vp     - \eps \o ^0   -\eps  \O^0  )\w ( dz+ \m d\bz   +\mo dt  +\c )  \nn\\  
    &&\quad \quad +
    (\ds)  ( \m d\bz   +\mo dt  +\c ) 
     +    \exp -\vp   \  (  \o^z +\ {\red{ \Omega ^z}} ))\w ( Ndt +c^0)  
     \Big ) 
     =0
     \nn\\
 \Th^\bz&=&    \tilde d  \tilde e ^\bz +  \eps \tilde \o  ^0\w \tilde e^\bz    - \eps   \tilde  \o^\bz \w \tilde e^0  = 
    \exp\vpb
    \Big (  
    ( \tilde d  \vpb     + \eps  \tilde \o ^0  + )\w\Eth^\bz   +\tilde d  \Eth^z -  \eps   \exp -\vpb   \    \tilde \o^\bz \w \tilde e^0  
     \Big ) 
     \nn\\
     \= 
      \exp\vpb
    \Big (  
    (( \ds ) \vp     +\eps \o ^0  +  \eps \O^z )\w ( d\bz+ \mb dz   +\mob dt  +\cb )  \nn\\&& \quad\quad     +(\ds) (   \mb dz   +\mob dt  +\cb )  
     -  \eps   \exp -\vpb   \  (  \o^\bz  +\O^\bz))\w ( Ndt +c^0)  
     \Big )
     =0.
   \eea

   The   components  with  form-degree 0 and  ghost  number  1      
   in  $T^z=0$ and $ T^\bz  =0 $  provide         \bea
   \label{svp}
    s\vp\= \eps  \Omega ^0     +\p \c    +    \c\pa_z\vp    -  \eps \o^0_z\c   +\eps\exp-\vp  \  \o^z_z\  c^0
    \nn
    \\
    s\vpb  \= - \eps \Omega^0     +\bp \cb    +    \cb\pa_\bz\vpb    + \eps \o^0_\bz\cb   -\eps\exp-\vpb  \  \o^\bz_\bz\  c^0.
      \eea
      This equation determines 
 $ \O^0 $  in function of the reparametrization ghosts as  displayed at the end of 4.6.2  when $\vp=\vpb$.
 In what follows,   It must be understood   that     the expressions of  $\O^z $ and $\O^z $ are those that are    also expressed at~the~end of~4.6.2.
  
  Since $\Eth^z\w\Eth^z=\Eth^\bz\w \Eth^\bz =0 $,   
  the  multiplication of  $\Th^z=0$  and  of  $\Th^\bz=0$ by $\Eth^\bz$   and   $\Eth^z$ implies
        \bea\label{bibi}
   ( dz+ \m d\bz   +\mo dt  +\c )  \w  (\ds)  ( \m d\bz   +\mo dt  +\c ) 
     +  \eps  \exp -\vp    ( dz+ \m d\bz   +\mo dt  +\c )  \w (  \o^z +\O^z) \w ( Ndt +c^0)  \=0
  \nn  \\      
 ( d\bz+ \mb dz   +\mob dt  +\cb ) \w   (\ds)  (  \mb dz   +\mob dt  +\cb ) 
     - \eps  \exp -\vpb    ( d\bz+
      \mb dz   +\mob dt  +\cb )  \w (  \o^\bz +\O^\bz) \w ( Ndt +c^0)   \=0.
    \nn\\ 
    \eea
   The components  with ghost number 1  
       and  ghost number 2  of \eqref{bibi}   imply   
     \bea\label{baba}
      s\m   \=  \bp \c +\c\p\m-\m\p \c
      +\red \eps\exp -\vp (\o ^z_\bz  -\m \o^z_z)c^0\nn
      \\
      \nn
       s\mo   \= \pa_0 \c +\c\p\mo -\mo\p\c  -\eps \exp-\vp  \big  (\ 
        N (\O^z  -c^z  \o^z_z  )
        -  (  \o^z_0    +\mo  \o^z_z    )c^0
        \ 
        \big)\nn\\
         s\c\=\c\p\c   -\red   \eps \exp - \vp (  {\red {\Omega }}^z -c^z\o^z_z) c^0
 \\\nn\\
      s\mb   \= \p \cb +\cb\bp\mb-\mb\bp \cb
      -\red \eps\exp -\vpb (\o _z^\bz  -\mb \o^\bz_\bz)c^0
      \nn\\
      \nn
       s\mob   \= \pa_0 \cb +\cb\bp\mob -\mob\bp\cb  +\eps \exp-\vpb  \big  (\ 
        N (\O^\bz  -c^\bz  \o^\bz_\bz  )
        -  (  \o^\bz_0    +\mob  \o^\bz_\bz    )c^0
        \ 
        \big)
        \nn
      \\
       s\cb \=\cb\bp\cb   -(\O^\bz -\cb\o^\bz_\bz)
      + \red   \eps \exp - \vpb (  {\red {\Omega }}^\bz -c^\bz\o^\bz_\bz) c^0.
      \eea
           One can verify the nilpotency of $s$ on all the fields by a  brute computation.  As already mentioned,  this nilpotency is    warranted by construction due to  compatibility of of  the geometrical  definition \eqref{magicalbrst}  of the BRST~symmetry 
with 
           the Bianchi identities  $DR=0$    and  $DT=R\w e$ of    both Poincar\'e  curvatures  $R $ and $T$.  
           The  direct verification that $s^2=0$ on all the Beltrami fields  from   \eqref{baba},  \eqref{sN} and  \eqref{svp} would be nothing but 
                      a (very) brute force method   
           to verify the closure and Jacobi identity  of the Lie algebra of  the  $SO(1,2)\times \Diff_3$ or $SO(3)\times\Diff_3$ local symmetry.   
           
                \subsection{ Possible choices of   BRST invariant  gauge fixings of the     $d=3$    Beltrami   metric  }

The above results allows one to investigate   various   possibilities for   the different  gauge fixings of $d=3$     gravity. 
Any given choice of consistent gauge functions    must be enforced by    adding    corresponding  BRST exact terms to the Einstein action. To~build such  terms one   introduces  appropriate BRST trivial    pairs  of antighost and Lagrange multiplier fields that are  adapted to the chosen gauge.  
 All choices of gauge,  provided they are consistent ones,  are equivalent, meaning that they provide  the same mean values for  the   physical observables. One given  choice can be more convenient than another one depending on the question one wishes to investigate.

A     natural choice  is inspired by the   conformal gauge choice of $d=2$ gravity. 
 By using     2 freedoms of the $d=3$ reparametrization symmetry,  one can  indeed  impose in a BRST invariant way the condition 
\bea\label{moduli}
\m =\gamma \quad\quad 
\mb = \bar \gamma.
\eea
  Here and elsewhere in the paper,   $\gamma $ and $\bar \gamma $ denote the  moduli of the $d=2$ leaf $\Sigma_2$. The expression of the  moduli~$\gamma$   depends on the  genus  $g$ 
  of $\Sigma_2$. In fact,  as it is well known in  the context of  the  covariant string quantization, for any given value of g, one can express
 \bea\label{genus}
 \gamma=\sum _{k=1}^{g-3} \lambda_k f^k(z,\bz),
 \eea  
 where the   $g-3$  constants $\lambda_k $  must be integrated  over  fundamental   domains of $\Sigma_2$.  The functions 
 $ f^k(z,\bz)$ build a  $g-3$ dimensional  basis  of the  quadratic differentials for the Riemann surfaces of  genus $g$.
 In this construction the modular invariance must be  taken into account for fixing the range of integration of the constants $\lambda_k $'s to fundamental domains and avoid  replica of modular copies. 
  (See~\cite{bb} for some   explanations    of this way  of taking into account   in a  BRST invariant way    the   global zero modes of the ghosts that occur when one gauge fix   the $d=2$ reparametrization symmetry of a Riemann   surface.)
 
The third     freedom of    the $d=3$      reparametrization symmetry  has to  be   further gauge fixed  to complete     \eqref{moduli}. 
One~can for instance       gauge fix the  scale invariant      function  $\N$ in \eqref{bel} as  \bea\label{rest}
 \N=1.
\eea
Since  $  N=   \exp \Phi \N $ and   $\sqrt {-g} =\exp\Phi (\mmb )$,    the latter condition  amounts to the  following     constraint between     the  lapse of     the $ d=2 $ ADM  leaves  and the  volume element  of the $d=3$     Lorentzian space 
\bea\label{ggf}
\sqrt {-g} =     (1-\gamma \bar \gamma)N.
\eea 
This gauge  choice corresponds    to  a  choice     of    coordinates  such that  the metric  reads as  
\bea \label{3dgauge}
ds^2=
\exp \Phi\Big  (   \eps dt^2 +   ( dz+\gamma d\bz+\mo  dt)(  d\bz+\gamma  d\bz+\mbo  dt \Big   ),
\eea
genus by genus.
At the quantum level, the   three  gauge conditions for the reparametrization invariance   \eqref{genus} and  \eqref{rest}       can be imposed in a BRST invariant way.  To do so, one   introduces  the 
 relevant BRST trivial doublets  of antighosts and Lagrange multipliers   for defining and     adding to the classical action
  a ghost number zero  s-exact action     according to  the standard method $ S_{cl} \to  S_{cl} +s(... )$ that expresses      the BRST invariant action in this gauge. The corresponding path integral must include an  integration  over the parameters  $\lambda$ varying in fundamental domains.
  
   The  BRST invariant action   for the gauge  fixed metric \eqref{3dgauge}    defines   a $d=3$     QFT  where the       classical field degrees of freedom  that propagate   are the field  $\Phi $   that compose 
   the conformal factor  
   and both components
  of  the leaf  shift vector field~$\mo$~and~$\mob$. The dynamics also involves  compensating   ghost antighost propagations  that maintains the  BRST symmetry Ward identities for all  correlators.

Another different  three dimensional gauge choice  that is maybe worth being studied  corresponds to   the   following    class  of  gauge functions
\bea \mo = \alpha  \p\m     \quad\quad 
  \mob = \bar \alpha  \bp\mb \quad\quad  
 \N = 1,
\eea
where  $\alpha$ and $\bar \alpha$ are a pair of   constant  parameters. 
This  gauge  choice eliminates the shift fields  $\mo,\mob$ in function of the field $\m,\mb$  and provides    a QFT  where the   remaining   propagating classical  fields  are  the Beltrami differentials $\m$ and $ \mb$, the field  $\Phi$ that define the conformal factor of the leaves and  all   ghost and antighost fields that are relevant to maintain the BRST~invariance. Such a gauge fixing is the analogous of the Yang-Mills regularized Coulomb  gauge. (See the third and  fourth reference of \cite{luca}.)


\section{    $d=4$    Beltrami   gravity }

  \def\t {\tau}
\def\mut{\mu^z_\t}
\def\mubt{\mu^\bz_\t}
\def\M{M} 
 There are     $ \frac{d(d-3)}{2}$  gravitational physical  degrees of freedom  that possibly propagate 
 in a   $d$-dimensional manifold~${\cal M}_d$.
  $d=4$    is thus the  lowest possible dimension     for the 
  existence of  a non trivial Spin 2   propagating graviton with two physical particle excitations.     The~way  to   pass from   the   $d=3$   Beltrami parametrization   
    to that in  $d=4$   Lorentzian  dimensions      must  parallel  what is done  in the previous section
     to  pass from the 
   $d=2$   to    the   $d=3$  cases.    This  section~confronts however   a  novel task   that is of  parametrizing      both    non trivial     physical   propagating degrees of freedom of  the $d=4$  graviton  in function of the   Beltrami fields.

  The  six generators of the $d=4$ Lorentz symmetry  $SO(3,1)$   split into   three generators for  the      rotations and   three generators for   the    Lorentz boosts.   It is     tempting enough to  
     stick with the  notational  conventions of the three-dimensional   Beltrami parametrization     (with  $\eps=1$)  to parametrize    
     the  Euclidean ADM   leaves  $ \Sigma_3$ of   ${\cal M}_4$.  Consequently,   one denotes in this  section the  third spatial dimension of $ \Sigma_3$ as     $x^0=t$ and  one   defines  
   the new four dimensional  Lorentz time coordinate as $\t$.  It follows that the   four Lorentzian  coordinates used in this section are  
 $x^\mu = (z,\bz, t,\t )$. The leaf of leaf sub-foliation  of  the  three dimensional Euclidean~ADM leaves~$  \Sigma^{ADM}_3 $~of~${\cal M}_4$ is  thus by  the  Riemann surfaces $\Sigma_2$ with   coordinates $z,\zb$  and  
   the  ``radial"    one-dimensional space~$\Sigma_1$~with     real   coordinate~$t=x^0$, according to
   $  \Sigma_3^{ADM}=   \Sigma_2\times \Sigma_1$  .

 This notation helps for an adequate use   of the  three dimensional results of the  previous section  but it~may~appear as   a bit     confusing.   
  In fact,  for the further  generic  construction of   the~$d$-dimensional  Beltrami parametrization,   the Euclidean sub-leaf $\Sigma_{d-3}$ is a priori multi-dimensional    with an inner $SO(d-3)$ symmetry and the use of the   single index  such as $0=t$ must be   abandoned.  Thus, in  the next section  6, one  denotes  the   spatial~coordinates    of   the    sub-foliated  ADM  leaves  $\Sigma^{ADM}_{d-1}=\Sigma_2\times \Sigma_{d-3}$    as   
   the complex  coordinate    $z,\zb$  for  $\Sigma_2$~and~the  d-3  real~coordinates    $x^3,..,x^i,... ,x^{ d-1 }$  for  $\Sigma_{d-3}$, while~$\t$~universally   stands  for     the Lorentz time coordinate.

  The four dimensional case      is quite special among all other cases (but $d=8$ to some extend) because of the~possibility of  irreducibly  decomposing  any given~two-form~into  the sum of  its selfdual and anti-selfdual parts, a property that applies for decomposing the upper pair of  Lorentz antisymmetric indices     of the  four dimensional   Spin connection. The  $d=4$  Lorentzian   Einstein action  can be  consequently expressed 
  as   a quadratic form    either    in       the selfdual part  or in the  antiselfdual part   of  the  Spin connection  according to  
 \bea\label{E4} 
\frac 1 4 \int \epsilon _{abcd}  e^a\w e^b \w R^{bc}\sim
i   \eta_{ab}    \int      \o^{ac^-} \w e_c  \w \o^{bd^-} \w e_d \sim  i   \eta_{ab}  \int       \o^{ac^+} \w e_c  \w \o^{bd^+} \w e_d,
 \eea
 modulo the  (complex)~boundary terms
$   \pm i \int d( \o^{ab^\pm} \w e_a\w     e_b)$. The definition of  both selfdual and antiselfdual components of the  four dimensional  Spin connection $\o^{ab}$ is  
   $\o^{ab^\mp }
 \equiv \frac  12( \o^{ab }   \mp   \frac i 2 \epsilon  ^{abcd}\o_{cd})$       such that   $  \frac 1 2 \epsilon ^{ab}_{\ \  cd}  \o^{cd\pm }= \pm i   \o^{ab\pm}$.  Each expression in the right hand side of    \eqref{E4} is the Lorentz invariant  combination  of four independent $4$-forms, each one being the  exterior square   of   the~$2$-form~${\cal F}^{a\pm}_2\equiv \o^{ac^\pm} \w e_c$ that is valued in the fundamental representation of
 $SO(1,3)$. To prove  \eqref{E4}, one uses $R^{ab} =  R^{ab^+}  +R^{ab^-}  $,
 $  \demi\    \epsilon^{ab}_{\ \ cd  }    R^{cd  ^\pm  }   =\pm  i  R^{ab   ^\pm  }$ and then 
    the  Bianchi identity  $DT=R\w e=0$ and a part integration on the term~$d\o$~stemming from  $R=d\o+\o\w\o$.

One uses complex coordinates for the Riemann surface $\Sigma_2$ but 
it is   often  convenient in intermediate computations  of the four dimensional case  to consider   the real indices $i,j,k$ for  the 3  real coordinates of $\Sigma_3$ with  the    convention that        the non vanishing components of the  fully antisymmetric  four dimensional Lorentz invariant  tensor   $\epsilon ^{abcd }  $ are  such that 
$\epsilon ^{ ijk\tau } =-
\epsilon ^{\t ijk  }
= -\epsilon ^{ ijk}_{\ \ \ \tau }=  \epsilon ^{ijk} =\epsilon_i ^{  \ jk}  $  equates $1$ (respectively $-1$)  if $ijk $ is an even (respectively odd)  permutation  of~$1,2,3$.  All other components  of the tensor  $\epsilon ^{abcd }  $ such that at least   two of their  four indices   are equal  obviously vanish.

  The vierbein  of    $d=4$     gravity    is    expressed  as
  $ e^a  = ( e^i,e^\t)$ in real coordinates and 
   $ e^a  = (\e,\eb, e^0,e^\t)$  when one uses   complex coordinates for  the component $\Sigma_2$ of the ADM    leaves  $\Sigma_3=  \Sigma_2 \times  \Sigma_1$.
  The  four dimensional 1-form Spin~connection  can   be expressed   as  $\o^ {ab} \equiv  ( \o^{ij}, \o^{\t i})$  or equivalently as     
 $ \o^ {ab}    =  ( \o^{i}, \o^{\t i})$ with    $\o^{ij} \equiv i\epsilon ^{ijk} \o^k $.  The~twenty four ~components 
of the Spin~connection on the basis of one-forms  $dx^\mu$  can thus be denoted  as 
\bea   
 \o^{ab} _\mu  = (\o^{ij}_\mu, \o^{\t i}_\mu)\sim (\o^{i}_\mu, \o^{\t i} _\mu)
 \equiv
 (\o^{z}_\mu, \o^{\zb}_\mu,  \o^{0}_\mu;   \o^{\t z}_\mu, \o^{\t \zb}_\mu,  \o^{\t 0} _\mu) . 
\eea

A quick computation shows that, when one decomposes     $\o^{ab}$   into  its 
selfdual and anti-selfdual components  $\o^{ab\pm }
 =\frac i 2( \o^{ab }   \mp   \frac i 2 \epsilon  ^{abcd}\o_{cd})$, the   selfdual and anti-selfdual properties of   
 $\o^{ab\pm }$
  simply satisfy      
$ 
    {\omega^{ij  ^\pm}  =\mp i  \epsilon _{i j  k } \    \omega^{\tau  k ^\pm}   } 
$. One     conveniently defines 
   $  { \o^{i ^\pm } 
\equiv 
 \omega^{i  }    \pm   \omega^{k\tau} }$.  With 
   the convention $   \omega^{ij} \equiv i  \epsilon^{ijk}\omega^k $ one finds that 
   the the   selfdual and anti-selfdual properties of   
 $\o^{ab\pm }$ simply mean $   \o^{i ^\pm } =0$
and 
    the Einstein action    \eqref{E4}  $ I_{Einstein}= \frac 1 4 \int \epsilon _{abcd}  e^a\w e^b \w R^{bc} $ 
can be suggestively expressed  either   as    
\bea\label{L+-}
  I_{Einstein} =  \int    d^4 x  L^-({\o^{i^-}},e^j,e^\t)  \equiv
i   \int      \o^{ac^-} \w e_c  \w    \o_a^{\ d^-} \w e_d = 2i \int  (e_i \w  e_j\w \o^{i  -}\w \o^{j -} 
 - \red  i    \epsilon _{ijk}  e^{\t}  \w e^{k} \w
  \o^{i  -}\w  \o^{j  -} )
\eea modulo
the    boundary term
 $ -  i     \int d(  \epsilon_{ijk}    \o^{i^-} \w  e^j\wedge e^k)$
or as  \bea I_{Einstein}=
\label{L-+}
 \int   d^4 x   L^+(\o^{i^+},e^j, e^\t)  \equiv
  i     \int       \o^{ac^+}\w e_c  \w \o_a^{\ d^+} \w e_d 
  = 2i\int    (
- e^i\w  e^j \w\omega^{i+}\w\omega^{j+}+i\epsilon_{  i j k } e^\tau\w\omega^{i^+}\w\omega^{j^+}\w  e^k )  
\eea
modulo  the    boundary term
$   i   \int d(  \epsilon_{ijk}    \o^{i^+} \w  e^j\wedge e^k)$.

The presence of the term $ e^\tau \w  \epsilon_{  i j k } \omega^{i^\pm}\w\omega^{j^\pm }\w  e^k$  in 
 both   \eqref {L+-}  or  \eqref {L-+} 
  quite  transparently illustrates     
the intertwining  between   the   $d=4$   Einstein action    and the  $d=3$  Einstein action  under the form    \eqref{Ea}.

 Both formula for $L^+$   and  $L^-$  in \eqref{L+-}~and \eqref{L-+} and their simplicity for covariantly separating  the Lorentz time 
 and space directions  rely on 
 the generic  selfdual and antiselfdual  decomposition $6=3\oplus 3$ of  any given    four dimensional  Lorentz antisymmetric  tensor~$M^{ab}$.  They   are  the   gravitational analogs of both    Lorentzian   Yang-Mills formulae 
$\int d^4x F^{\mu\nu} F_{\mu\nu}  = 2   \int d^4x  (F_{\t i} \mp i   \epsilon _{ijk}  F^{ ij})^2
$ modulo  the (complex) boundary terms $\pm 2i   \int   \epsilon ^{\mu\nu\rho\sigma }F_{\mu\nu}F_{\rho\sigma}$~\footnote{ The occurrence and the usefulness of selfduality   in gravity and Yang Mills also occurs for $ d=8$ but in a somehow restricted way.  It     involves the   octonionic selfduality 
instead of the quaternionic one.  But then the $SO(7,1)$  symmetry must be restricted by considering manifolds with 
      $  Spin(7)$, $SU(4) $ 
 or $G_2 $~$\subset SO(7,1)$ holonomy \cite{Baulieu:1986hw}. The  existence and the role   of the   complex solutions 
    of  the Lorentzian four dimensional   selfduality equations   have  always been a   challenging subject.
More about    the Chern class  type  boundary terms $  \int d( \o^{ab^\pm}  e_a\wedge e_b)$  will appear in a separate publication. 
}. 
The  perspectives offered   by    the underlying selfduality properties revealed   by   \eqref{L+-}  and  \eqref {L-+}   is a further motivation for  a detailed $d=4$  Beltrami decomposition of the vierbein and its  associated   Spin connection.  Appendix~A~displays 
the decomposition  of the vanishing  torsion  equations that determine the Spin connection  in terms of the Beltrami fields.

\subsection {Beltrami  vierbein and  $d=4$     Beltrami metric in  $ z,\bz, t,\t$ coordinates}

The  six local  freedoms offered by  the $SO(1,3)\subset \Diff_4  \times SO(1,3)$    Lorentz gauge symmetry    allow      
 the    covariant    parametrization   
 of the sixteen  components  of an arbitrary given    vierbein  $ e^a_\mu$ in function of the  ten fields  that determine its    Beltrami  expression.
The  replication of the method used in   three  dimensional  case suggests  that one    postulates  as a first step  the following~$ z,\bz, t,\t$ dependent  field decomposition for the   Beltrami  vierbein $(\e,\eb, e^0,e^\t)$ :
 \def\aa{{\red{a}}}\def\aa{{ {a}}} 
 \def\aa{{\red{a}}}
 \def\aa{{ {a}}}
\bea\label{4bein}
\pbM
 \Eo \cr
 \Eob\cr
{\cal  E}^t  \cr
  {\cal  E}^\t 
\peM\equiv
\pbM
 \exp -\vp\   \e \cr
  \exp -\vpb\  \eb\cr
\frac{1}N {e^t}     \cr
  \frac  1 \M {e^\t}   
\peM
=\pbM
\bM
1 &\m
\cr
\mb &1
\eM
& 
\bM
\mo
&
\mut
\cr
\mbo
&
\mubt
\eM\cr
\bM
0
&
0
\eM
&
\bM
1
&
\aa
\eM
\cr
\bM
0
&
0
\eM
&
\bM
-\aa
&
1
\eM
\peM
\pbM
dz\cr
d\bz\cr
dt\cr
d\t
\peM.
\eea
This  exhausts  five among  the six local Lorentz freedoms, leaving a $U(1)\times\Diff_4   \subset SO(1,3)\times \Diff_4 $ covariance for   the  eleven  Beltrami  fields
$\m,\mo,\mut, \vp, \mb,\mbo,\mubt, \vpb,a,N,\M $.  By  the  further elimination of  the $U(1)$  freedom,  one can impose    $\vp=\vpb
 \equiv \demi\Phi$,
which provides  
a   $  \Diff_4$  covariant  Beltrami  parametrization of the vierbein  that only depends on  the  same ten fields that  are         to    parametrize    the  four  dimensional Beltrami metric. 

The~decomposition~\eqref{4bein}  is   consistent with that  done in section 4      for $d=3$. The   three-dimensional Beltrami matrix   \eqref{trois}  is  indeed recovered  by
restricting  the  $4\times4$  Beltrami matrix  in   \eqref{4bein}   to its upper-left  cornered $3\times3$~matrix~;  and the upper-left  cornered   $2\times2$  matrix   of the latter  
is nothing but the  bidimensional Beltrami matrix \eqref{MuMu}.    The  antisymmetric  dependence  on the field~$a$~in \eqref{4bein}   is a further subtleties absent in    the    three  dimensional  case.  Its origin is rooted in     the generic Beltrami formulation  of  section 6 but it  cannot be detected for $d=3$ where the sub-leaf $\Sigma_{d-3}$ reduces to a point. It is in fact    best  to  denote  
\bea
a\equiv   \mu^0_\t.
\eea
Indeed,  it is to  be shown  that the three fields  $      \mu^z_\t,   \mu^\bz_\t,  a=\mu^0_\t $      compose  the four dimensional ADM  shift vector.

 The~gauge fixing  that leads   to  \eqref{4bein}   is nothing but  a   clear  four dimensional   generalization  of  the process    done respectively  in section 3 and  4  
in dimensions 
     two and   three for defining the Beltrami parametrization. 
The~gauge fixing  of the $SO(1,3)    \subset SO(1,3)\times \Diff_4 $     Lorentz gauge symmetry   that covariantly imposes  the six conditions    
$e^\t=Nd\t$ and $\vp =\vpb$
     determines   a trivial Faddeev--Popov determinant.
The latter~generates     a Gaussian   algebraic dependence
 on the  six Lorentz ghosts and    antighosts   in quantum field theory\footnote{ It must be noted  that 
the     gauge  fixing of the   Lorentz   gauge   invariance of gravity  that is used in this work    deeply  differs  from   the      choice $e^a_\mu=e^\mu_a$ that is     often    presented  in the literature for  defining perturbative gravity.}.     The~Lorentz~ghost and antighost  dependence   can  then be    integrated out from the
   BRST invariant     gauge fixed Einstein action.  They   play no  dynamical  role in  the  path integral computations of correlation functions but they can be used as classical sources  to write Ward identities that control  the Lorentz gauge symmetry in a reparametrization invariant way,   which is useful when the gravitational  theory  is for instance coupled to spinors.    
   When one proceeds to 
    a    BRST invariant gauge fixing  for  the   six~constraints 
 $e^\t=Nd\t$ and $\vp=\vpb$,     the six four-dimensional     Lorentz ghosts   $\Omega^{ab}\sim ( \Omega^{i},\Omega^{\t i}  )$ satisfy  the algebraic equations of motion of the  Lorentz antighosts    that enforces  the six   constraints between the Lorentz ghost and the $ \Diff_4$ ghosts. The  $d=4$  formulae     are   simple generalisations  of those    detailed right after  \eqref{sN}  for $d=3$ and there is no need~to~display~them in this~paper.
 
The vierbein  Beltrami parametrization \eqref{4bein}       therefore   introduces the following 
     four dimensional  Weyl invariant    Beltrami $4\times 4$~matrix  :
 \bea\label {4bm}  {\cal M}= \pbM 
\bM
1 &\m
\cr
\mb &1
\eM
& 
\bM
\mo
&
\mut
\cr
\mbo
&
\mubt
\eM\cr
\bM
0
&
0
\eM
&
\bM
1
&
\aa
\eM
\cr
\bM
0
&
0
\eM
&
\bM
-\aa
&
1
\eM
\peM.\eea    Its  inverse is  
\bea  \label{4pminv}
{\cal M} ^{-1}=
\pbM
\frac 1\mmb\invMu & 
- \frac 1\mmb\invMu
 \pbM
 \mo&\mut
\cr 
 \mbo & \mubt
 \peM \frac 1{1+\aa^2}
 \pbM
 1&-\aa\cr\aa&1
 \peM
\cr
\bM
0
&
0\cr 
0
&
0
\eM
&
\frac 1 {1+\aa^2}
\pbM
1
& 
-  \aa    \cr
\aa
&1
\peM
\peM.
\eea
 The       Weyl non invariant    fields that compose the Beltrami vierbein  in  \eqref{4bein}  are       $\vp $,   $\vpb$,    $M$ and $N$.
 Their dependence  can be arranged as the components  of a  diagonal matrix.
It follows that 
  the   four dimensional       metric  $ g_{\mu\nu}  =e^a_\mu   \eta
  _{ab}   e^b_\nu$~can be   decomposed  as the following product  of   
  $4\times4$ matrices
  \bea \label{ds2d=4}
\pbM
\bM
1 &\mb
\cr
\m &1
\eM
& 
\bM
0
&
0
\cr
0
&
0
\eM\cr
\bM
\mo
&
\mob
\eM
&
\bM
1
&
-\aa
\eM
\cr
\bM
\mut
&
\mubt
\eM
&
\bM
\aa
&
1
\eM
\peM  
\pbM
\exp\vp
&0&0&0
\cr
0&\exp\vpb&0&0
\cr
0&0&N&0\cr
0&0&0&M
\peM
\begin {pmatrix}
   0  & 1 &0&0
   \cr
 1  & 0 &0&0
 \cr
 0 & 0 &2&0
\cr
 0 & 0 &0&-2
\end{pmatrix}
\pbM
\exp\vp
&0&0&0
\cr
0&\exp\vpb&0&0
\cr
0&0&N&0\cr
0&0&0&M
\peM
  \pbM
\bM
1 &\m
\cr
\mb &1
\eM
& 
\bM
\mo
&
\mut
\cr
\mbo
&
\mubt
\eM\cr
\bM
0
&
0
\eM
&
\bM
1
&
\aa
\eM
\cr
\bM
0
&
0
\eM
&
\bM
-\aa
&
1
\eM
\peM.\nn\\
\eea

  The   reparametrization invariant  infinitesimal  line element in the Beltrami parametrization  for $d=4$ is therefore
\bea\label{ds}
 \quad\quad ds^2
 =
  -  2M^2  (d\t -\aa dt )^2 + 2N^2 (dt+\aa d\t)^2 + ( dz+\m d\bz+\mo  dt+\mu^z_\t d\t ) \exp \Phi (  d\bz+\mb d\bz+\mbo  dt +\mu^\bz  _\t d\t ).\quad\quad\quad  \hskip 1,3mm
\eea

   The four dimensional       Beltrami metric      \eqref{ds}   is parametrized by the ten  Beltrami fields 
    \bea \m,\mb,\mo,\mob,    \mu^z_\t,     \mu^\bz_\t,    a\equiv\mu^0_\t, M, N,    \Phi. \eea
    The     antisymmetric    dependence  of the Beltrami matrix  \eqref {4bm}      on the  field     $a\equiv \mu^0_\t$   makes   the Beltrami metric  formula  \eqref{ds}  subtly different than     the   four dimensional      ADM~formula   $ds^2= - {\cal N }^2 d\t^2   +(dx^i+\beta^i d\t)    g_{ij} (dx^j+\beta^j  d\t)$.      
     Setting $a=0$  from the beginning  would be  an over-gauge fixing of the Lorentz symmetry  for a $\Diff_4$~invariant gauge fixing of the sixteen components of a generic vierbein. 
   One~must    generically     break the  four dimensional    reparametrization invariance  in a BRST invariant way to impose $a$ =0.

    In fact   the      excitations of   the   Weyl invariant   fields   $\mu^z_0, \mu^\bz_0$ 
   describe both  degrees of freedom of     the    propagating traceless and transverse graviton  in   the case    $d=4$     while    
      the same fields carry 
   no  gravitational physical degree of freedom    for  $d=3$.  Indeed,  for  $d=3$, both    $\mu^z_0$ and $ \mu^\bz_0$  in \eqref{ds3333}    have   no conjugate momentum as can be observed  from      the tree dimensional  Einstein action \eqref{Einstein3d}. They  can be identified   as    both components of   the shift vector of  the  ADM~leaves~$\Sigma_2$~of~${\cal M}_3$ but  for $d=4$  the shift vector
of  the leaves $\Sigma_3$   of   of  the leaves $\Sigma_3$  of   ${\cal M}_4$ is  composed 
by    the  other three fields $\mut, \mubt$ and $a \equiv\mu^0_\t $   in \eqref{ds}.   
(The    generic   $d$-dimensional   formulae    for  the  exact correspondence  between the   ADM   vector shift components and the  Beltrami fields $\mu_\t^m$, $m=z,\bz,i$ is to be further written  in section 6.    In fact,    \eqref{shift} indicates   that if the $\mu_\t^m$'s vanish the same happens for the ADM shift vector components.)

One possible     gauge fixing  of  the $\Diff_4$  invariance  of   \eqref{ds}  is     by 
the four   conditions $\mu^z_\t=\mu^\bz_\t =0$ and 
 $\m=    \gamma, \ \mb= \bar \gamma $ with no prejudice on the genus of $\Sigma_2$ by using  \eqref{genus} for $\gamma$. It implies  
\bea\label{dsrr}
  ds^2
=
  -  2 (M-\mu_\t^0)   ^2 d\t ^2 +  2 (N+\mu_\t^0  )^2dt ^2 + 2 \exp \Phi  ||dz+\gamma   d\bz+\mo  dt   ||^2.
\eea
\eqref{dsrr} can be presumably used to describe   in a  compact form  systems that consists   of  perturbative gravitons  whose degrees of freedom are both fields $\mo$  and $\mob$  in a gravitational background.
 
As for static  configurations, one may impose   $dz+\m d\bz+\mo  dt+\mu^z_\t d\t =dz+\m d\bz$ and $\mu_\t^0=0$.  The  static~Beltrami~metrics   read as  
\bea
\label{genussigma}
  ds^2=-  2M^2   d\t  ^2 + 2N^2 dt^2 + ( dz+\m d\bz) \exp \Phi (  d\bz+\mb d\bz   ).
\eea
One may compute the Ricci tensor for this metric and examine various possibilities with  $\m= \gamma$ as in  \eqref{genus},
\bea
\label{genussigma1}
  ds^2=-  2M^2   d\t  ^2 + 2N^2 dt^2 + ( dz+\gamma d\bz) \exp \Phi (  d\bz+\bar\gamma d\bz   ).
\eea

\subsection{Examples}
The  case $\m=\gamma=0$ identifies   $\Sigma_2$ as a sphere.     Solving the Einstein equation in this case  determines the  spherical symmetric Schwarzschild  solution with  flat space   boundary condition at spatial infinity. One gets     $MN=1$
 and   $\Phi=0$ with 
$\exp \Phi dzd\bar z  =    t^2 (d\theta^2   +  sin^2\theta d\phi^2)^2$.

Solving the Ricci flat condition for  the  static metric  \eqref{genussigma1}   seems    a doable  task  when the  topology  of    the   sub-manifolds $\Sigma_2$  is restricted to  that of a  torus.  It is so  when the moduli~$\gamma $~are    complex  constants  $\gamma\neq 0$. An~encouraging   signal  is the existence of the so called       axisymmetric Weyl metric  \cite{Weyl}  with two Killing vector fields  $\xi^\t = \pa_\t$   
and $\xi^t = \pa_t$ that read as 
\bea\label{weylmetric} ds^2=-  \exp 2 \Psi (x,y) d\t  ^2 +  x^2 \exp -2 \Psi (x,y)    d t  ^2       + \exp \big(2\kappa  (x,y) -2 \Psi (x,y)\big)  (dx^2+dy^2), \eea
so that  \eqref{weylmetric} appears  as a particular case    of \eqref{genussigma1} with $\gamma=cte$.  This can be formally seen by locally redefining   the complex  coordinate  $z+\gamma$ into  the real  coordinates~$x,y$~according to $dx^2 +dy^2 \equiv  || dz+\gamma d\bz||^2$ and then  by   expressing    the three fields   $M,N$ and $\Phi$  in   \eqref{genussigma1}     as functions of  both      so-called Weyl  metric potentials    $ \Psi  ( x, y )$ and $ \kappa  ( x, y )$  in 
\eqref{weylmetric}.
 Another   open    question  is   wether analytical  solutions  of \eqref{genussigma1} can be found when  
   $\Sigma_2$ has  genus equal to two.

\subsection {  $d=4$    Beltrami metric with   light cone  coordinates $z,\bz, \t^+,\t^-$}
  \def\bb{{\color{black}{\bar a}}}
One  can    rotate 
 the Lorentz time $\t$ and the third spatial coordinate $x^0=t$
  into   the    light-cone coordinates \bea
\t^\pm=\t\pm t.
\eea   The   $d=4$  Beltrami parametrization metric    \eqref{ds}  becomes then   the      sum of~two~interestingly factorized terms  : 
\bea\label{dslc}
  ds^2
=
  -  { \cal N}^2  (d\t ^+   +\mu^+_ - d\t ^-   ) (d\t ^-   +\mu^-_+      d\t ^+   ) 
    +  \exp \Phi  ( dz+\m d\bz+\mu^z _+  d\t^+  +\mu^z _-  d\t^-    )(  d\bz+\mb d\bz+ \mu^\bz _+    d\t^+  +\mu^\bz _-  d\t^-      ).
    \nn\\
\eea
In fact,  when one uses the light cone coordinates, the $4\times4$  Beltrami matrix      \eqref {ds2d=4} is   
\def\MMMM{  
\pbM
\bM
1 &\m
\cr
\mb &1
\eM
& 
\bM
\mu^z_+
&
\mu^z_-
\cr
\mu^\bz_+
&
\mu^\bz_-
\eM\cr
\bM
0
&
0
\eM
&
\bM
1
&
\mu^-_+
\eM
\cr
\bM
0
&
0
\eM
&
\bM
\mu^+_-
&
1
\eM
\peM
}
\def\MLCE4t{  
\pbM
\bM
1 &\mb
\cr
\m &1
\eM
& 
\bM  
   0
&
   0
\cr
  0
&
   0
\eM\cr
\bM
\mu^z_+  
&
\mu^\zb _+
\eM
&
\bM
1
&
\mu^+_-
\eM
\cr
\bM
\mu^z_-
&
\mu^\bz_-
\eM
&
\bM
\mu^-_+
&
1
\eM
\peM
}
\def\idlc{\pbM0,1,0,0\cr1,0,0,0\cr0,0,0,-\demi \cr0,0,-\demi,0\peM}
\bea 
\MLCE4t
\pbM
\exp\vp
&0&0&0
\cr
0&\exp\vpb&0&0
\cr
0&0&\cal N&0\cr
0&0&0&\cal N
\peM
\idlc
\pbM
\exp\vp
&0&0&0
\cr
0&\exp\vpb&0&0
\cr
0&0&\cal N&0\cr
0&0&0&\cal N
\peM
\MMMM.
\nn\\
\eea  
There is    a   mapping   between the Beltrami fields   $M,N,a$ and  $\mo,\mut,\mob,\mubt  $   in  \eqref{ds}  and  the fields   ${\cal N},   \mu^+_,\mu^-_+  $ and   $\mu ^z_+, \mu ^z_-, \mu ^\bz_+, \mu ^\bz_-   $   in \eqref{dslc}. It reads  as follows :
 \bea 
 \pbM1,\aa\cr-\aa,1\peM\to  \pbM1,\mu^+_ -\cr\mu^ -_+,1\peM, 
 \quad
 N\to {\cal N},
 \quad
    M \to \cal N\ \ {\rm and}
 \quad
\pbM0,1,0,0\cr1,0,0,0\cr0,0,2,0\cr0,0,0,-2\peM
\to  \idlc .
\eea
 (The    last    arrow   is  for   the change of the  Lorentz flat  metrics when one  uses light cone coordinates $
 (t,\t)\to \t^\pm$.)
 
  The    light cone coordinate redefinition   of the   Beltrami     matrix  \eqref{4bm} is     
\bea\label{lcv}
  \MMMM.
\eea 
One has therefore the   following  expression for the  light cone    Beltrami vierbein 
 \bea\label{4beinlc}
\pbM
 \Eo \cr
 \Eob\cr
{\cal  E}^+  \cr
  {\cal  E}^- 
\peM\equiv
\pbM
 \exp -\vp\   \e \cr
  \exp -\vpb\  \eb\cr
\frac{1}{\cal N}   {e^+}     \cr
  \frac  1 {\cal N} {e^-}   
\peM
=
 \MMMM 
\pbM
dz\cr
d\bz\cr
d\t^+\cr
d\t^-
\peM.
\eea

The above expressions suggest   a formal  four dimensional   parallel  between  the 
complex coordinates  $z,\zb$  of~$\Sigma_2$ and the   light cone coordinates~$\t^+,\t^-$.  Indeed, \eqref{dslc}~involves~the     single ``light cone lapse function" $\cal N$, the conformal factor  $\exp \Phi$  and     the   eight  Weyl invariant  fields 
$    \m,  \mb,  \mu^z_+, \mu^z_-,  \mu^{\bz}_+, \mu^{\bz}_-, \mu^+_ -,   \mu^ -_+  $ with a quite striking symmetry between the $z,\bz$ and  $+,-$ indices\footnote{The   $SO(4)$  versus  $SU(2)\times SU(2)$ and      $SO(1,3)$ versus $SL(2,C)$   correspondences  help to enlighten     the four-dimensional  Beltrami results.}.

Moreover, the  gauge choice  for the $d=4$  reparametrization symmetry $ \mu^z _+= \mu^z _-       =   \mu^\bz _-=  \mu^\bz _-   =  0$
provides the following  suggestive expression for the $d=4$ metric
\bea\label{dslcgf}
  ds^2
=
  -  { \cal N}^2  (d\t ^+   +\mu^+_ - d\t ^-   ) (d\t ^-   +\mu^-_+      d\t ^+   ) 
    +  \exp \Phi  ( dz+\m d\bz )(  d\bz+\mb dz     ).
\eea
  A more mathematically oriented  publication will discuss      some  properties of 
the   four dimensional  light cone    Beltrami vierbein   and metric   in  \eqref{4beinlc} and   \eqref{dslc}. The Beltrami  parametrization as expressed  in~\eqref{4bein}~seems  however  physically  handier than that in \eqref{4beinlc} since    it offers    a genuine distinction between the space   and time~coordinates. 
On the other hand \eqref{4beinlc} could be of interest for studying  gravitational solutions for the~$(2,2)$~signature.
   
   \subsection{ $d=4$   Beltrami  Spin connection}
   
 Computing   the   four dimensional   Spin connection
  is a mere generalization  of  what is done in section 4   for   the  three dimensional  case. 
  On often chooses  $\Eo ,
 \Eob,
{\cal  E}^t ,
  {\cal  E}^\t $ as a basis of exterior forms giving  the      decomposition  
\bea\red   d   = {\cal E}^\t {\cal D}_\t+  {\cal E}^0{\cal D}_0  + \Eo \Dz  +  \Eob    \Dbz
\eea
for    the  exterior differential operator $d =   d\t  \pa_\t+    dt\pa_0  + dz\p  + d \bz \bp $ 
and 
$ \o \equiv {\cal E}^\t {\o  }_\t+  {\cal E}^0{\o  }_0  + \Eo \o_Z  +  \Eob   \o  _{\bar Z}$ for the Spin connection.
     The     derivatives
 ${\cal D}_\t, {\cal D}_0,   \Dz ,      \Dbz$   are made explicit  in the formulae \eqref{derD}   of   Appendix A.

One    defines  the following    
   $SO(3) \subset SO(1,3)$ invariant
  matricial  decomposition     of the      Spin connection     $\o$
\bea\label{matrixo}
\ol ^{intrinsic} \equiv
\pbM
\o^z\cr
\o^\bz\cr
\o^0
\peM
\equiv 
\pbM
\o^z_Z
&
\o^z_\bZ
&
\o^z_0
&
\o^z_\t
\cr
\o^\bz_Z
&\o^\bz_\bZ 
&
\o^\bz_0
&
\o^\bz_\t
\cr
\o^0_Z
&
\o^o_\bZ
&\o^0_0
&\o^0_\t
\peM
\pbM
\Eo\cr
\Eob\cr
{\cal E}^t
\cr
{\cal E}^\t
\peM
\quad
\quad
\o^{extrinsic}\equiv 
\pbM
\o^{\t z}\cr
\o^{\t \bz}\cr
\o^{\t 0}
\peM
\equiv 
\pbM
\o^{\t z}_Z
&
\o^{\t z}_\bZ
&
\o^{\t z}_0
&
\o^{\t z}_\t
\cr
\o^ {\t \bz}_Z
&
\o^{\t \bz}_\bZ 
&
\o^{\t \bz}_0
&
\o^{\t \bz }_\t
\cr
\o^{\t 0}_Z
&
\o^{\t 0}_\bZ
&\o^{\t 0}_0
&\o^{\t 0}_\t
\peM
\pbM
\Eo\cr
\Eob\cr
{\cal E}^t
\cr
{\cal E}^\t
\peM.
\nn\\
\eea
All components of $\o$ must be computed    as  the solution of the    vanishing torsion four  conditions
\bea\label{T4}  
\begin{matrix}
 T^0&=&    de^0 \textcolor{\orange}{-}\demi \o^z \w e^\bz    \textcolor{\orange}{+} \demi   \o^\bz \w  e^z  -\o^{0\t} \wedge  e^\t   =0 \cr 
  T^\t&=&   \red  de^\t \textcolor{\green}{+} {\red {\demi} }\o^{\t z }\w e^\bz    + \demi   \o^{\t \bz }    \w  e^z  +\o^{\textcolor{\green}{\t 0}} \wedge  e^0=0  \cr 
 T^z&=&     de^z \textcolor{\orange}{-} \o ^0\w e^z    \textcolor{\orange}{+}   \o^z\w  e^0   -\o ^{z \t}  \w e^\t  =0    \cr
 T^\bz &=&       de^\bz \textcolor{\orange}{+} \o^0 \w e^\bz \textcolor{\orange}{-} \o^\bz \w e^0  -\o ^{\bz \t}  \w e^\t =0  .
 \end{matrix} 
 \eea  
{\eqref{T4}  provide  24 independent   equations linear in the components of  $\o$ displayed in   \eqref{matrixo}.   They  are      displayed in   Appendix A with  a unified notation that apply for both cases when one uses the $(z,\bz,t,\t )$ and 
 $(z,\bz,\t^+,\t^- )$  coordinates. Their   restriction 
for  the simpler case $d=3$        is      solved  in Appendix B. The  complete  four dimensional resolution  of \eqref{T4} and  the computation  of the     $d=4$ Einstein action as in~\eqref{L+-}~and~\eqref{L-+}
      is   quite more involved  than doing the analogous  work in the   case $d=3$. It  deserves further  work and  is to  be    published  in a separate     article specifically devoted  to the various aspects of the four dimensional~case.

\section { Generic    Beltrami parametrization   for  the leaf of leaf foliated  $d$-gravity}
Consider now   the generic    $d$-dimensional  case. 
The experience gained in $d=2,3$ and $4$ dimensions  clearly suggests   that,    to  possibly determine for any given value of $d>2$    a    Beltrami  parametrization  for 
the    $d^2$ components of the   generic   $d$-bein  of a $d$-dimensional  Lorentzian manifolds ${\cal M}_d$, 
one must   proceed   by      covariantly   gauge  fixing   its
$SO(1,d-1) \subset  SO(1,d-1)\times \Diff_d  $ Lorentz gauge  symmetry.
Since the $SO(1,d-1)$ gauge symmetry offers
         $\frac {d(d-1)} 2$   local   freedoms,  the number of       Beltrami $d$-dimensional fields is expected to    be 
         $\frac {d(d+1)} 2$.  Each one   of the Beltrami fields  is to   be  classified      by its Weyl symmetry and by its significance  in gravitational theories, independent of  the details of  the model     one wishes to build. Moreover    there must be a recursive   inclusion    of    the   $(d-1)$-dimensional   Beltrami metric  into the 
        the   $d$-dimensional one.
         
{  Before explaining      the details of the  generic   formulation of the Beltrami parametrization in  any given dimension of the manifold ${\cal M}_d$ one should 
 note that, once the Beltrami $d$-bein is     defined,  the     difficulty for   computing   the   Spin connection   associated to the Beltrami parametrization   is  basically the same   whether or not     gravity is coupled to matter fields  (and possibly to   auxiliary fields in supergravity).    
    Indeed, whichever the  gravity matter couplings are,   the Spin connection  $\o$ is systematically  defined as  the solution of  a linear constraint  
    that      differs from  the  pure gravity  vanishing  torsion    constraint   $de+\o(e)\wedge e=0$  by   the addition of  
    a  Lorentz  covariant~2-form. Indeed, the Spin connection equation  generally  reads      as follows
\bea \label{lorentzT}
de^a +   (\o^{a}_{b}-  e^cG_{cb}^{a}({ \rm matter\ and\  auxiliary\  fields })   )  \w e^b=0.
\eea     
       $G_{abc} $  is     a Lorentz covariant tensor according to the property  that 
  $G_3=\frac 1 6 G_{abc}  e^a\w e^b\w e^c$ is a $3$-form
   and it  locally depends     on matter   and/or auxiliary  fields \footnote{For  instance,   in the new minimal $d=4$     supergravity with gravitino $\Psi =\Psi_\mu dx^\mu$,  $G_3=\frac 1  6 G_{abc} e^a\w e^b\w e^c$  is nothing but the curvature of the auxiliary field 2-form $B_2$,   with
$G_3=dB_2 +\frac i 2 \Psi\gamma^a \w  \Psi \w e_a$ \cite{ml}.}.
Thus, if  one is able to compute the solution  $\o(e)$ of     the    linear   equation   $de+\o(e) \w e=0 $ as a  function of the~$\frac {d(d+1)} 2$~components of  the
Beltrami $d$-bein,     the solution of    \eqref{lorentzT} for any  given   $G^{abc}\neq 0$   is  
 the trivial  shift of this pure gravity solution  $\o^{a}_{b}(e) \to \o^{a}_{b}(e)+  e^cG_{cb}^{a}$.
 The~3-form~$G_3$~is~often 
       a complicated  function of the fields that couple to gravity.   However,  both   Lorentz symmetry of~$  G_{cb}^{a}$  and    $\Diff_d$~covariance of  the Beltrami  parametrization   warrantee the consistency of this computation of the Spin connection for~$G_3\neq0$.

}

\subsection {$d$-dimensional   Beltrami vielbein }
  This  section  defines  a  notation for the coordinate indices    that is  generically     more appropriate     than that     used 
  in sections 4 and 5 for   the  dimensions~$d=3 $ and~$ 4$.
 The    $d-1$ spatial  coordinates of  the   sub-foliated   ADM leaf~$\Sigma^{ADM}_{d-1}=  \Sigma_{2} \times  \Sigma_{d-3}  $  are 
      the    complex       coordinate  $z, \bz$  for $\Sigma_{2}$
and   the   $d-3 $  real  coordinate  of ~$\Sigma_{d-3}$ that are now denoted as~$x^i$~($i=3,... ,d-1$). 
The~Lorentz time coordinate is      called~$\tau$. 

The  $d$ coordinates  of the pseudo-Riemannian manifold ${\cal M}_d$ are  thus denoted    as    $(  z,\bar z, x^i,\t$).

\def \diagd{
\pbM
\exp \frac{ \Phi}{ 2}   &  0&0    &\ldots&0    &\ldots& 0 &0
\cr
0&
\exp \frac{ \Phi}{ 2}   &  0&  \ldots  &0     &\ldots& 0&0
\cr
0  &  0&N^3   &\ldots&0     &\ldots  & 0&0
\cr
\ldots
\cr
0  &  0&0   &\ldots&N^j  &  \ldots  & 0& 0&
\cr
\ldots
\cr
0  &  0&0   &\ldots&  0 &  \ldots  & N^{d-1}& 0&
\cr
0  &  0&0   &\ldots&  0 &  \ldots  & 0& N&
\cr
\peM
}

\def \MBd{
\pbM
\bM  
1&\m&\mu^z_3 & \mu^z_4  &\ldots&\mu^z_i&\ldots \mu^z_{d-1}& \mut
\cr
\mb &1&\mu^\bz_3 &\mu^\bz_4  &  \ldots&\mu^\bz_i&\ldots \mu^\bz_{d-1}& \mubt
\cr
 0 &0&1   &\mu^3_4  &\ldots&\mu^3_i&\ldots \mu^3_{d-1}& \mu^3_\t
 \cr
 0 &0&\ldots   &\ldots&\ldots&\ldots \ldots& \ldots
 \cr
 0 &0&\mu^j_3  &   \mu^j_4   &\ldots&1&\ldots \mu^j_{d-1}& \mu^j_\t
 \cr
 0 &0&\ldots   &\ldots&\ldots&\ldots \ldots& \ldots
 \cr
 0 &0&\mu^{d-1}_3   &   \mu_4^{d-1}    &\ldots&\mu^{d-1}_i&\ldots 1& \mu^{d-1}_\t
 \cr
 0 &0&\mu^\t_3  &\mu^\t_4  &\ldots&\mu^\t_i&\ldots \mu^\t_{d-1}& 1
\eM
\peM
}
\def\ngrav{N_{metric }}
 \def\nbel{N_{beltrami}}
  \def\ndiag{N_{diagonal}}
   \def\nphys{N_{physical}}

Now comes the definition of the 
$d$ one-forms      
$ e^a$ that compose the 
 ``Beltrami  $d$-bein". 
They  
  are 
       obtained  by the action  on the   $d$-vector  $(dz,d\bar z, dx^3,...dx^i, .... , dx^{d-1},d\t)$ of   a matrix that is  the product   
    of two  $d \times d$   squared matrices $ M_{diag}^{(d)} $ and   ${  M}   _ {(d)}  $. Both matrices are   of  a very different nature.
        $ M_{diag}^{(d)} $ is a~diagonal  matrix 
  whose    $d-1$ independent elements   concentrate the     dependence
  on the non Weyl  invariant fields of the Beltrami parametrization ; 
in contrast,   the 
    ``Beltrami  matrix"   ${  M}   _ {(d)}  $   is composed of the  Weyl invariant fields  of the Beltrami parametrization and it 
    reduces  to  the unit matrix when ${\cal M}_d$  is flat. 
    The Beltrami $d$-bein  $e^a$  is in fact   parametrized   by   $\frac {d(d+1)} 2 $   fields   defined as  follows :
   \bea\label {nbein}
   e^a
   =
 \pbM
\e\cr \eb\cr e ^3\cr \ldots  \cr e^i \cr    \ldots \cr   e ^{d-1}\cr   e ^\t
\peM
=
M_{diag}^{(d)} 
 \pbM
\Eo\cr \Eob\cr {\cal E} ^3\cr \ldots  \cr{\cal E} ^i \cr    \ldots \cr   {\cal E} ^{d-1}\cr   {\cal E} ^\t
\peM.
 \eea
 
    The   non Weyl invariant diagonal matrix  $ M_{diag}^{(d)} $  is parametrized by  the $d-1$     fields
    $\Phi, N^i$ and $N$ and reads as 
    \bea\label{nwbeld}
M_{diag}^{(d)}
\equiv\diagd.
 \eea
 
     The alternative   basis of one-forms 
$
\Eo, \Eob, {\cal E} ^3, \ldots ,{\cal E} ^i,    \ldots,   {\cal E} ^{d-1},   {\cal E} ^\t
$~in \eqref {nbein}  is defined by the    $d$-dimensional    Weyl invariant    Beltrami     $d\times d$ matrix  ${  M}   _ {(d)}  $ such that  
\bea \label{clef}
\pbM
\Eo\cr \Eob\cr {\cal E} ^3\cr \ldots  \cr{\cal E} ^i \cr    \ldots \cr   {\cal E} ^{d-1}\cr   {\cal E} ^\t
\peM
\equiv 
{ \cal  M}_{(d)}  
\pbM
dz\cr d\bz\cr dx ^3\cr \ldots  \cr dx ^i \cr    \ldots \cr   d x ^{d-1}\cr   d \t
\peM
\eea 
with 
\bea\label{beld}  
{ \cal M}_{(d)}  \equiv \MBd .\eea
The    $(d-2)\times (d-2)$ squared  sub-matrix that is cornered in    the   right   bottom sector  of  the   Beltrami  matrix ${\cal M}_{(d)}$   is     antisymmetric, that is,
\bea  \label{skew}
\mu^i_\t =-\mu_i^\t 
\quad   \quad 
\mu^i_j =-\mu^j_i \quad   \quad 3\leq    i,j \leq d-1.
\eea
%

       ${\cal M}_{(d)}$ is  the  $d$-dimensional generalization  of   the
$2\times 2$ matrix
\eqref{MuMu} for $d=2$, of the   
 $3\times 3$ matrix
\eqref{trois}     for~$d=3$ and of the 
$4\times 4$ matrix  \eqref{4bein}  for $d=4 $, patiently     constructed in sections 3,4 and~5.  

 The   generic~antisymmetry property   \eqref{skew}  cannot be detected for   $d=3$   since    $\Sigma_{d-3}$ reduces to a point in this case.
The $a$ field dependence of  \eqref{4bein}  is  early signal     that reveals the necessity of the  antisymmetry condition 
 \eqref{skew} of the Beltrami matrix. (\eqref{beld}  justifies   the consistency  of  expressing   $a\equiv \mu^0_\t$ in the notation of section 5 where the third spatial coordinate  is called $x^0$.)

  The generic  structure   
  of the   Beltrami     $d$-matrix     \eqref{beld} suggests   quite evidently   that     the well known 
  parametrization of Riemann surfaces modulo Weyl transformations  by the 
  $d=2$ Beltrami differential    $dz+\m d\bz$  is  the tip of an iceberg but  its  existence is   revealed by  the not so trivial   
  leaf of leaf foliation process   of   manifolds~${\cal M}_d$ according to $\Sigma^{ADM}_{d-1}
  = \Sigma_{2}\times \Sigma_{d-3}$  that only makes sense for   $d  \geq3$ .

  The covariant reduction   of    the    $d^2$ independent 
   components of a generic  $d$-bein in function of     the  $\frac { d (d+1)  } 2$~ Beltrami fields that   parametrize the Beltrami $d$-bein   \eqref{nbein}  is  a covariant gauge  fixing of the  $ \frac { d(d-1)  }2$ freedoms offered by  
 the    local Lorentz symmetry $SO(1,d-1 )\subset  SO(1,d-1 )\times \Diff_d $. The  remaining~$ \Diff_d $~symmetry   transforms consistently all     these  $\frac { d (d+1)  } 2$   Beltrami fields.   
 If one uses the language   of  the BRST  invariant  quantum field theories, the gauge fixing  of a general  $d$-bein~$e^a_\mu$  down to its associated    Beltrami $d$-bein  \eqref {nbein}   provides  a trivial Faddeev--Popov  determinant that is consistent  with  the $\Diff _d$ symmetry as  a mere generalization of  what is done in  the previous sections for  $d=2,3,4$.

 The   BRST formalism  introduces as many  independent  anticommuting  reparametrization  vector  ghost    fields~$\xi^\mu$ as  there are local  parameters for the  infinitesimal reparametrizations  transformations.
   The point is that  the~$d$~independent  ghost    fields $\xi^\mu$  can    be redefined    into   the  $d$  Beltrami ghosts $c^a,\  a=(z,\bz,i,\t)$.    The~generic~correspondence  between $\xi^\mu$ and $c^a$  is  to be   given in  the further written  formula \eqref{belgh}  that generalizes   in  any given arbitrary~dimension 
 both     formulae  \eqref{belg} and 
     \eqref{tilde} for $d=2$ and $d=3$.  The experience  gained in these previous sections  is such that one 
     one can  thus safely anticipate
        that  using  of the Beltrami ghost fields    $c^a$  in place   of    the ghosts  $\xi^\mu$ greatly simplifies   the geometrical   determination  of the  BRST transformation laws 
        of all   the classical 
         $\frac { d (d+1)  } 2$ Beltrami
          fields   and thereby   the determination of their infinitesimal   reparametrization transformations.
        Moreover, the   
      BRST invariance of  the   $\frac { d (d-1)  } 2$  constraints  
      that define the Beltrami $d$-bein   \eqref{nbein}    provide     algebraical constraints whose expressions imply that the $d$-dimensional  Lorentz ghosts are  themselves constrained to  well-defined 
          local functionals of the reparametrization~ghosts~$c^a$~\footnote{ These effective  Lorentz ghosts can be used to study   the local Lorentz invariance in the Beltrami formulation for instance for the study
        of gravitational anomalies.}.
          All~needed~technical aspects  for   computing  the BRST transformation laws  of the Beltrami fields   are  cautiously  explained 
              in~section~4~for~$d=3$. Their $d$-dimensional generalization for $d>3$ is quite straightforward  so that    the further  section  6.6 that is concerned by  this question is  to be  very~short. 
      
 \subsection{Counting the degrees of freedom  of  $d$-gravity in the      Beltrami parametrization   }

The  number of the  Weyl invariant fields  that compose the $d$-dimensional 1-forms  $
\Eo, \Eob, {\cal E} ^3, \ldots ,{\cal E} ^i,    \ldots,   {\cal E} ^{d-1},   {\cal E} ^\t
$,~that is, the number of fields  that   parametrize   the Beltrami  $d\times d$ matrix  ${\cal M}_{(d)}$      \eqref{beld},
 is
    \bea \nbel =2(d-1) + \frac { (d-2)(d-3) }2
=\frac { d^2- d +2}2.\eea
This counting  takes into account the  antisymmetry relations \eqref{skew}. The  number of independent  Weyl  non invariant fields   fields    $\Phi, N^3,... ,N^{d-1},N$    that  parametrize  $M_{diag}^{(d)}$  is 
 \bea
\ndiag= d-1.\eea 
  One can thus check  the  number of  the    Beltrami fields that parametrize  the Beltrami $d$-bein  \eqref{nbein}   matches the number of components of a generic $d$-metric according to  
 \bea
 \nbel + \ndiag=  \frac { d^2- d +2}2 +d-1
 =  \frac { d(d+1)  }2  = \ngrav.
 \eea

  \subsection{Beltrami expression    of     the   $\frac {d(d-3)} 2$   physical propagating  gravitational degrees of freedom }
 Classically, the number of physical propagating  degrees of freedom of  $d>2 $ gravity is  well known to be  
 \bea
 \nphys=  \frac {d(d-3)} 2.
 \eea
 York \cite{York} has proved that   these  gravitational physical degrees of freedom can be consistently represented 
  at the classical level  
   by the equivalent  classes of the $d-1$ inner metrics  of   spatial ADM leaves, defined modulo  the Weyl$\times \Diff_{d-1}$ symmetry.      
    An    elementary  consistency check   of the  correctness of this proposition  is  by the counting  $\frac {d(d-1)} 2-1-(d-1)= \frac {d(d-3)} 2$
    or $\frac {(d-2)(d-1)} {2}-1 = \frac {d(d-3)} 2$.

This paper  equivalently    postulates  that the  quantum observables      correspond  to  expectation values of   functionals of  a given subset of the    Weyl invariant  fields that parametrize   of the Beltrami~$d$-bein.

 This leaves no choice but   to select the fields that compose this subset    among the  $  \nbel=\frac { d^2- d +2}2$  components of  ${\cal M}_{(d)}$ while    leaving aside the fields  that compose the  ADM shift vector.
 
 This postulate   implies    that  one checks that all   the 
 non Weyl invariant Beltrami fields that 
 that compose    the diagonal matrix ${\cal M}_{diag}$    have no   physical dynamic.  They are   $\Phi $, which determines the conformal factor
 $\exp \Phi$,~$N$ and the    various rescaling factors $N^i$    (whose interpretation  will be further  clarified    after the writing of the Beltrami metric  in  \eqref{ds2d}). 

The number of the Weyl  invariant fields  that are displayed     in  ${\cal M}_{(d)}$
  is such that
  \bea
  \nbel -\nphys =\frac { d^2- d +2}2  -\frac { d(d-3) }  2=d+1.
  \eea
  This indicates that    $d+1$  fields  among the   Weyl invariant Beltrami  fields that parametrize the Beltrami matrix~${\cal M}_{(d)}$      must be considered as unphysical~ones. 
  
  The  fact that  $ d+1=d -1  +2$ is suggestive enough to identify  these  $d+1$   fields. They are  respectively  the~$d-1$~fields  $\mu^z_\t, \mu^\bz_\t, \mu^i_\t$  
and both  components  $\mu^z_\t, \mu^\bz_\t$ of the Beltrami differential  of $\Sigma_2$.  

  The     QFT explanation of this    generalizes that given for the  $d=2,3$ and $4$ cases. The $d-1$~fields~$\mu^z_\t, \mu^\bz_\t, \mu^i_\t$~are nothing but  a parametrization of the   ADM    shift  1-form field, as made explicit  in the further equation 
  \eqref{shift} obtained    by comparing  the  Beltrami metric \eqref{ds2d} (to be shortly established)   and  the ADM formula
  $ds^2=   -{\cal N }d\tau^2 +(dx^m +  \beta^m d\t )  g_{mn}  (dx^n+  \beta^n d\t ) $, where $m,n=(z,\zb,i)$.    It follows  
 that   the  excitations  of the $d-1$~Weyl invariant  fields $\mu^z_\t, \mu^\bz_\t, \mu^i_\t$   cannot be considered as parts of the physical degrees of freedom since we know from the ADM analysis that the shift fields don't have  canonical momenta at the classical level (as well as the lapse~$\cal N$~in~\eqref{lapse}).

As for  the fields  $\m$~and~$\mb$,~they are  the Beltrami differentials of $\Sigma_2$ at    fixed $\t$ and $x^i$.    They   characterize the     Riemann surfaces~$\Sigma_2$~that     sub-foliate each   ADM leaf    decomposed according to  $\Sigma^{ADM}_{d-1}= \Sigma_{2}\times \Sigma_{d-3}$.      They~can be~actually~gauge fixed by using    two freedoms that one may  covariantly  choose among  those of the  $\Diff_{d}$~symmetry,  
     according~to   the   same equation  as   in  \eqref{genus}, namely
   \bea\label{modn}
  \m=\gamma
  \quad\quad
    \mb= \bar \gamma,
 \eea
where    $
 \gamma=\sum _{k=1}^{g-3} \lambda_k f^k(z,\bz)
 $. Because the fields   $\m$ and $\mb$ can be gauge fixed as  the  moduli  of  $\Sigma_2$,  it is  consistent not   to count them as parts of the physical degrees of freedom of $d$-gravity.
 
The proposal   made in this paper is   thus that the $\frac { d(d-3) }  2$  physical propagating   degrees of freedom  of    $d$-dimensional gravity must  be    identified  with  the     Beltrami   fields $\mu^z_i, \mu^\bz_i, \mu^j_i$
that compose    the  following  $(d-1)\times (d-1)$ subpart of the Beltrami matrix \eqref{beld} 
 \bea\label{belp}
 { M}_
  {  Physical \  dofs    }
 =
 \pbM
\bM  
1&\gamma&\mu^z_3 & \mu^z_4  &\ldots&\mu^z_i&\ldots \mu^z_{d-1} 
\cr
\bar\gamma &1&\mu^\bz_3 &\mu^\bz_4  &  \ldots&\mu^\bz_i&\ldots \mu^\bz_{d-1} 
\cr
 0 &0&1   &\mu^3_4  &\ldots&\mu^3_i&\ldots \mu^3_{d-1} 
 \cr
 0 &0&\ldots   &\ldots&\ldots&\ldots \ldots& \ldots
 \cr
 0 &0&\mu^j_3  &   \mu^j_4   &\ldots&1&\ldots \mu^j_{d-1} 
 \cr
 0 &0&\ldots   &\ldots&\ldots&\ldots \ldots& \ldots
 \cr
 0 &0&\mu^{d-1}_3   &   \mu_4^{d-1}    &\ldots&\mu^{d-1}_i&\ldots 1& 
\eM
\peM
 \eea
where 
$\mu^i_j=-\mu_i^j$ and $\m$ has been gauge fixed equal  the moduli  $\gamma$ of $\Sigma_2$.
One can     check that the number   of  the independent  fields in   ${  M}_
  {  Physical \  dofs    }
$   is   $\frac{d(d-3)}2$    according to     $2(d-3) +\frac{(d-3)(
d-4)}2=  \frac{d(d-3)}2$ or equivalently $2+3+... +(d-2)= \frac{(d-2)(d-1)}2 -1 = \frac{d(d-3)}2$. 

The   definition of  the  $d$-dimensional propagating gravitational  physical states as  
   the  $ \frac{d(d-3)}2$ generalized  Beltrami differential  components  $\mu^m_j$, $m= (z, \bar z, i,j=3,... ,d-1)$ that  compose the matrix  \eqref{belp}  at fixed moduli~$\gamma$ is      an  alternative and interesting 
  proposition.
   The   perturbative  excitations  of  these fields      correspond to the   traceless and transverse excitation of  $d$-metrics.

 An   attractive  feature  of this definition  of the  physical gravitational degrees of freedom is of  relying on a geometrically  well-identified      subset  of the fundamental local fields of the theory.  It  may render  easier  a  formal definition of the mean values of the gravitational physical observables within a  (yet to be  defined)  path~integral  formalism where the measure is expressed in terms of the Beltrami fields, as a generalization  of  the bidimensional case where the classical part of the
 gravitational  measure is simply $[d\Phi] [d\m][d\mb]$. Moreover, the genus dependence of    the sub-foliating    Riemann  surfaces~$\Sigma_2$~is encoded in    \eqref{modn} that  expresses the gauge~fixing of $\m$ and $\mb$.   
Following this partial  gauge fixing of the  $\Diff_d$ invariance, the left upper  $2\times2$~sub-matrix  $\pbM  1\gamma\cr \bar \gamma 1  \peM$ is what is left   from  the internal metric  of    $\Sigma_2$~modulo      Weyl invariance.  This~partial~gauge fixing  might  be   a way to systematically encode the    relevant topological information about the  the spatial sub-manifold 
  $ \Sigma^{ADM}_{d-1}$ in a path integral  formulation. This  point    certainly deserves more  clarifications.  
  
One can  conclude  this discussion    with  a simple remark     concerning the  dimensional reduction.   The Kaluza--Klein compactification of the genuine $d$-gravity  defines a
  $(d-1)$~dimensional gravitational 
   theory with  a  $(d-1)$-dimensional metric coupled to a    1-form   gauge field $A_\mu dx^\mu$ and a scalar field  $\phi$. $\phi$   carries one physical degree of freedom.  Since   the number of physical degrees of freedom of the $d$-dimensional  graviton is $  N^{d}   _{\rm physical  \ graviton} =\frac {d(d-3)}2$ and since  the compactification   conserves the number of physical degrees of freedom, one has
  \bea\label{compact}
  N^{d}   _{\rm physical  \ graviton}  -  N^{d-1}   _{\rm physical  \ graviton} = d-2= N^{d-1}   _{\rm physical \ gauge  \ field}  +1
  \eea             
  This implies    that  $N^{d-1}   _{\rm physical \ gauge  \ field}=d-3$. One finds  therefore that 
 the  counting of the    physical degrees of freedom of the graviton consistently  predicts that  a  $d$-dimensional  gauge field    carries  $ N^{d}   _{\rm physical \ gauge  \ field}=     d-2$  physical degrees of freedom.  
 This simple  counting  gives  the way to express the dimensional reduction of a  Beltrami $d$-bein into a Beltrami $(d-1)$-bein completed with a $(d-1)$ gauge field and a scalar, both related to the Beltrami fields in~$d$~dimensions. 
 The argument can be repeated for    further compactifications.  
   \eqref{compact}  is   a simplistic    but  quite physical  justification for the  consistency of the  definition of   the gravitational degrees of freedom as the above well-defined subset of the Beltrami  fields.

  \subsection {A  ``physical" gravitational  gauge choice}
    
Once  $\m$ and $\mb$ are   gauge fixed to be   equal to the moduli     $\gamma$ and $\bar \gamma$, of $\Sigma_2$    one must 
go on and complete  the gauge fixing 
 of  the   $d-2$  freedoms     of  the the $\Diff_d$  symmetry of ${\cal M}_d$.
 A   suggestive  completion  of  the  gauge fixing    is  by further imposing    $\Phi=0$ and the 
 $d-3$ conditions  $N^i=1$. One then gets
 \bea\label{pchoice}
 {  M}^{(d)}_{diag}=
 \pbM1  &  0&0    &\ldots&0    &\ldots& 0 &0
\cr
0&
1  &  0&  \ldots  &0     &\ldots& 0&0
\cr
0  &  0&1  &\ldots&0     &\ldots  & 0&0
\cr
\ldots
\cr
0  &  0&0   &1   &  \ldots  & 0& 0&
\cr
\ldots
\cr
0  &  0&0   &\ldots&  0 &  \ldots  & 1& 0&
\cr
0  &  0&0   &\ldots&  0 &  \ldots  & 0& N&
\cr
\peM.
 \eea
 The remaining  field dependence  of the  gauge fixed Beltrami $d$-bein is through  the lapse  $N$,   the      components   $\mu^z_\t, \mu^\bz_\t, \mu^i_\t $   of the ADM  shift fields and      the    $\frac {d(d-3)}2$     propagating physical Beltrami   fields.  
 The  $d$-dimensional   Beltrami metric will be        shortly  made explicit in this gauge.
  
  %
%
%
%
  \subsection {Beltrami $d$-metric and its possible gauge fixings} 
 The   reparametrization invariant  $d$-metric associated  to the  $d$-bein    $ e^a_\mu $  in  \eqref{nbein} is 
$g_{\mu\nu}   =\eta_{ab}    e^a_\mu   e^b_\nu $.   This yields
 the following  formula for   the Beltrami $d$-metric of ${\cal M }_d$ : 
\bea \label{ds2d}
\demi ds^2\=-N^2 \Big ( 
d\t   +\sum_{i=3}^{d-1}  \mu^\t _i dx^i \Big )^2\nn\\
&&+
\exp \Phi  \Big   { |} \Big { |}
dz+\m d\bz +\mu^z_3 d x^3 +\ldots \mu^z_{d-1}  dt^ {d-1}+ \mut d\t
\Big { |}\Big{|}
^2
\nn\\
&&
+\sum_{i=3}^{d-1} \sum_{j\neq i, j=3}^{d-1}    {N^i}^2 
 \Big
(
\mu^i_3dx^3
+... 
+\mu^i_{j-1}dx^{j-1}
+dx^i
+\mu^i_{j+1}dx^{j+1}+
... 
+
\mu^i_{\t }d\t
\Big
)^2, 
\eea
(One must enforce  the antisymmetric relations $ \mu^i_\t =-\mu_i^\t $ and 
$
\mu^i_j =-\mu^j_i  $ in this formula.)

 \eqref{ds2d}  is     quite       basic.   It allows   to precisely relate the Beltrami  fields  $  N$ and    $\mu_\t ^a\equiv \mu_\t^z, \mu_\t^\bz, \mu_\t^i$  to  the    time   lapse $\cal N$ and  the shift one-form $\beta_m$ defined by the ADM metric. 
 Indeed, the comparison         of   the Beltrami metric \eqref{ds2d}  to  the   ADM metric
$ds^2= -{ \cal N}^2     +   (dx^m+ \beta^m  d\t  )g_{mn}(dx^n+\beta^n  d\t  )
$ for $m=z,\bz,i$ provides the following formula for      $\cal N$  and $\beta_m$
         \bea\label{lapse}{\cal N}^2= N^2    -  \exp \Phi \mu^z_\t \mu^\bz_\t     -   \sum_i N_i^2     {\mu^i_\t}^2 \eea   
         \bea
         \label{shift}
         \beta_z=\exp\Phi    ( \mu_\t^\bz        +\mb \mu^z_\t) \quad   
       \beta_\bz    = \exp\Phi (   \mu_\t^z  +\m  \mu^\bz_\t)  \quad 
      \beta_i=   (N^2- {N^i}^2) \mu_\t^i  -\exp \Phi (\mu^z_i \mu^\bz_\t  +\mu^\bz_i  \mu^z_i  ) 
      - \sum_{j \neq i} {N^j}^2 \mu^j_i \mu^j_\t   .\quad 
  \eea

 These  relations  between the   ADM lapse and  shift fields $N$  and $\beta_m$ and the  Beltrami  fields $N$ and   $\mu^\beta_\t$   indicate that the possible excitations  of the latter carry no propagating  physical degrees of freedom.     The     $d-3$  fields  $N^i$    appear    as  dilatation factors for the $d-3 $ coordinates in $\Sigma_{d-3}$,   and their excitations     are neither expected to be  counted as parts of the  physical gravitational fields  degrees of freedom like 
the~conformal~factor~$\Phi$~\cite{York}.     The~fate~of the  fields    
   $\m$ and $\mb$  is to  be gauge fixed equal  to the moduli  of  $\Sigma_{2}$  as in \eqref{moduli}.      
   The  remaining   $\frac {d(d-3)}2$  Weyl invariant      fields   $\mu^a_i= ( \mu^z_i, \mu^\bz_i, \mu^j_i)$  that compose the    Beltrami   metric    \eqref {ds2d}   must    be  therefore identified with the  propagating physical degrees of freedom of gravity  as already claimed  at the level of the Beltrami $d$-bein.

 One possible way   to compute  the Einstein action in function of the Beltrami fields is by expressing it as    a quadratic form  in  the  Spin connection that generalizes the three and four dimensional formulae  \eqref{Einstein3d} and~\eqref{E4}.~In a  presumably   less illuminating way,   one might go ahead and    compute   the Christophel symbols for the Beltrami  metric  \eqref{ds2d} and their derivatives and  then the  Einstein action  by its standard expression in terms of these entities. It is not   obvious  that  the current algebraic general relativity  softwares   can be straightforwardly adapted   to compute  the Einstein action under a satisfying  form       such as  \eqref{Einstein3d},  with   \eqref{ds2d}  as an input.

One can now address the question of the gauge fixing of the Beltrami metric.
The understanding of the nature of the Beltrami fields  suggests two natural choices for gauge fixing the remaining 
local $d$ freedoms under the  $\Diff_d$ symmetry of the Beltrami metric.
 
 The first one  is     
to impose       the $  d= (d-3) +2+1$  gauge conditions
  $\mu^i_\t =-\mu_i^\t =0$,   
$\m =\gamma$,  $\mb =\bar \gamma$, 
$N=1$.   This defines the following gauge fixed  metric :
 $$
  ds^2=
  -d\t^2   +
\exp \Phi      { ||}  
dz+\gamma d\bz +\mu^z_3 d x^3 +\ldots+ \mu^z_{d-1}  dt^ {d-1}+ \mut d\t
||
^2$$ \bea
+\sum_{i=3}^{d-1} \sum_{j\neq i, j=3}^{d-1}    {N^i}^2 
  (
\mu^i_3dx^3
+... 
+\mu^i_{j-1}dx^{j-1}
+dx^i
+\mu^i_{j+1}dx^{j+1}+
... 
+\mu^i_{d-1}dx^{d-1}
)^2.
\eea
The spatial lapse interpretation of the  $d-3$ fields $N^i$  will be clarified in a separate publication by using this~gauge.

Alternatively one can use   the  $d$ gauge   conditions   made of  $\Phi=0$,  
  $N^i =1$,       
$\m =\gamma$ and   $\mb =\bar \gamma$.
(A possible variant  is  by  replacing the condition  $\Phi =0$  by the   unimodular gauge condition.)  
This   gauge choice   is   that already      already suggested at the level of the Beltrami~$d$-bein in  \eqref{pchoice} and provides the following expression for the metric :
\bea \label{ds2dgf}
\demi ds^2\=-N^2 \Big ( 
d\t   +\sum_{i=3}^{d-1}  \mu^\t _i dx^i \Big )^2
 +
  \Big { |}
dz+\gamma d\bz +\mu^z_3 d x^3 +\ldots \mu^z_{d-1}  dt^ {d-1}+ \mut d\t
\Big { |}\Big{|}
^2
\nn\\
&&
+\sum_{i=3}^{d-1} \sum_{j\neq i, j=3}^{d-1}    
 \Big
(
\mu^i_3dx^3
+... 
+\mu^i_{j-1}dx^{j-1}
+dx^i
+\mu^i_{j+1}dx^{j+1}+
... 
+
\mu^i_{\t }d\t
\Big
)^2.
\eea
The  gauge fixed metric  \eqref {ds2dgf}  is  only  function of       the   $\frac {d(d-3)}2$  gravitational    physical  propagating degrees of freedom~$\mu^M_i$~and of   $\mu^m_\t$  and $N$.   $\mu^m_\t$  and $N$   compose  the ADM shift vector and the   lapse function. 
 The~Einstein~Lagrangian   reduces in this gauge to  a     polynomial  function  of   the physical  fields $ \mu^z_i, \mu^\bz_i, \mu^j_i$
 and their derivatives with an          algebraic dependence  on both  ADM          lapse and shift  functions $\cal N$ and  $ \beta_m$ expressed in \eqref{lapse} and \eqref{shift}.   
 A solution such that $\beta _m=0$ and ${\cal N}=1$ is for $\mu^m_\t=0$. If this  constraint is a consistent one,   one reaches 
   a     situation where  one ends up with a gauge fixed formulation for the Einstein action   whose dynamics only depends on the physical degrees of freedom  $\mu^z_i, \mu ^\bz_i, \mu^i_j$,  geometrically well defined from the beginning, with  their own classical momenta $p^z_i, p^\bz_i, p^i_j$. This  choice of gauge  formally suggest  the existence of a  covariant physical Hamiltonian phase space with $\frac {d^2(d-3)^2}9$ components.  This   sustains the appellation of  a  "physical gauge choice" that was already done in section 6.4.
 
{ {More ingenious gauge fixings  of the  fields    selected  by the Beltrami parametrization  might suggest  new      hints      for a better    definition of the quantum gravity rules   keeping in mind   that  the gravitational physical degrees of freedom    are   identified  as  a well-defined  subset  of  the     fields  that parametrize the metric   \eqref{belp}.
 It is however quite clear  that the fundamental questions about a consistent definition of quantum gravity remain as they are 
  within   the Beltrami  parametrization  independent of      its mathematical appeal and       physical motivations. }}

  \subsection{BRST symmetry}
  
     Terms that involve the    ghosts    must be systematically added to the gauge fixed  Einstein action to ensure  that  the whole action is BRST invariant for any given gauge choice.  
   The method       explained  in   sections  3 and 4 that   geometrically determines
   the BRST symmetry transformations on all Beltrami fields and their ghosts  
   in the cases 
     $d=2$ and $d=3$         
     generalize straightforwardly  for all   values of the space dimension $d$.
     One can     indeed    generically unify
$ {\cal E}^a   \to   {\cal E}^a   +c^a    $,   $a=z,\bz$, $   3  \leq i \leq   d$. The relation  between the Beltrami  ghost   $c^a$  and  the $d$-dimensional   standard reparametrization   ghost  $\xi^\mu$ is
\bea \label{belgh}
c^a\equiv\exp  i_\xi \ {\cal E}    ^a
\eea 
and   the determination of  the BRST symmetry  transformation laws   acting  all on fields of  the generalized  Beltrami $d$-dimensional  parametrization of the metric 
   \eqref{nbein}    and their ghosts  proceeds  by expanding in  form degree and ghost  number the $d$-dimensional  horizontality equations that simply generalize    \eqref{magicalbrst}.

\section {Spaces with  special holonomy ${\cal G} \subset   SO(d-1,1)$}
The    Beltrami  $d$-parametrization   \eqref{ds2d}     can be further  simplified          if the     space-time has a spatial holonomy.  The~resulting   larger freedom    allows one to  further simplify the form  of the Beltrami   $d$-bein.
 This section exemplifies the general process  with  the case     of    eight dimensional   manifolds with  holonomy $G_2 \subset  SO(7,1) $.
 \def\M7{M}
  \def\mt7{a}
  
The   formula 
 \eqref{nbein}  defines  the    Beltrami   eightbein for a general eight dimensional manifold ${\cal M}_8$  as follows :
\bea\label{8bein}
 \pbM
\e\cr \eb\cr e ^3\cr e^4  \cr e^5 \cr    e^6 \cr   e ^{7}  \cr e ^\t
\peM
=
\pbM
\exp\Phi&0&0&0 &0 &0 &0 &0
\cr
0&\exp\Phi &0&0 &0 &0 &0 &0\cr
0&0 & N^3  &0 &0 &0 &0 &0
\cr0&0 & 0  &N^4 &0 &0 &0 &0
\cr
0&0 & 0  &0 &N^5 &0 &0 &0\cr
0&0 &0  &0 &0 &N^6 &0 &0
\cr
0&0 & 0 &0 &0 &0 &\M7 &0
\cr
0&0 & 0 &0 &0 &0 &0 &N
\peM
\pbM
1&\m&\mu^z_3&\mu^z_4 &\mu^z_5 &\mu^z_6 &\mu^z_7 &\mu^z_\t
\cr
\mb&1&\mu^\bz_3&\mu^\bz_4 &\mu^\bz_5 &\mu^\bz_6 &\mu^\bz_7 &\mu^\bz_\t
\cr
0&0&1 &\mu^3_4 &\mu^3_5 &\mu^3_6 &\mu^3_7 &\mu^3_\t
\cr
0&0&-\mu_4^3   &1 &\mu_4^5 &\mu^4_6 &\mu^4_7 &\mu^4_\t
\cr
0&0&-\mu_5^3    &-\mu_5^4  &1 &\mu^5_6 &\mu^5_7 &\mu^5_\t
\cr
0&0&-\mu_6^3   &-\mu_6^4 &-\mu_6^5 &1  &\mu^6_7 &\mu^6_\t
\cr
0&0&-\mu_7^3  &-\mu_7^4 &-\mu^7_5 &-\mu_7^6 &1 &\mt7
\cr
0&0&-\mu_\t^3 &-\mu_\t^4 &-\mu_\t^5 &-\mu_\t^6 &-\mt7 &1
\peM
\pbM
dz\cr d\bz\cr dx ^3\cr dx^4  \cr dx^5 \cr    dx^6 \cr   dx ^{7}  \cr d\t
\peM.\nn\\
\eea
  One may   consider the   class of  the eight  dimensional   manifolds with   holonomy $G_2  \subset SO(1,7)$,   $G_2$ being the simplest  exceptional    rank~$2$~group  with    $14$ generators.  
   This extends  the  number of covariant  constraints   that one can impose  to  the~$64$~components of a general $8$-bein from 
  the  $28$   coming from the    $SO(1,7)$~Lorentz  gauge freedoms  and  resulting to  \eqref {8bein}    to  the higher value $42=14+28$. 
The    remaining  $14$~freedoms   that result  from    the $G_2$ holonomy of the manifold  allow   further  covariant constraints for  the matrix elements of \eqref{8bein}, allowing   to express    the   Beltrami eightbein as  follows :
\bea
\label{gh2}
 \pbM
\e\cr \eb\cr e ^3\cr e^4  \cr e^5 \cr    e^6 \cr   e ^{7}  \cr e ^\t
\peM
=
\pbM
\exp\frac \Phi  2  &0&0&0 &0 &0 &0 &0
\cr
0&\exp\frac \Phi  2 &0&0 &0 &0 &0 &0\cr
0&0 & N^4  &0 &0 &0 &0 &0
\cr0&0 & 0  &N^4 &0 &0 &0 &0
\cr
0&0 & 0  &0 &N^6 &0 &0 &0\cr
0&0 &0  &0 &0 &N^6 &0 &0
\cr
0&0 & 0 &0 &0 &0 &\M7 &0
\cr
0&0 & 0 &0 &0 &0 &0 &N
\peM
\pbM
1&\m&\mu^z_3&\mu^z_4 &\mu^z_5 &\mu^z_6 &\mu^z_7 &\mu^z_\t
\cr
\mb&1&\mu^\bz_3&\mu^\bz_4 &\mu^\bz_5 &\mu^\bz_6 &\mu^\bz_7 &\mu^\bz_\t
\cr
0&0&1 &\mu^3_4 &0&0&0&0 
\cr
0&0&-\mu_4^3   &1 &0 &0 &0 &0
\cr
0&0&0    &0&1 &\mu^5_6 &0 &0
\cr
0&0&0 &0&-\mu_6^5 &1  &0 &0
\cr
0&0&0 &0 &0 &0 &1 &\mt7
\cr
0&0&0&0 &0 &0 &-\mt7 &1
\peM
\pbM
dz\cr d\bz\cr dx ^3\cr dx^4  \cr dx^5 \cr    dx^6 \cr   dx ^{7}  \cr d\t
\peM.
\eea
{
The  corresponding    Beltrami  $ d=8$  metric  is    
\bea
\label{ds2=8s}
\demi ds^2\= -  N^2 (d\t -\mt7 dx^7 )^2 + \M7 ^2 (dx^7+\mt7 d\t )^2 
\nn\\
&&
+{N^4}^2\big  (\  {dx^3}^2+{dx^4}^2   + {\mu^3_4}^2 ( {dx^3}^2-{dx^4}^2  )  \  \big )
  +{N^6}^2\big   ( \ {dx^5}^2+{dx^6}^2   + {\mu^5_6}^2 ( {dx^5}^2-{dx^4}^2 )
\nn\\
&&+
\exp \Phi  \Big {|}
 \Big{ |}
dz+\m d\bz
 +\mu^z_3 d x^3 
+\mu^z_4 d x^4
+\mu^z_5 d x^5 
+\mu^z_6 d x^6 
+\mu^z_7 d x^7 
 + \mut d\t
  \Big {|}
  \Big {|}
 ^2.
\eea
This formula exhibits  some  analogy    with he  generic   four dimensional  Beltrami metric     
   \eqref{ds}.
   
One can furthermore use the  eight
 gauge freedoms of the remaining  $d=8$ reparametrization symmetry.   
          {   A possible gauge choice for the $\Diff_8 $ invariance is 
$a=0, \ 
\M7
 =   
 N 
  =
N^4
=N^6,  
  \m   =\gamma, 
   \mb   q
   =\bar\gamma, 
 \mu_4^3
  =
  \mu_6^5=    0
  $.
This~defines a coordinate system in which  the  $ d=8$ metric with holonomy $G_2$  is  
   \bea
\label{ds2=8r}
\demi ds^2
=N^2(-
d\t  ^2   +{dx ^7 
} ^2  
+
 \big(
 {dx^3}^2+
 {dx^4}^2   + 
   {dx^5}^2+
   {dx^6}^2  ) \big)
\nn\\
+ 
\exp \Phi
 \Big {|}
 \Big{ |}
dz+\gamma d\bz
 +\mu^z_3 d x^3 
+\mu^z_4 d x^4
+\mu^z_5 d x^5 
+\mu^z_6 d x^6 
+\mu^z_7 d x^7 
 + \mut d\t
  \Big {|}
  \Big {|}
 ^2.
\eea
This fully gauge fixed  metric \label{ds2=8r}    can be compared  to   the gauge fixed 
four dimensional    metrics~\eqref{dsrr}.
}
 
}

\section{Conclusion }

This  work   presents a    generalization of  the bidimensional  Beltrami parametrization for gravity and    theories coupled to gravity    that is   valid in  all   dimensions $d>2$.     The  Beltrami parametrization  for $d\geq 3$ is made possible by 
 a covariant sub-foliation    of   the   ADM leaves~$\Sigma^{ADM}_{d-1} \sim \Sigma_{d-3}\times \Sigma_{2} $~of  
 Lorentzian $d$-dimensional  manifolds~${\cal M}_d$.  The~found  expressions of   the~$d>2$ Beltrami  $d$-bein and associated Beltrami $d$-metric  derive   from  a    covariant gauge fixing 
of the~$\frac {d(d-1)}{2}$~local  freedoms offered by  the     Lorentz~gauge symmetry $SO(d-1,1)  \subset  SO(d-1,1)\times \Diff_d $ symmetry
in ${\cal M}_d$.
 The  fields that  compose    the  generalized  $d$-dimensional  
   Beltrami  vielbein are    neatly and covariantly separated according to their Weyl weights. They
  fall in different    categories, each one  having its   distinct    gravitational interpretation. 
  The   generic formula       \eqref{ds2d}   defines the       ``Beltrami parametrization"  that holds true for arbitrary dimension~$d\geq2$. It exhibits    a non trivial   $z \leftrightarrow\zb$  symmetry where $z$ and $\bz$ are the  complex coordinates of the sub-foliating Riemann surface~$\Sigma_{2}$. As noted in the introduction, the notion of  a generalized Beltrami parametrization, as it  is defined in this article, must be  stricto sensu taken in a local~sense.
    
  The  sub-foliation  of the ADM leaves by  the Riemann surfaces $\Sigma_{2}$ provides a  suggestive  definition of  the  gravitational  propagating physical degrees of freedom as a  well identified  subset  of the Beltrami fields.   They arise  as  a generalization of the bidimensional Beltrami differential according~to 
   $$\m,\mb \to
    \m, \mb,  \    \mu^z_i   ,      \mu^\bz_i,    \mu^i_j, \    \mu^z_\t,   \mu^\bz _\t,\    \mu^i_\t$$    where  $\mu^i_j=- \mu_i^j   $,
  $3\leq i,j\leq d-1$.
   The~excitations  of $\mu^z_i $, $\mu^\bz_i $ and $\mu^i_j $ can be at least  perturbatively identified as the $\frac {d(d- 3)}2$ physical degrees of freedom of the graviton.  $\m$ and $\mb$ have no physical excitations and can be gauge fixed as moduli of $\Sigma_2$. The rest of the fields  that parametrize  the Beltrami metric are   the conformal factor,     $d-2$   generalized  lapse functions (one    time lapse for  $\t$  and  $d-3$ dilatation functions for  the  coordinates $x^i$) and~$d-1$  Weyl invariant fields that  are related to  the  ADM shift vector.

   The  Beltrami $d$-bein and  the associated    metric can be further simplified 
  for 
    spaces with a given holonomy, as exemplified by the~$d=8$~formula 
 \eqref{ds2=8s}.   

  The  paper is  written  in a bottom to top approach. It   starts from   
  the      well-known  bidimensional situation that is ~generalized   till  one  obtains  a satisfying formulation for the generic  $d$-dimensional  Beltrami metric. 
   It~could have been alternatively presented  in a~top  to   bottom approach   by putting    section~6~in first position and considering    
 afterward  the  $d=2,3,4$~dimensional cases  as applications.  Some readers might prefer
     such a  more formal   presentation.        But,   the generic case  presented in  section~6~was truly obtained  by a   trial and error construction   of  the $d=3$      and $d=4$      cases as    cautious     generalizations of   the   solidly established    $ d=2$ Euclidean case.
      The accumulation of      all the details gathered in section 4  for the three dimensional case  (with zero  physical degrees of freedom) and in section    5  for the four~dimensional case    (with two   physical degrees of freedom that  have a large enough space    to 
      possibly propagate)  were in fact necessary to get  a global  understanding of the      general case for  all values of  $d$.


     One can maybe go further in the use of the $d$-dimensional  Beltrami parametrization than    being only concerned  by the mere propagation of gravitational degrees of freedom. Indeed,   part of the quantization  program  consists in  functionally integrating  over all possibilities for the sub-foliating surfaces  $\Sigma_{2}$  with
     $\Sigma^{ADM}_{d-1} \sim \Sigma_{d-3}\times \Sigma_{2} $. To do so one may consider the Beltrami fields as the fundamental  gravitational fields for defining the measure of the path integral. This presents a clear  advantage  since a subset of the Beltrami fields   directly    defines   the physical gravitational degrees of freedom.
      It~might be that one can make  firstly    this  reduced  part     
        of the functional  integration 
            with   an  appropriate BRST invariant gauge fixing  of    $\Sigma_{2}$ that   possibly  takes  into account   the number of holes in each ADM leaf $\Sigma_{d-1}^{ADM}$ and  by doing  only  afterward the remaining part  of the path integral that concerns the local dynamic. This picture requires more thinking.
           One may furthermore 
             question wether the  Riemann surfaces~$\Sigma _2 $~that   sub-foliate   the ADM leaves of $d$-gravity  
           can be identified as remnants of the worldsheets of an underlying string theory with ${\cal M}_d$ as a target space.   A   more refined string theory  zero 
         limit  might  exist that could possibly  offer more    information  than  the   well-known perturbative diagrammatic  link between string theory and perturbative gravity around a     classical    gravitational background,  a  possibility  that  might be the subject  of~a~separate publication. There,  an    additional path integration over  the gravitational background of string theory will be tentatively  incorporated  in the Polyakov path integral for allowing   a more sophisticated  string theory gauge fixing   that     would   hopefully consistently   identify   the string worldsheets  with  the   sub-foliating surfaces~$\Sigma _2 $~of the $d$-gravity that the present work    introduces  from the strict point of view of the Einstein~gravity. 

%

 \vskip 1cm
 
  \noindent {\bf Note :}
 This paper  refers to  a very limited number of published works. The author thanks in advance the readers who would inform him about references   where   ideas similar to those presented in this work have been discussed and  plans to add some of them in future versions.

  \vskip 1cm

 \noindent {\bf Acknowledgements:}
 It is   a  real pleasure to thank John Iliopoulos for many discussions on the subject and Vadim~Briaud 
 and Tom Wetzstein for    precise and careful readings of this article.

 \vskip 1cm

\def \et{{\hat e}}
\def \ot{{\hat \omega}}
\def \Ot{{\hat \Omega}}
\def \O{{  \Omega}}
\def \O{{ \hat  \Omega}}

\def \Et{       { \cal  E  }    }
\def \Eth{      \hat  { \cal  E  }    }
\def \Th{       { \hat  T  }    }
\def \ds{d+s}
\def\dt{  {\hat d}}

\def\t {\tau}
\def\mut{\mu^z_\t}
\def\mubt{\mu^\bz_\t}
\def\M{N_\t}

\def\M{M}
         

\def\Mo
{
\pbM
\mo&\mut
\cr
\mob&\mubt
\peM
}

\def\Moinv
{
\pbM
\mo&-\mut
\cr
-\mob&\mubt  
\peM
}

\def\mmbo{\mo\mubt-\mbo\mut}

\def\mmo{\mob\mut-\mo\mubt}

\def\Mot
{
\pbM
\mo&\mob
\cr
\mut&\mubt
\peM
}

\def\Motinv
{
\pbM
\mo&-\mob
\cr
-\mut&\mob
\peM
}
%
%

\def \idd { \pbM  0&1\cr-1&0 \peM}
      \def\M{\bar N}
\appendix
%
%
\section {Appendix :  $d=4$  Beltrami Spin connection equations  }
\def\M{M}
This appendix  computes   the 24  linear equations   that  determine the Spin connection~$\o(e) $ for   $d=   4$~when the vierbein is expressed  as   in \eqref{4bein}. One has from section 5
\bea\label{basis}
\pbM
\Eo \cr \Eob\cr dt \cr d\t
\peM
\equiv
\pbM
\Mu& \Mo
\cr
\pbM
 0&0\cr0&0
\peM
&
\pbM
 N&\aa\cr\bb& \M
\peM
\peM
\pbM
dz \cr d\bz \cr dt\cr d\t
\peM.
\eea
The    $d=3$     case  is   obtained by restricting the   4X4  matrix  \eqref {basis} to the  $3\times 3 $  matrix in its upper left  corner  and the  $d=2$ case   is when it is reduced to  the $2\times 2$  matrix $\Mu$ in its left top corner.
 One uses $d=  dz\p+d\bz  \bp   +\pa_0 dt +\pa_\t d\t = \Eo \Dz +\Eob \Dbz +   {\cal E}^0 \Do    +  {\cal E}^\t   {\cal D}_\t 
$ with
\bea
\label{derD}\red
\pbM
\Dz \cr \Dbz\cr \Do  \cr \Dt
\peM
=
\pbM
\frac 1 \mmb
\invMu
&
,
 \pbM
 0&0\cr0&0
\peM
\cr
\frac 1 \mmb
  \frac   1 {N\M-\aa\bb}
\pbM
\M&-\bb\cr -\aa&N
\peM
\pbM
\mo&\mu^z_\t \cr \mbo&\mu^\bz_\t
\peM
\pbM
1&-\mb \cr -\m&1
\peM
,
&
\frac   1 {N\M-\aa\bb}
\pbM
\M&-\bb\cr -\aa&N
\peM
\peM
\pbM
\p \cr \bp  \cr \pa_0\cr \pa_\t
\peM.
\eea

The  24 components of  the $d=4$     Beltrami Spin connection $\o(e)$ solve   the  torsion free conditions $T^0=T^z=T^\bz=T^\t=0$. Once they are determined, one can compute  Einstein action  as a quadratic    form in the $\o(e)$'s.  The  equation $T^\t=0 $    becomes irrelevant in $d=3$ as well as $T^0=T^\t=0$    in $d=2$. The   case $d=2$ is trivial and   directly   solved  in section 2. 
The 9 torsion zero  equations for the case the case $d=3$ are    solved in Appendix~B.

\eqref{basis}  implies     $\Eo= dz+\m d\bz+\mu^z_0 dt +\mu^z_ \t   d\t$  
 and 
 $\Eo= d\bz+\mb dz+\mu^\bz_0 dt +\mu^\bz _\t d\t$.
Since     $e^0 =   Ndt 
   + \aa d\t $ and $e^\t= \bb   dt  +  \M   d\t      $
   one has   $de^0 =   dN\w  dt
   +d \aa\w d\t $ and $de^\t= d\bb \w dt  +  d\M  \w d\t      $.  
   
In the coordinate system $x^\mu= (\t,t,z,\bz$ one has $a=-\bar a =\mu^0_\t$ and $M\neq N$;  in the light cone 
coordinates  sytem $x^\mu= (\t^+,\t^-, z,\bz$  and
$a=\mu^+-$,      $a=\mu^-_+-$, $M=N\equiv \cal N$.    Thus the use of $a$ and $\bar a \neq a$   covers both cases for the coordinates   
$(\t,t,z,\bz)$ and $(\t^+,\t^-,z,\bz)$.

\def\t {\tau}
\def\mut{\mu^z_\t}
\def\mubt{\mu^\bz_\t}
\def\M{N_\t}
\def\M{\bar N}
 \def\M{M}

  \noindent {\bf {    $\bullet$  Consequences of  $T^\t=0$}}
  
   The 2 form  $T^\t$  writes 
\bea
 T^\t&\equiv &    de^\t     + {\red{\demi}} \o^{\t z}   \w e^\bz + {\red{\demi}}   \o^{\t \bz}   \w e^z +  \o^{\t 0}   \w e^0  
 \nn\\
 \=
  (\Eo\Dz  \M +\Eob\Dbz \M  + {\red{ \Do}} \M dt )      \w d\t  
  + (\Eo\Dz  \bb +\Eob\Dbz \bb  +  {\red{ \Dt}} \bb d\t )      \w dt
   \nn\\
&& -     {\red{ \demi }}  {\exp \vpb } \   \Eob  \w   \o^{\t z }
-       {\red{ \demi }}  {\exp \vp } \   \Eo  \w   \o^{\t \bz }
+
   \o^{\t0}\w
   ( N dt +\aa d\t)
    \nn\\ 
 \= {  \Eo  \w  \Eob}
(    
 {\red{ \demi }} \exp \vpb\ \o^{\t z}_\Z - {\red{ \demi }} \exp \vp\ \o_\bZ^{\t \bz})
     \cr 
&& 
+\Eo\w d\t  (     \Dz \M   - 
{\red{ \demi }} 
{\exp \vp}{ }\  \o^{\t \bz}_\tau  
 +\o^{\t0}_Z\aa   )
+\Eob\w d\t  (    \Dbz \M   - {\red{ \demi }} {\exp \vpb}{ }\  \o^{\t z}_\tau  
 +\o^{\t0}_\bZ\aa      )
\cr
&&
+\Eo\w dt  (     N \o^{  \t 0}_Z  -   {\red{ \demi }} {\exp \vp}{ } \ \o^{\t\bz }_0   + \Dz \bb  )
+\Eob\w dt  (     N \o^{  \t 0}_\Zb   -{\color{black} \frac{1}{2}}{\exp \vpb}{ } \ \o^{\t z }_0    +\Dbz \bb   )
\cr 
&& +\textcolor{\green}{dt \w d\t} ( {\red{ \Do }}  \M  {\color{black} -}  {\red{ \Dt }}  \bb {\color{black} -} N \o^{\t 0}_\t      {\red{ + }}  \aa  \o^{\t0}_0  ).
 \eea
 The condition   $T^\t=0$ implies    { {
 \bea    
 \exp \vpb\ \o^{\t z}_\Z  {  -}\exp \vp\ \o_\bZ^{\t \bz}\=0
\cr
   N \o^{ 0 \t }_Z   +    {\red{ \demi }}  {\exp \vp}{ } \ \o^{\t\bz }_0    \= \Dz \bb
   \cr
    N \o^{ 0 \t }_\Zb   +    {\red{ \demi }}  {\exp \vpb}{ } \ \o^{\t z }_0     \=  \Dbz \bb \nn 
     \eea
     \bea
         {\red{ \demi }}   {\exp \vp}{ } \   \o^{\t \bz}_\tau    -   \o^{\t0}_Z\aa       \=     \Dz \M   
  \cr
       {\red{ \demi }}   {\exp \vpb}{ } \     \o^{\t z}_\tau    -\o^{\t0}_\bZ\aa    \=    \Dbz \M 
\cr 
N  \o^{\t 0}_\t  - \aa    \o^{\t0}_0     \=        {\red{ \Do }}   \M   -    {\red{ \Dt }}   {\color{black} \bar{a}}
 \eea
 }}

\noindent {\bf {    $\bullet$ Consequences of  $T^0=0$. }}

 The 2 form  $T^0$  writes 
 \def\blue{\color{black} }
    \bea
 T^0&\equiv&    de^0 \textcolor{\orange}{-} \demi   \o^z \w e^\bz     \textcolor{\orange}{+} \demi   \o^\bz \w e^z-\o^{0\t} \w e ^\t
 \nn\\ \=   
 (\Eo\Dz  N +\Eob\Dbz N  +   {\red{ \Dt }}  N d\t )      \w dt  
 +(\Eo\Dz  \aa +\Eob\Dbz \aa  +  {\red{ \Do }}  \aa dt )      \w d\t 
\textcolor{\orange}{+}   \frac  {\exp \vpb } {  2}   \Eob  \w   \o^z  
\textcolor{\orange}{-} \frac{\exp \vp } {2} \Eo\w
   \o^\bz 
      -\o^{0\t}   \w  (\bb dt +\M d\t)
      \nn\\
 \=\demi  {    \Eo  \w  \Eob}
(    
\exp \vpb\ \o^z_Z +\exp \vp\ \o_\bZ^\bz)
     \cr 
&& 
+\Eo\w dt  (     \Dz N   \textcolor{\orange}{-} \frac{\exp \vp}{2} \o^\bz_0    -\bb\o^{0\t}_Z    )
+\Eob\w dt  (     \Dbz N   \textcolor{\orange}{+} \frac{\exp \vpb}{2} \o^z_0     -\bb\o^{0\t}_\Zb  )
\cr
&&
+\Eo\w d\t  ( -    \M \o^{ 0 \t }_Z   \textcolor{\orange}{-} \frac{\exp \vp}{2} \o^\bz_\t      +\Dz \aa   )
+\Eob\w d\t  ( -   \M \o^{ 0 \t }_\bZ   \textcolor{\orange}{+} \frac{\exp \vpb}{2} \o^z_\t    +\Dbz \aa 
   )\cr 
&& +d\t \w dt ( {\red{ \Dt }}  N
-  {\red{ \Do}} \aa
{\color{black} -}\M \o^{\t 0}_0    -\bb \o^{0 \t }_\t   ).
\eea
 The vanishing conditions   of the projections of $T^0=0$ on  $\Eo\w \Eob,\  \Eo \w dt,\   \Eob \w dt $  and   $\Eo\w d\t, \ \Eob \w d\t, \ d\w dt$ imply  
 \bea
 \label{To3}
  \exp -\vp\ \o^z_Z +\exp -\vpb\ \o_\bZ^{{\color{black} \textcolor{\green}{\bz}}}
\=0 
 \nn\\ 
  \o^z_0
  \=
   2  {\exp -\vpb}  \   (\textcolor{\orange}{-} \Dbz N   \textcolor{\orange}{+} \bb\o^{0\t}_\bZ)
 \nn\\ 
  \o^\bz_0\=
   2 {\exp -\vp} \  (\Dz N  \textcolor{\orange}{-} \bb\o^{0\t}_Z)
   \eea
\bea
  \label{To4}
 \M \o^{ 0 \t }_Z   \textcolor{\orange}{+} \frac{\exp \vp}{2} \o^\bz_\t   \=   \Dz\aa
\nn\\
 \M \o^{ 0 \t }_\bZ  \textcolor{\orange}{-} \frac{\exp \vpb}{2} \o^z_\t  
\=   \Dbz\aa   
\nn\\
 \M  \o^{\t 0}_0  + {\red{ \bar a }}       \o^{0 \t }_\t  \=    {\red{ \Dt }} N {\color{black} -}  {\red{ \Do }}  \aa\
\eea 

\vskip 1cm
      \noindent {\bf {    $\bullet$ Consequences of   $T^z=T^\bz  =0 $    } }
      
   Expanding  both  2-form equations     $T^z=T^\bz $   is slightly more involved than  doing it for  $T^0=T^\t   =0 $.  
The  following   relation  is useful to  relate the components of  forms expressed  either on the basis of $1$-forms $(dz, d\bz, dt,\t)$ or  on the basis   of $1$-forms
 $(\Eo,\Eob, dt,d \t)$: 
\bea
\pbM dz\cr d\bz\peM=\frac 1  \mmb \invMu
\Big  [  
\pbM \Eo  \cr \Eob \peM 
 -\Mo
  \pbM dt  \cr d\t \peM 
  \Big ]\nn. 
\eea
 One has therefore  
\bea\label{ouf}
dz\w d\bz    \=
\frac 1{2(\mmb)^2}
 ^t
\Big [  
\pbM \Eo  \cr \Eob \peM 
 -\Mo
  \pbM dt  \cr d\t \peM 
 \Big ] \invtMu
\idd 
 \invMu
\Big [  
\pbM \Eo  \cr \Eob \peM 
 -\Mo
  \pbM dt  \cr d\t \peM 
 \Big ] \nn\\
  \=
    \frac 1{2(\mmb)} 
 ^t
 \Big[  
\pbM \Eo  \cr \Eob \peM 
 -\Mo
  \pbM dt  \cr d\t \peM 
 \big  ] 
\idd 
\Big [  
\pbM \Eo  \cr \Eob \peM 
 -\Mo
  \pbM dt  \cr d\t \peM 
  \Big ] \nn\\
  \=
    \frac 1{2(\mmb)}
 \Big[  
 ( \Eo  \   \Eob ) 
 -  
  ( dt   \  d\t ) \Mot
 \big  ] 
\idd 
\Big [  
\pbM \Eo  \cr \Eob \peM 
 -\Mo
  \pbM dt  \cr d\t \peM 
  \Big ] \nn\\
  \=
  \frac 1{\mmb}
 \Big[   \Eo\w \Eob +
(\mmbo) dt\w d\t
-\Eo\w (\mob dt +\mubt d\t)
  +\Eob\w  (\mo dt +\mut d\t ) 
  \Big ],
   \nn\\
     d\t \w dz \=     \frac{ d\t} {\mmb}\w      (\Eo-\m\Eob   - \red  (    \mu^z_0   -  \m \mbo)  dt)
     \nn\\
     d\t \w d\bz \=     \frac{ d\t} {\mmb}\w      (\Eob-\mb\Eo   - \red  (    \mu^\bz_0   -  \mb \mo)  dt    )
     \nn\\
     dt \w dz \=     \frac{ dt} {\mmb}\w      (\Eo-\m\Eob- \red  (    \mu^z_\t    -  \m \mu ^\bz_\t)  d\t   )
     \nn\\
     dt \w d\bz \=     \frac{ dt} {\mmb}\w      (\Eob-\mb\Eo- \red  (    \mu^\bz_\t    -  \textcolor{\green}{\mb} \mu ^z_\t)  d\t  ).
     \eea
The   following matricial identities are  useful     to derive \eqref{ouf}
$$
(A,\bar A) \idd 
\pbM B\cr \bar B
 \peM =A\bar B -  \bar A B
\  \rm{and }
\
\pbM
 a&c\cr   b&d
 \peM
  \idd 
\pbM
 a&b\cr   c&d
 \peM
 =(ad-bc)\idd.
$$
Then,  $T^z$ and $T^z$ read
 \bea
 T^z&\equiv&    de^z \textcolor{\orange}{-} \o ^0\w e^z    \textcolor{\orange}{+}   \o^z \w e^0 -\o^{z\t}  {\color{black} \wedge} e^\t= 
    \exp\vp
    \Big (  
    ( d \vp     \textcolor{\orange}{-} \o ^0 )\w\Eo   +d\Eo  \textcolor{\orange}{+}    \exp -\vp       \o^z \w( Ndt+\aa d\t)
      -   \exp -\vp     \o^{z\t}     \w (\bb dt +\M d{\color{black} \t})    \Big )
     \nn\\
     \= 
    \exp\vp
    \Big (  
    ( d \vp     \textcolor{\orange}{-} \o ^0 )\w\Eo 
     \textcolor{\orange}{+} \exp -\vp      \o^z \w (Ndt  +\aa d\t)
     \   - \exp -\vp  \   \o^{z\t}\w    ( \M   d\t      +\bb dt)
    \cr&&
    + dz\w d\bz  \   \p \m 
    +
     dt\w d\bz  \   (\pa_o \m- \bp \mo)
    -dt\w  dz   \   \p \mo    
     +    
    d\t \w d\bz  \ (  \pa_\t \m   -\bp \mu^z_\t)
      -d\t\w  dz   \   \p \mu^z_\t
   +
      dt\w   d  \t
       \   (\pa_0 \mu^z_\t- \pa_\t \mo)  
       \big).
       \nn
     \eea
     \bea
 T^\bz&=&    de^\bz \textcolor{\orange}{+} \o ^0\w e^\bz    \textcolor{\orange}{-}   \o^\bz \w e^0 -\o^{\bz\t} {\color{black} \wedge}  e^\t= 
    \exp\vpb
    \Big (  
    ( d \vpb     \textcolor{\orange}{+} \o ^0 )\w\Eob   +d\Eob  \textcolor{\orange}{-}    \exp -\vpb       \o^\bz \w( Ndt+\aa d\t)
      -   \exp -\vpb     \o^{{\color{black} \zb}\t}     \w (\bb dt +\M dt)    \Big )
     \nn\\
     \= 
    \exp\vpb
    \Big (  
    ( d \vpb     \textcolor{\orange}{+} \o ^0 )\w\Eob 
     \textcolor{\orange}{-} \exp -\vpb      \o^\bz \w (Ndt  +\aa d\t)
     \   - \exp -\vpb  \   \o^{\bz\t}\w    (\bb dt+  \M   d\t      )
    \cr&&
    + d\bz\w dz  \   \bp \mb 
    +
     dt\w dz  \   (\pa_o \mb- \p \mo)
    -dt\w  d\bz   \   \bp \mob    
     +    
    d\t \w dz  \ (  \pa_\t \mb   -\p \mu^\bz_\t)
      -d\t\w  d\bz   \   \bp \mu^\bz_\t
   +
      dt\w   d  \t
       \   (\pa_0 \mu^\bz_\t- \pa_\t \mob)  
       \big).
       \nn
     \eea
The condition $T^z=0$ is
      \bea
     0\=  
    ( d \vp    \textcolor{\orange}{-} \o ^0 )\w\Eo 
    \nn\\ &&
     \textcolor{\orange}{+} \exp -\vp      \o^z \w (Ndt  +\aa d\t)
         \nn\\ &&
     \   - \exp -\vp  \   \o^{z\t}\w    ( \M   d\t      +\bb dt)
    \nn\\ &&
    + \frac 1{\mmb}
 \Big[   \Eo\w \Eob +
(\mmbo) dt\w d\t
-\Eo\w (\mob dt +\mubt d\t )
  +\Eob\w (\mo dt +\mut d\t ) )  
  \Big ]  \   \p \m 
        \nn\\ &&
    +
      \frac{ dt} {\mmb}\w      (\Eob-\mb\Eo-  \red ( \mu^\bz_\t  -\mb  \mu^z_\t   )d\t)\   (\pa_o \m- \bp \mo)
         \nn\\ &&
    -  \frac{ dt} {\mmb}\w      (\Eo-\m\Eob-    \red ( \mu^z_\t - \m \mu^\bz_\t)    d\t) \   \p \mo    
        \nn\\ &&
     +    
      \frac{ d\t} {\mmb}\w      (\Eob-\mb\Eo- \red (   \mu^\bz_0  - \mb \mo  )    dt) \ (  \pa_\t \m   -\bp \mu^z_\t)
        \nn\\ &&
       -\frac{ d\t} {\mmb}\w      (\Eo-\m\Eob-\red (\mu^z_0 -  \m  \mbo     dt)
 \   \p \mu^z_\t
          \nn\\ &&
   +
      dt\w   d  \t
       \   (\pa_0 \mu^z_\t- \pa_\t \mo)  
       \big),
      \eea
    that is
      \bea  \label{Tz}
     0  \=\blue\Eo\w\Eob
       ({\red -\Dbz }\vp \textcolor{\orange}{+} \o^0_{\red \bZ} +\frac{ \p\m} \mmb)
   \nn\\     &&+
   \blue
       \Eo\w d t
      \Big  (-{\red {\Do}} \vp \textcolor{\orange}{+} \o^0_0  \textcolor{\orange}{+} \exp-\vp\Big ( N \o_Z^z \textcolor{\orange}{-} \bb  \o_Z^{{\color{black} z\t}}  \Big )      +\frac 1 \mmb  
       (  \p\mo
       -
       \mob \p\m
       +\mb (\pa_o \m- \bp \mo)
       )\Big)
       \nn\\
         &&+ \blue  \Eob\w  d t  
         \Big(    \exp-\vp  ( N \o^z_\bZ  -  \bb \o^{{\color{black} z\t}}_\bZ )    
         +\frac 1\mmb(  - \pa_o \m+\bp \mo +\mo\p\m
-\m\p\mo)
\Big)
       \nn\\
            &&+
      \Eo\w\d\t
       \Big( 
       -   {\red {\Dt}} 
       \vp  \textcolor{\orange}{+} \o^0_\t  \textcolor{\orange}{+}  \exp-\vp   (     {\red{\textcolor{\orange}{-}  \M  \o^{z\t} _z}}
       + \aa \o^z_Z    )+
       \frac 1 \mmb
       (  \p\mut
       - \mubt \p\m
       +  \mb  \textcolor{\green}{(\partial_{\t} \m- \bp \mu^z_\t)}
         ) 
       \Big)
       \nn\\
           &&+
            \Eob\w\d\t 
           \Big(\exp-\vp
(
            \aa \o^z_\bZ  - {\red{ \M \o^{ z\t}_\bZ}}
           )
           +     \frac 1 \mmb      (
            -   \pa_\t \m +\bp \mu^z_\t 
                + \mut  \p\m 
           -\m\p\mut)
       \Big)
\nn\\
           &&+ \d t\w d\t
           \Big(
           \exp-\vp
           ( \aa  \o^{ z}_0   \textcolor{\orange}{-} {\red{ N  \o^{ z}_\t    -\M \o^{z\t }_0
             +\bb \o^{z\t }_\t  )}}
             +\red    \pa_0 \mu ^z_\t  -  \pa_\t \mu^z_0
              \nn\\  && \quad  \quad\quad    
             +
            \red  \frac{  1} {\mmb} \red
        \big[ \p\m(\mo\mu^\bz_\t - \mu^z_\t   \mu^\bz_0\textcolor{\green}{)}
        -( \pa_0\m-\bp\mo) )(\mu^\bz_\t  -\mb \mu^z_\t )
        +\p\mo(\mu^z_\t -\m\mu^\bz_\t )\nn\\&&
        \red\quad\quad\quad\quad\quad\quad\quad\quad\quad
        +(\pa_\t\textcolor{\green}{\m}-\bp\mu^z_\t)(\mbo-\mb \mo)
        -
        \p\mu^z_\t (\mo-\m\mbo) \big]
            \Big).
  \eea

  \vskip 1cm
  Analogously, one has
    \bea
 T^\bz&=&    de^\bz \textcolor{\orange}{+}  \o ^0\w e^\bz    \textcolor{\orange}{-}  \o^\bz \w e^0 -\o^{\bz\t}  {\color{black} \wedge} e^\t= 
    \exp\vpb
    \Big (  
    ( d \vpb     \textcolor{\orange}{+}   \o ^0 )\w\Eob   +d\Eob  \textcolor{\orange}{-}    \exp -\vpb       \o^\bz \w( Ndt+\aa d\t)
      -   \exp -\vpb     \o^{{\color{black} \zb}\t}     \w (\bb dt +\M dt)    \Big )
     \nn\\
     \= 
    \exp\vpb
    \Big (  
    ( d \vpb  \textcolor{\orange}{+}   \o ^0 )\w\Eob 
     \textcolor{\orange}{-} \exp -\vpb      \o^\bz \w (Ndt  +\aa d\t)
     \   - \exp -\vpb  \   \o^{\bz\t}\w    (\bb dt+  \M   d\t      )
    \cr&&
    + d\bz\w dz  \   \bp \mb 
    +
     dt\w dz  \   (\pa_o \mb- \p \mo)
    -dt\w  d\bz   \   \bp \mob    
     +    
    d\t \w dz  \ (  \pa_\t \mb   -\p \mu^\bz_\t)
      -d\t\w  d\bz   \   \bp \mu^\bz_\t
   +
      dt\w   d  \t
       \   (\pa_0 \mu^\bz_\t- \pa_\t \mob)  
       \big),\nn
     \eea
and  the condition $T^\bz=0$ implies 
 \bea\label{Tbz}
    0  \= 
    ( d \vpb    \textcolor{\orange}{+}    \o ^0 )\w\Eob 
    \nn\\ &&
     \textcolor{\orange}{-} \exp -\vpb      \o^\bz \w (Ndt  +\aa d\t)
         \nn\\ &&
     \   - \exp -\vpb  \   \o^{\bz\t}\w    ( \M   d\t      +\bb dt)
    \nn\\ &&\blue
    + \frac 1{\mmb}
 \Big[   \Eob\w \Eo +
(\mmo) dt\w d\t
-\Eob\w (\mo dt +\mut d\t )
  +\Eo\w (\mob dt +\mubt d\t ) )  
  \Big ]  \   \bp \mb 
        \nn\\ &&
    +
      \frac{ dt} {\mmb}\w      (\Eo-\m\Eob-{\color{black} (}\mu^z_\t {\color{black} - \m\mu^\bz_\t)} d\t)\   (\pa_o \mb- \p \mob)
         \nn\\ &&
    -  \frac{ dt} {\mmb}\w      (\Eob-\mb\Eo-{\color{black} (}\mu^\bz_\t {\color{black} -\mb\mu^z_\t)} d\t) \   \bp \mob    
        \nn\\ &&
     +    
      \frac{ d\t} {\mmb}\w      (\Eob-\mb\Eo-{\color{black} (}\mu^\bz_0 {\color{black} - \m\mu^\bz_0)} dt) \ (  \pa_\t \m   -\bp \mu^z_\t)
        \nn\\ &&
       -\frac{ d\t} {\mmb}\w      (\Eo-\m\Eob-{\color{black} (}\mu^z_0 {\color{black} -\mb\mu^z_0)} dt)
 \   \p \mu^z_\t
          \nn\\ &&
   +
      dt\w   d  \t
       \   (\pa_0 \mu^\bz_\t- \pa_\t \mob)  
       \big)
       \nn\\
       \=\blue\Eob\w\Eo
       (-{\color{black} \Dz} \vpb\textcolor{\orange}{-}\o^0_{\color{black} Z} +\frac{ \bp\mb} \mmb)
   \nn\\     &&+
   \blue
       \Eob\w d t
      \Big  (-{\color{black} \mathcal{D}_{0}} \vpb\textcolor{\orange}{-}\o^0_0   -\exp-\vpb\Big ( N \o_\bZ^\bz +\bb  \o_\bZ^{{\color{black} \bz \t}}  \Big )      +\frac 1 \mmb  
       (  \bp\mob
       -
       \mo \bp\mb
       +\m (\pa_o \mb- \p \mob)
       )\Big)
       \nn\\
         &&+ \blue  \Eo\w  d t  
         \Big(   \textcolor{\orange}{-} \exp-\vpb  ( N \o^\bz_Z  \textcolor{\orange}{+} \bb \o^{{\color{black} \bz \t}}_Z )    
         +\frac 1\mmb(  - \pa_o \mb+\p \mob +\mob\bp\mb
-\mb\bp\mob)
\Big)
       \nn\\
            &&+
            \red
       \Eob\w\d\t
       \Big( 
       -{\color{black} \mathcal{D}_{\t}} \vpb{\color{black} \textcolor{\orange}{-}}\o^0_\t-  \exp-\vpb   (       \M  \o^{{\color{black} \bz \t}} _{\textcolor{\green}{\bZ}}
       \textcolor{\orange}{+} \aa \o^\bz_\bZ    )+
       \frac 1 \mmb
       (  {\color{black} \bp}\mubt
       - \mut \bp\mb
       +  \m  \textcolor{\green}{(\pa_\t \mb- \p \mu^\bz_\t)}
         ) 
       \Big)
       \nn\\
           &&+
           \red
            \Eo\w\d\t 
           \Big(\textcolor{\orange}{-}
           \exp-\vpb
(
           \aa \o^\bz_Z  \textcolor{\orange}{+}  \M \o^{{\color{black} \bz \t}}_Z
           )
           +     \frac 1 \mmb      (
            -   \pa_\t \mb +\p \mu^\bz_\t 
                + \mubt  \bp\mb 
           -\mb\bp\mubt)
       \Big)
\nn\\
           &&+\red d t\w d\t
           \Big(
          \textcolor{\orange}{-} \exp-\vpb
           (  \aa  \o^{ \bz}_0   -N  \o^{ \bz}_\t  
             \textcolor{\orange}{+}\M \o^{{\color{black} \bz \t}}_0
             \textcolor{\orange}{-}\bb \o^{{\color{black} \bz \t}}_\t  ) {\color{black} +\pa_0\mu^\bz_\t - \pa_\t\mu^\bz_0 }
              \nn\\  && \quad  \quad\quad   \red
             +
             \frac{  1} {\mmb} 
        ({\color{black} -\pa_\bz\mu^\bz_0(\mb\mu^z_\t-\mu^\bz_\t) + \pa_\bz\mu^\bz_\t(\mb\mu^z_0-\mu^\bz_0) - \pa_\bz\mb(\mu^z_0\mu^\bz_\t-\mu^\bz_0\mu^z_\t) - (\pa_z\mu^\bz_0-\pa_0\mb)(\m\mu^\bz_\t-\mu^z_\t) } \nn\\ &&
       {\color{black}  + (\pa_z\mu^\bz_\t-\pa_\t\mb)(\m\mu^\bz_0-\mu^z_0)}) 
            \Big)
  \eea
 
The result of this  Appendix  is  thus the following  system  of   twenty four linear  independent  equations that  determine  the components of the  four dimensional Spin connection  when the vierbein is expressed in the   Beltrami parametrization : 
 \def\aa  {     {\color{red} {a}} } 
  \def\aa  {     {\color{black} {a}} } 
  
  \def\bb  {     {\color{red} {\bar a}} } 
    \def\bb  {     {\color{black} {\bar a}} } 
  \def\mut  {     {\color{red} {  \mu ^z  _\t }} } 
    \def\mut  {     {\color{black} {  \mu ^z  _\t }} } 
   \def\mubt  {     {\color{red} {  \mu ^\bz  _\t }} } 
            \def\mubt  {     {\color{black} {  \mu ^\bz  _\t }} } 
         \def\mubt  {     {\color{red} {  \mu ^\bz  _\t }} } 
         \def\mubt  {     {\color{black} {  \mu ^\bz  _\t }} } 
 \def \blue{\color{black}}
 \bea   \label{eqlin}
      \blue
      \o^0_Z\= 
        {\color{black} \textcolor{\orange}{-}}\DD_z {\color{black} \vpb}  
   \nn\\   
   \blue
      \o^0_\bZ \=
       \textcolor{\orange}{+} \DD_\bz \vp
      \nn\\
     \blue
     \o^0_0   +\exp-\vp  (  N \o_Z^z \textcolor{\orange}{-} \bb  \o_Z^{{\color{black} z\t}} )\= 
      {\color{black} \mathcal{D}_{0}}\vp   \textcolor{\orange}{-} \DD_z \mo  
       \nn\\
       \blue
     \o^0_0   +\exp-\vpb  (  N \o_\bZ^\bz \textcolor{\orange}{+} \bb  \o_\bZ^{{\color{black} \bz\t}} )\=
      {\color{black} \textcolor{\orange}{-} \mathcal{D}_{0}}\vpb   \textcolor{\orange}{+} \DD_\bz \mbo
       \nn\\
         \blue
           \exp-\vp  ( N \o^z_\bZ  \textcolor{\orange}{-}  \bb \o^{{\color{black} z\t}}_\bZ ) \=  
             \DD_o \m
       \nn\\
       \blue
           \exp-\vpb  ( N \o^\bz_Z  {\color{black} \textcolor{\orange}{+}}  \bb \o^{{\color{black} \bz\t}}_Z ) \=  
         -   \DD_o \mb
       \nn\\
            \red
       \o^0_\t+  \exp-\vp   (   \textcolor{\orange}{-}   \M  \o^{{\color{black} z\t}} _{\color{black} Z}
       +{   \aa \o^z_Z  }  )\=  
         {\color{black}\mathcal{D}_{\t}} \vp  \textcolor{\orange}{-} \DD_z\mut       \nn\\
       \o^0_\t {\color{black} +}  \exp-\vpb   (   \M  \o^{{\color{black} \bz\t}} _{\color{black} \bZ}
       + \aa \o^\bz_\bZ    )\=
        {\color{black} \textcolor{\orange}{-} \mathcal{D}_{\t}} \vpb  {\color{black} \textcolor{\orange}{+}}\DD_{\textcolor{\green}{\bz}}\mubt   
       \nn\\
           \red
           \exp-\vp
(
           \aa \o^z_\bZ  \textcolor{\orange}{-}  \M \o^{{\color{black} z\t}}_\bZ
           )\=
            {\color{black} }  \DD_\t \m 
\nn\\
           \exp-\vpb
(
           \aa \o^\bz_Z  \textcolor{\orange}{+}  \M \o^{{\color{black} \bz\t}}_Z
           )\=  
           -
              \DD_\t \mb 
\nn\\
\red
            \exp-\vp
           ( {\blue   \aa  \o^{ z}_0 }  \textcolor{\green}{-} N  \o^{ z}_\t     \textcolor{\orange}{+} \M \o^{\t z}_0
             \textcolor{black}{-} \bb \o^{\t z}_\t  )\= 
                \red
                     \DD_\t \mo
       \textcolor{\green}{-}
       \DD_0 \color{black} \mu^z_\t
       \nn\\
           \exp-\vpb
           (  \aa  \o^{ \bz}_0   \textcolor{\green}{-} N  \o^{\bz}_\t    \textcolor{\orange}{-} \M \o^{\t \bz}_0
             \textcolor{\orange}{+} \bb \o^{\t \bz}_\t  )\= 
                \red
                    \textcolor{\green}{-}\DD_\t   \mob
       \textcolor{\green}{+}
       \DD_0    \color{black} \mu^\bz_\t
  \eea
 \bea
 \blue \exp- \vp\ \o^z_Z +\exp -\vpb\ \o_\bZ^\bz
\= 
 0 
 \nn\\ \blue
  \o^z_0
  \= 
   2  {\exp -\vpb}  \   (\textcolor{\orange}{-} \Dbz N   \textcolor{\orange}{+} \bb\o^{0\t}_\bZ)
 \nn\\ \blue
  \o^\bz_0\= 
   2 {\exp -\vp} \  ( \Dz N  \textcolor{\orange}{-} \bb\o^{0\t}_Z)
   \nn\\
 \red
 \M \o^{ 0 \t }_Z   \textcolor{\orange}{+}  \frac{\exp \vp}{2} \o^\bz_\t   \=   \Dz\aa
\nn\\\red
 \M \o^{ 0 \t }_\bZ   \textcolor{\orange}{-} \frac{\exp \vpb}{2} \o^z_\t  
\=   
  \Dbz\aa   
\nn\\\red
 \M  \o^{\t 0}_0  {\color{black} +}  \bar{a}        \o^{\textcolor{\green}{0 \t}  }_\t  \=  
  {\color{black} \mathcal{D}_{\t}} N \textcolor{\green}{-}  {\color{black} \mathcal{D}_{0}} \aa\
\eea 
 {\red{
 \bea    
\exp -\vp\ \o^{\t z}_\Z -\exp -\vpb\ \o_\bZ^{\t \bz}\=
 0
\cr
   -N \o^{ \t0 }_Z   +{\color{black} \frac{1}{2}}{\exp \vp}{ } \ \o^{\t\bz }_0    \=  
     \Dz \bb
   \cr
   - N \o^{\t 0  }_\Zb   +{\color{black} \frac{1}{2}}{\exp \vpb}{ } \ \o^{\t z }_0     \=
    {\color{black} +} \Dbz \bb
     \cr 
      {\color{black} \frac{1}{2}}{\exp \vp}{ } \   \o^{\t \bz}_\tau    -   \o^{\t0}_Z\aa       \=  
        \Dz \M   
  \cr
   {\color{black} \frac{1}{2}}{\exp \vpb}{ } \     \o^{\t z}_\tau    -\o^{\t0}_\bZ\aa    \=
      {\color{black} +}  \Dbz \M 
\cr 
N  \o^{\t 0}_\t  - \aa    \o^{\t0}_0     \=   
  {\color{black} \mathcal{D}_{0}} \M   -{\color{black} \mathcal{D}_{\t}} \aa.
 \eea
 }}
The action on all fields of   the derivation  operation $\DD= \pa+... $  that figures   in \eqref{eqlin}   involves the derivatives  $\mathcal{D}_z, \mathcal{D}_\bz,   \mathcal{D}_0, \mathcal{D}_\t $  defined in \eqref{derD}.
To make it explicit,  one must  look at the details      of the        twenty four  independent $d=4$ vanishing  torsion conditions for $T^z=T^\bz=0$ that have been computed above in \eqref{Tz} and \eqref{Tbz}. The method of inverting     the above 24 equations to determine the $d=4$ Spin connection 	is not so simple 	and  will be published elsewhere   for both cases $a=-\bar a=\mu^0_\t, N\neq M$
 and $a= \mu^+_-,    \bar a=\mu_+^-, N=M \equiv\cal N$ in a paper specifically dedicated to the four dimensional case. In what follows we reduce them and obtain the~9(equations that determine the three dimensional case.

\section {Appendix :   Computing  the $d=3$       Beltrami  Spin connection}

The 24  four dimensional    Spin connection equations  can be obviously  reduced  to the  simpler system of the~9~linear equations that    determine  the expression of the Spin connection in $d=3$ dimensions.    Both    Euclidean and Lorentz   three dimensional cases are respectively determined  by  the values   $\eps =1$ or $\eps=-1$  as expressed   in section 4.
 One   gets linear   equations 
for the 9  three dimensional   Spin connection components  $\o_Z^z,  \ \o_\bZ^z,  \ \o_0 ^z,\
\o_Z^\bz,  \ \o_\bZ^\bz,  \ \o_0 ^\bz,\ 
\o_Z^0,  \ \o_\bZ^0,  \ \o_0 ^0  
 $. They derive from the $d=4$ equations 
  $\o^z,\o^\bz, \o^0$ solve  where one   leaves aside the $\t$ dependance.  The   Euclidean and Lorentz  are solved at once  by replacing 
$\o$ into $\eps \o$ in  both    equations for $T^z=T^\bz=0$ as it is obvious from  \eqref{TRz}.   
  It is useful to define the following derivation operation $\nabla$ for $d=3$:
\bea   \label{nablamuo} \nabla_\bz \mo  &  \equiv  &  \bp \mo + \mo  \p\m 
    -\m  \p \mo   
    \nn\\
    \nabla_z \mob    &\equiv &    \p \mob + \mob  \bp\mb 
    -\mb  \bp \mob
    \nn \\ \red 
      \nabla_z \mo  &\equiv  & \p \mo          -\red  \mb \bp \mo   \red +\mb \pa_0 \m-  \mbo\p\m
      \nn \\\red
      \nabla_\bz \mob  &\equiv  & \bp \mob         \red   - \m \p \mob     \red +\m  \pa_0 \mb-  \mo\bp\mb .
    \eea
    Notice that   $  \nabla_\bz \m\equiv   \bp \mo + \mo  \p\m 
    -\m  \p \mo  $ formally equates a  leaf  holomorphic reparametrization transformation of the Beltrami differential $\m$ with a parameter   $\mo$ (as can be verified from  \eqref{smu}).  
  
  The   $d=3$  condition $T^0=0$  is
 \bea\label{To}
   \bM
 \exp {\color{black} -\vp}\ \o^{{\color{black} z}}_{{\color{black} Z}} +\exp {\color{black} -\vpb}\ \o_{{\color{black} \bZ}}^{{\color{black} \bz}}
=0
\cr  
\o^z_0=
-     {\exp -\vpb}  \  2  \Dbz N
\cr
\o^\bz_0=
    {\exp -\vp}  \   2\Dz N.
\eM
\eea
The    6  conditions stemming from 
  $T^\bz=T^z=0$ are
\bea\label{Spinc3}
      \o ^0_\Zb \=  \Dbz \vp          -\frac   { \p\m}{\mmb}
       \nn\\
      \o ^0_Z       \= -\Dz {\color{black} \vpb}          +\frac   { \bp\mb}{\mmb}
     \nn\\
     \o_\Zb^z  \=   {\frac 1 N} \frac { \exp \vp  }
       {\mmb }  (\pa_o \m-  \nabla_\zb \mo )
      \nn\\
      \o_\Z^\bz  \=  -  {\frac 1 N}  \frac{ \exp \vpb   }
      {\mmb  }   (\pa_o \mb - \nabla_z {\color{black} \mob} )
      \nn\\
     \o^z_Z 
      \=  \frac { \exp -\vp }{ N }  \Big(         {  }  {\color{black} \mathcal{D}_{0}} \vp          {  -}  \frac {\nabla_z\mo}{\mmb }+       \o ^0_0\Big)
      \Big)\nn\\\o^\bz_\Zb 
     \=  \frac { \exp -\vpb }{ N }  \Big( {  -} {\color{black} \mathcal{D}_{0}} \vpb          {  +} \frac{\nabla_\bz\mob }{\mmb} +       \o ^0_0\Big)
      \Big).
      \eea
 Combining      the first equation in \eqref{To}  with   the sum and the difference of both   last   equations in \eqref{Spinc3} implies 
\bea
\o^0_0\= {\color{black} - \Big(} \demi  \frac{\nabla_z\mo   -  \nabla_\bz\mob }\mmb   {\blue-} \demi {\color{black} \mathcal{D}_{0}}( \vp-     \vpb ) {\color{black} \Big)}  
\nn\\
  \o^z_Z 
      \=  \frac { \exp \vp }{2 N }  \Big(          {\color{black} \mathcal{D}_{0}}(\vp+ \vpb )        -  \frac{  \nabla_z\mo  +    \nabla_\bz\mob}{\mmb}   
      \Big)
      \nn\\\o^\bz_\Zb 
     \=  \frac { \exp \vpb }{2 N }  \Big(   -      {\color{black} \mathcal{D}_{0}}(\vp+ \vpb )          +   \frac{ \nabla_\bz\mob  +        \nabla_z\mo }{\mmb}
      \Big).
      \eea
      One gets therefore
       \bea
 \label {Tva} 
 \o^a_\mu=
    {\color{black} 
    \begin{pmatrix} \frac{\epsilon e^{\phi}}{2N}(\mathcal{D}_{0}(\phi+\Bar{\phi})-\frac{\nabla_{z}\mu_{0}^{z}+\nabla_{\Bar{z}}\mu_{0}^{\Bar{z}}}{1-\mu_{\Bar{z}}^{z}\mu_{z}^{\Bar{z}}}) & \frac{\epsilon}{N}\frac{e^{\phi}}{{1-\mu_{\Bar{z}}^{z}\mu_{z}^{\Bar{z}}}}(\partial_{0}\mu_{\Bar{z}}^{z}-\nabla_{\Bar{z}}\mu_{0}^{z}) & -2e^{-\Bar{\phi}}\mathcal{D}_{\Bar{z}}N \\ \frac{\epsilon}{N}\frac{e^{\Bar{\phi}}}{{1-\mu_{\Bar{z}}^{z}\mu_{z}^{\Bar{z}}}}(\nabla_{z}\mu_{0}^{\Bar{z}} - \partial_{0}\mu_{z}^{\Bar{z}}) & \frac{\epsilon e^{\Bar{\phi}}}{2N}(-\mathcal{D}_{0}(\phi+\Bar{\phi})+\frac{\nabla_{z}\mu_{0}^{z}+\nabla_{\Bar{z}}\mu_{0}^{\Bar{z}}}{1-\mu_{\Bar{z}}^{z}\mu_{z}^{\Bar{z}}}) & 2e^{-\phi}\mathcal{D}_{z}N \\ \epsilon(\frac{\partial_{\Bar{z}}\mu_{z}^{\Bar{z}}}{1-\mu_{\Bar{z}}^{z}\mu_{z}^{\Bar{z}}} - \mathcal{D}_{z}\Bar{\phi}) & \epsilon(\mathcal{D}_{\Bar{z}}\phi - \frac{\partial_{z}\mu_{\Bar{z}}^{z}}{1-\mu_{\Bar{z}}^{z}\mu_{z}^{\Bar{z}}}) & \frac{\epsilon}{2}(\mathcal{D}_{0}(\phi-\Bar{\phi}) + \frac{\nabla_{\Bar{z}}\mu_{0}^{\Bar{z}} - \nabla_{z}\mu_{0}^{z}}{1-\mu_{\Bar{z}}^{z}\mu_{z}^{\Bar{z}}}) \end{pmatrix}}
\eea
where $\Phi=\vp+\vp$. 
       This concludes the quite simple determination  proof  of the      $d=3$     Spin connection    components  
      $\o_Z^z,  \ \o_\bZ^z,  \ \o_0 ^z,\
\o_Z^\bz,  \ \o_\bZ^\bz,  \ \o_0 ^\bz,\ 
\o_Z^0,  \ \o_\bZ^0,  \ \o_0 ^0  
 $
         as     displayed in~\eqref{Tva3} under the form   \def\DD{\mathbb{D}}
\bea
 \label {ointuitif}
\o^a_\mu=
{\color{black} \begin{pmatrix} \frac{\epsilon}{2N}\mathbb{D}_{0}\Phi & \frac{\epsilon}{N}\mathbb{D}_{0}\mu_{\Bar{z}}^{z} & -2\mathbb{D}_{\Bar{z}}N \\ -\frac{\epsilon}{N}\mathbb{D}_{0}\mu_{z}^{\Bar{z}} & -\frac{\epsilon}{2N}\mathbb{D}_{0}\Phi & 2\mathbb{D}_{z}N \\ -\frac{\epsilon}{2}\mathbb{D}_{z}\Phi & \frac{\epsilon}{2}\mathbb{D}_{\Bar{z}}\Phi & \frac{\epsilon}{2} \nabla.\mu_{0} \end{pmatrix}}  .
\eea

        \begingroup\raggedright
    
    \endgroup

  \end{document}